\documentclass[twocolumn]{aastex62}
\usepackage{color} 
\usepackage{tabularx} 
\usepackage{epsfig}
\usepackage{amsmath} 
\usepackage{amssymb} 
\usepackage{bm}
\usepackage{graphicx}
\usepackage{wasysym}
\usepackage{multirow}
\usepackage{threeparttable}
\graphicspath{ {./graphic/} }
\newcommand{\Ha}{H$\alpha$}
\newcommand{\Hb}{H$\beta$}
\newcommand{\Oiii}{[OIII]}
\newcommand{\Oii}{[OII]}
\newcommand{\dq}{$DQ $}
\newcommand{\qz}{$Q_z $}
\newcommand{\signm}{$\sigma_{NMAD}\ $}
\definecolor{purp}{HTML}{8904B1}
\newcommand{\Azrange}{1.55\leq z_{grism} \leq 1.9}
\newcommand{\Bzrange}{1.8\leq z_{grism} \leq 2.2}
\newcommand{\AIfrac}{$15_{-4.4}^{+9.4}\%$ }
\newcommand{\AIIfrac}{$12_{-3.5}^{+7.8}\%$ }
\newcommand{\AIIIfrac}{$14_{-4.3}^{+9.1}$ }
\newcommand{\BIfrac}{$21_{-5.5}^{+9.3}\%$ }
\newcommand{\BIIfrac}{$19_{-5.1}^{+8.7}\%$ }
\newcommand{\BIIIfrac}{$24_{-5.8}^{+8.9}$ }

\newcommand{\fieldIfracs}{$6.7^{+1.4}_{-1.0}\%$}
\newcommand{\fieldIIIfracs}{$6.9^{+1.4}_{-1.0}\%$}

\newcommand{\Afraccor}{$11_{-3.2}^{+8.2}$}
\newcommand{\Bfraccor}{$18_{-4.5}^{+7.8}$}

\newcommand{\fieldfraccor}{$5.0_{-0.8}^{+1.1}\%$}

\shortauthors{Watson et al.}

\begin{document}

\title{Galaxy Merger Fractions in Two Clusters at $z\sim2$ Using the Hubble Space Telescope}

\author{Courtney Watson}
\affil{George P. and Cynthia W. Mitchell Institute for Fundamental Physics and Astronomy, Department of Physics \& Astronomy, Texas A\&M University, College Station, TX 77843, USA}
\affil{Department of Astronomy, Yale University, New Haven, CT 06520, USA}

\author{Kim-Vy Tran}
\affiliation{George P. and Cynthia W. Mitchell Institute for Fundamental Physics and Astronomy, Department of Physics \& Astronomy, Texas A\&M University, College Station, TX 77843, USA}
\affiliation{School of Physics, University of New South Wales, Sydney, Australia}
\affiliation{Australian Astronomical Observatory}

\author{Adam Tomczak}
\affiliation{Department of Physics, University of California, Davis, One Shields Ave., Davis, CA 95616, USA}
\affiliation{Visiting astronomer, Cerro Tololo Inter-American Observatory, National Optical Astronomy Observatory, which are operated by the Association of Universities for Research in Astronomy, under contract with the National Science Foundation}

\author{Leo Alcorn}
\affiliation{George P. and Cynthia W. Mitchell Institute for Fundamental Physics and Astronomy, Department of Physics \& Astronomy, Texas A\&M University, College Station, TX 77843, USA}
\affiliation{LSSTC Data Science Fellow}

\author{Irene V. Salazar}
\affiliation{Department of Physics and Astronomy, Louisiana State University, Baton Rouge, Louisiana 70803, USA}

\author{Anshu Gupta}
\affiliation{School of Physics, University of New South Wales, Sydney, Australia}

\author{Ivelina Momcheva}
\affiliation{Space Telescope Science Institute, 3700 San Martin Drive, Baltimore, MD 21218, USA}

\author{Casey Papovich}
\affiliation{George P. and Cynthia W. Mitchell Institute for Fundamental Physics and Astronomy, Department of Physics \& Astronomy, Texas A\&M University, College Station, TX 77843, USA}
\affiliation{Visiting astronomer, Cerro Tololo Inter-American Observatory, National Optical Astronomy Observatory, which are operated by the Association of Universities for Research in Astronomy, under contract with the National Science Foundation}

\author{Pieter van Dokkum}
\affiliation{Department of Astronomy, Yale University, New Haven, CT 06520, USA}

\author{Gabriel Brammer}
\affiliation{Space Telescope Science Institute, 3700 San Martin Drive, Baltimore, MD 21218, USA}
\affiliation{Cosmic Dawn Center, Niels Bohr Institute, University of Copenhagen, Juliane Maries Vej 30, DK-2100 Copenhagen ?, Denmark}

\author{Jennifer Lotz}
\affiliation{Space Telescope Science Institute, 3700 San Martin Drive, Baltimore, MD 21218, USA}

\begin{abstract}

We measure the fraction of galaxy-galaxy mergers in two clusters at $z\sim2$ using imaging and grism observations from the {\it Hubble Space Telescope}.  The two galaxy cluster candidates were originally identified as overdensities of objects using deep mid-infrared imaging and observations from the {\it Spitzer Space Telescope}, and were subsequently followed up with HST/WFC3 imaging and grism observations.  We identify galaxy-galaxy merger candidates using high resolution imaging with the WFC3 in the F105W, F125W, and F160W bands.  Coarse redshifts for the same objects are obtained with grism observations in G102 for the $z\sim1.6$ cluster (IRC0222A) and G141 for the $z\sim2$ cluster (IRC0222B). Using visual classifications as well as a variety of selection techniques, we measure merger fractions of \Afraccor\ in IRC0222A and \Bfraccor\  in IRC0222B. In comparison, we measure a merger fraction of \fieldfraccor\ for field galaxies at $z\sim2$. Our study indicates that the galaxy-galaxy merger fraction in clusters at $z\sim2$ is enhanced compared the field population, but note that more cluster measurements at this epoch are needed to confirm our findings. 
\end{abstract}

\section{Introduction}

Our current model of galaxy formation is based on hierarchical merging where numerous small structures merge to form progressively larger systems \citep{Peebles1970}. Observational evidence of major and minor merging at all epochs \citep{Conselice2009,Lotz2008,Bundy2009,DeRavel2009,Lopez-Sanjuan2009,Williams2011,RyanJr.2008,Man2012,Man2016} confirms the importance of merging throughout cosmic time, especially at earlier epochs \citep{Kauffmann1996,Baugh1996}. In this picture of hierarchical formation, massive galaxies that dominate clusters also should continue to grow through accretion of satellite galaxies.  However, whether the galaxy merger fraction in clusters is enhanced or diminished relative to the field remains debated \citep{Lotz2013,Delahaye2017,Tran2005,Tran2008,VanDokkum1999,McIntosh2008}

Galaxy clusters in the local Universe contain hundreds to thousands of galaxies \citep{Zwicky1957,OmerJr.1965}, however mergers in local clusters are rare. Thus the ideal epoch for determining whether galaxy-galaxy merging is enhanced relative to the field is to study clusters early in their formation at $z>1$ when the clusters are dynamically young and the relative velocities are comparable to the galaxies' internal velocities.  We can then determine how galaxies interact with each other and their environment, and how these interactions affect their physical properties including star formation rates, sizes, and morphologies \citep{Brodwin2013,Papovich2012,Tran2010,Lotz2013,Delahaye2017}. However, identifying clusters at $z>1$ is challenging given increasing contamination by field galaxies along the line of sight and the rarity of clusters with increasing redshift.

N-body and hydrodynamical simulations of galaxy mergers predict an increase in the merger rate of field galaxies as a function of increasing redshift \citep{Rodriguez-gomez2015,Fakhouri2008}. However,  recent observational surveys find that the galaxy merger fraction appears constant, or even decreases, at $z\gtrsim 1$ \citep{Williams2011,RyanJr.2008,Man2012,Man2016}.  
Whether the merger fraction in galaxy clusters also increases with redshift is unclear \citep{Delahaye2017,Lotz2013, Lidman2013}.  Studies find evidence of mass growth through merging of red, luminous galaxies in groups and clusters up to $z\sim1$ \citep{Tran2005,VanDokkum1999,McIntosh2008}, but measurements of the galaxy merger fraction in clusters at $z>1$ is limited.

Studies that have investigated the merger rate within cluster environments in comparison to a field population often present findings that are in disagreement with each other. \cite{Lotz2013} and \cite{Coogan2018} found merger rates $\sim 3-10$ times higher than the field within a $z=1.62$ protocluster and $z=1.99$ cluster, respectively and \cite{Hine2016} found a 60\% enhancement of the merger rate in a $z=3.1$ proto-cluster compared to the field. However, \cite{Delahaye2017} sees no evidence of an increased merger fraction in comparison to the field for four clusters at $1.59<z<1.71$.  

Note each study uses different methods for identifying galaxy mergers that include close pair, morphological, flux ratio, or stellar mass ratio selection. There are also differences in the use of either photometric \citep{Lopez-Sanjuan2009,Man2012,Man2016,Conselice2009,Williams2011}  or spectroscopic redshifts \citep{Tran2008,DeRavel2009}, or some either/or combination \citep{Lotz2008,Lotz2013,Bundy2009,Delahaye2017} in the selection of galaxy mergers. The differences in the selection of merger candidates throughout the various studies makes it difficult to perform direct comparisons of merger rates. To compensate for this, we perform a broad analysis, selecting merger candidates using a variety of selection methods, allowing better comparison to the merger fractions measured in these previous studies.

In this paper we combine high-resolution near-infrared HST imaging with grism spectroscopy to measure the galaxy-galaxy merger fraction in two candidate galaxy clusters at $z\sim 2$ and investigate whether the merger rate is enhanced or diminished relative to the field population. In \S 2 we summarize our imaging and grism observations for our two clusters and field sample. \S 3 describes our analysis and the selection of merger candidates. \S 4 discusses our results and how we compare to previous studies. We assume a flat $\Lambda$CDM cosmology with $\Omega_M = 0.3$, $\Omega_{\Lambda} = 0.7$, and $H_0 = 70$km s$^{-1}$ Mpc$^{-1}$.

\begin{figure*}[ht]
\plotone{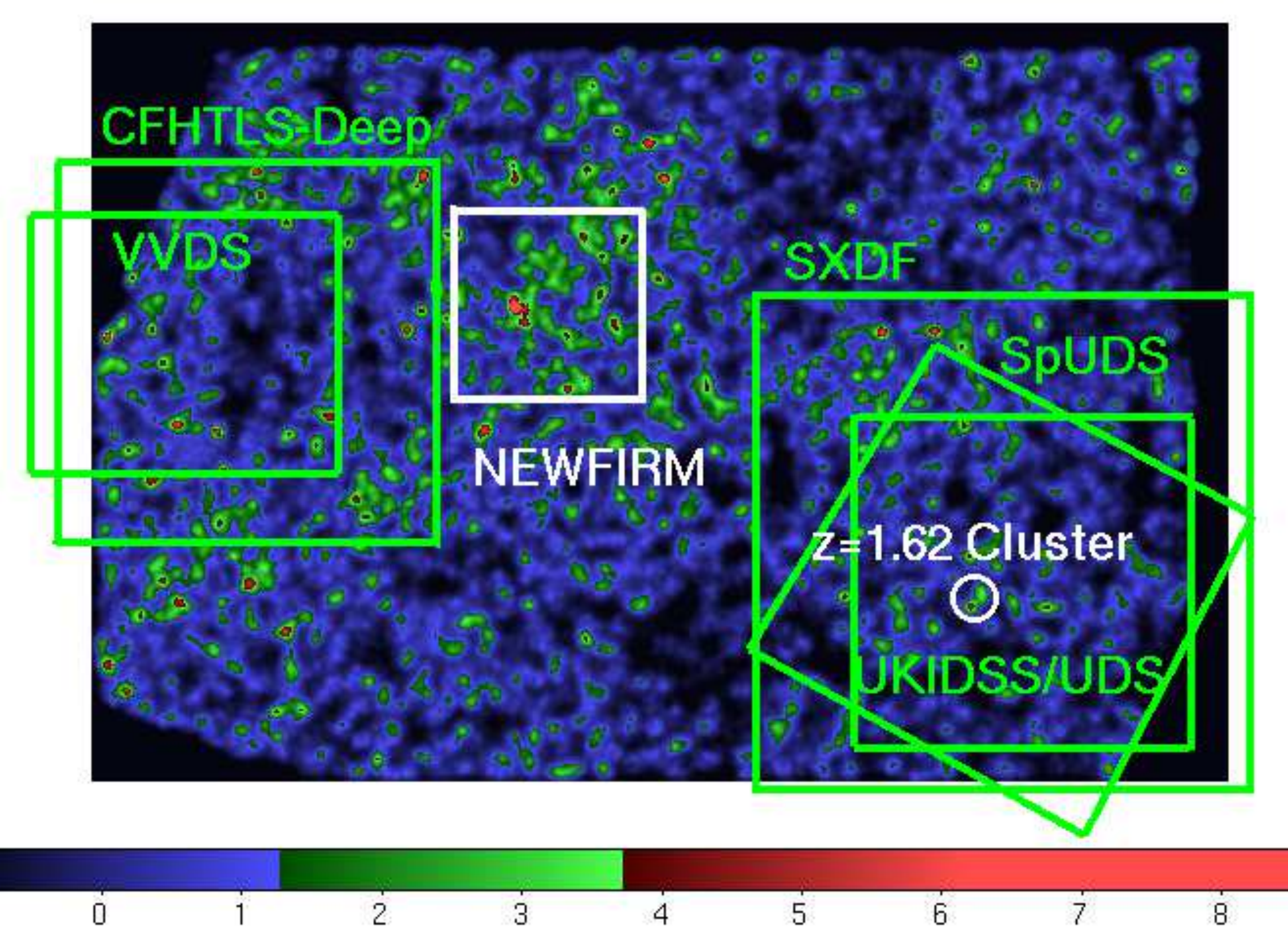}
\caption{Surface density of galaxies at $1.3 < z < 2.0$ in a $3^{\circ} \times 2^{\circ}$ portion of the XMM-LSS SWIRE field using {\it Spitzer}/IRAC. The scale indicates the number of standard deviations ($\sigma$) above the mean field surface density. The area targeted by the NEWFIRM Cluster Survey, which houses the largest overdensity ($>10 \sigma$) in the XMM-LSS SWIRE field, is indicated. The NEWFIRM field includes $\sim 10$ cluster candidates with significance equal to the $z=1.62$ UDS cluster.  \label{NCS}}
\end{figure*}

\section{NEWFIRM Cluster Survey (NCS)}

\subsection{XMM-LSS field : 1.6 $\mu$m bump \label{sec: XMMLSS}}

The NEWFIRM Cluster Survey (NCS) targeted a 28$\times$28\arcmin\ portion of the XMM Large Scale Structure Survey (XMM-LSS) field using data from the {\it Spitzer} Wide-Area Infrared Extragalactic survey \citep[SWIRE:][]{Lonsdale2003} (see Fig. \ref{NCS}). Galaxy cluster candidates at $z>1.3$ initially identified from the entire SWIRE field as overdensities of galaxies with red colors between $3.6\mu$m and $4.5\mu$m, indicating the peak in their star formation at $1.6\mu$m, were subsequently targeted for follow up NEWFIRM imaging. As discussed by \cite{Papovich2008} roughly $\gtrsim$ 90\%\ of galaxies at $z > 1.3$ have {\it Spitzer}/IRAC $([3.6]-[4.5]) > -0.1$ mag making this an effective criterion for selecting candidate galaxy structures at high redshift. In the initial search for overdensities using the SWIRE data, candidate galaxy clusters at $z>1.3$ were selected as overdensities with $\gtrsim$ 30 objects with {\it Spitzer}/IRAC colors $([3.6]-[4.5]) > -0.1$ mag within radii of 1.4' \citep{Papovich2008,Papovich2010}.

Preliminary galaxy density maps were made by counting the number of objects with $([3.6]-[4.5]) > -0.1$ mag and $r < 22.5$ mag within apertures of 1.4\arcmin\ radii (roughly 0.7 proper Mpc at $z = 1.5$) across the full SWIRE field. This process successfully preselected a massive ($>10^{14}$ $M_{\odot}$) $z = 1.62$ galaxy cluster which was later confirmed spectroscopically and found to coincide with a faint, diffuse X-ray source likely originating from a hot intra-cluster medium \citep{Papovich2010, Tanaka2010}. Within these density maps we isolated a 30$\times$30\arcmin\ subregion hosting 10 candidate overdensities at/above the significance of the aforementioned confirmed galaxy cluster \textbf{($>3\sigma$)} (see Fig. \ref{NCS}). The 10 candidate overdensities were subsequently targeted for follow-up with NEWFIRM.

The two galaxy associations studied in this work (IRC0222A and IRC0222B) were initially identified as candidate high-redshift overdensities in the SWIRE XMM-LSS field. In October 2010 we targeted this subregion for near-infrared photometric observations with the NOAO Extremely Wide-Field Infrared Imager (NEWFIRM\footnote{\url{https://www.noao.edu/ets/newfirm/}}). Tables \ref{newfirm1} and \ref{newfirm2} provide a summary of the NEWFIRM observations and photometry. Ground-based optical imaging in the {\it ugriz} filters for this field are drawn from the CFHT Legacy Survey (CFHTLS\footnote{\url{http://www.cfht.hawaii.edu/Science/CFHTLS/}}).

\begin{deluxetable}{ccc}
\tablecaption{NEWFIRM Observations \label{newfirm1}
}
\tablehead{\colhead{Observation Date} & \colhead{Filter\tablenotemark{a}} & \colhead{Exposure Time [hr]}}
\startdata
2010-10-15 & J1 & 1.0 \\
2010-10-15 & J2 & 1.0 \\
2010-10-15 & J3 & 2.4 \\
2010-10-15 & H & 0.8 \\
2010-10-16 & J1 & 1.1 \\
2010-10-16 & J2 & 1.3 \\
2010-10-16 & J3 & 1.0 \\
2010-10-16 & Ks & 0.9 \\
2010-10-17 & J1 & 1.5 \\
2010-10-17 & J2 & 1.0 \\
2010-10-17 & J3 & 1.2 \\
2010-10-17 & H & 1.2 \\
2010-10-18 & J1 & 1.0 \\
2010-10-18 & J2 & 1.3 \\
2010-10-18 & J3 & 1.2 \\
2010-10-18 & H & 1.0 \\
2010-10-18 & Ks & 0.7 \\
2010-10-19 & J1 & 1.3 \\
2010-10-19 & J2 & 1.7 \\
2010-10-19 & J3 & 2.7 \\
2010-10-19 & Ks & 0.3 \\
\enddata
\tablenotetext{a}{Name of photometric bandpass}
\end{deluxetable}

\subsection{Near-Infrared Imaging with CTIO/NEWFIRM}

As described above, the IRC0222A/B candidate clusters are initially identified as overdensities via the $1.6\mu$m bump technique. To further refine our cluster selection, we obtain near-infrared imaging with NEWFIRM. NEWFIRM has a custom set of 5 medium-bandwidth near-IR filters that provide higher spectral resolution which have been shown to improve the accuracy of photometric redshifts at $z>1$ \citep{VanDokkum2009, Whitaker2011, Bezanson2015, Straatman2016}.  The NIR filters span $\lambda$=1.0-1.8$\mu$m and pinpoint the location of the 4000 $\mathring{\text{A}}$ break.  With NEWFIRM imaging, we are able to  measure photometric redshifts with accuracies $\Delta z/(1 + z) \simeq 0.02$ for $1.3<z<2.0$. 

\begin{deluxetable}{ccc}
\tablecaption{NEWFIRM Photometry \label{newfirm2}}
\tablehead{\colhead{Filter\tablenotemark{a}} & \colhead{FWHM [arcsec]} & \colhead{Completeness\tablenotemark{b}}}
\startdata
J1 & 1.31 & 23.3 \\
J2 & 1.35 & 22.9 \\
J3 & 1.32 & 22.7 \\
H & 1.26 & 22.1 \\
Ks & 1.33 & 21.8 \\
\enddata
\tablenotetext{a}{Name of photometric bandpass.}
\tablenotetext{b}{80\% point source limiting magnitude [AB]}
\end{deluxetable}

Furthermore, with a 28$\times$28\arcmin\ field of view NEWFIRM is able to observe the full subregion with a single pointing (see Fig. \ref{NCS}). This field was observed for 5.9, 6.3, and 8.5 hours in each of the medium-band $J_1$, $J_2$, and $J_3$ filters, 3.0 hours in the broadband $H$ filter, and $1.9$ hours in the $K_{\rm s}$ filter. All NEWFIRM imaging data were reduced using the {\tt nfextern} software package, a suite of standard data reduction tasks tailored specifically for the NEWFIRM observations and written in {\tt IRAF}\footnote{Image Reduction and Analysis Facility: \url{http://iraf.noao.edu/}}. A guidebook for using {\tt nfextern} can be found at the website for NEWFIRM\footnote{\url{https://www.noao.edu/ets/newfirm/}}.

Combining these new near-IR images with the existing optical imaging from CFHTLS we follow similar procedures as used in recent extragalactic photometric surveys \citep[e.g.,][]{Whitaker2011, Skelton2014, Straatman2016}.
Photometry is measured on PSF-matched images in fixed circular apertures of $d=3$\arcsec\ for objects detected in the NEWFIRM $K_{\rm s}$ image.
We next use the SED-fitting software Easy and Accurate Redshifts from Yale \citep[{\tt EAZY}:][]{Brammer2008} in order to estimate photometric redshifts for all objects.

Using the NEWFIRM photometric redshifts, we generated new galaxy density maps in each narrow ($\Delta z$=0.1) redshift slice between $1 < z < 2.5$ with an overlap of $dz$=0.05. We determine galaxy overdensities by counting galaxies in circles with $r = 1.4$\arcmin\ across the full field of view. The near IR imaging confirms two overdensities at photometric $z \approx 1.6$ and $z \approx 1.9$, hereafter IRC0222A and IRC0222B, respectively. We obtain further observations of these two candidate galaxy clusters using the Hubble Space Telescope. 

\section{Observations \& Analysis} 

\subsection{Hubble Space Telescope Observations}

All near-infrared HST observations (GO proposal 12896; PI Tran) were obtained using the Wide Field Camera-3 (WFC3) where imaging is taken in three wide filters (F105W, F125W, and F160W). Table \ref{filter} summarizes the wavelength ranges, zeropoints, and exposure times for each WFC3 filter.  We work with two fields of view, each centered on the galaxy cluster candidates: IRC0222A (2:22:03.5, -4:12:06.0) at $z\sim 1.9$, and IRC0222B (2:22:17.3, -4:21:45.0) at $z\sim 2.0$, hereafter referred to as IRC-A and IRC-B, respectively. 

Our WFC3 grism observations were reduced using the pipeline developed by the 3D-HST team \citep{VanDokkum2011,Brammer2012a,Skelton2014,Momcheva2015}. The 3D-HST pipeline provides low-resolution, spatially resolved spectra through the combination of multi-wavelength imaging with the grism spectroscopy. They also employ the use of interlacing, rather than drizzling, which provides the highest resolution imaging. Objects are identified using the multi-wavelength reference images and every spectrum is extracted simultaneously to account for contamination between overlapping spectra. 

\begin{table}[h]
\centering
\caption{WFC3 Filter Information}
\begin{tabular}{crrr}
 \hline\hline 
\rule{0pt}{15pt} Filter& Wavelength&	Exposure& Zeropoint\\
$\quad$& Coverage ($\mu m$)&	Time ($sec$)&	(AB)\\ [0.2em]\hline
F105W&	$1.1-1.4$&	$2174$&	$26.2687$\\ 
F125W&	$1.2-1.6$&	$1324$&	$26.2303$\\ 	
F160W&	$1.4-1.7$&	$2624$&	$25.9463$\\ 	
G102&	$0.8-1.15$&	$24071$&	 $26.2687$\\ 	
G141&	$1.075-1.7$&	$19847$&	$26.4524$ \\ [1em] \hline
\end{tabular}
\label{filter}
\end{table}

\begin{figure*}[ht!]
\plottwo{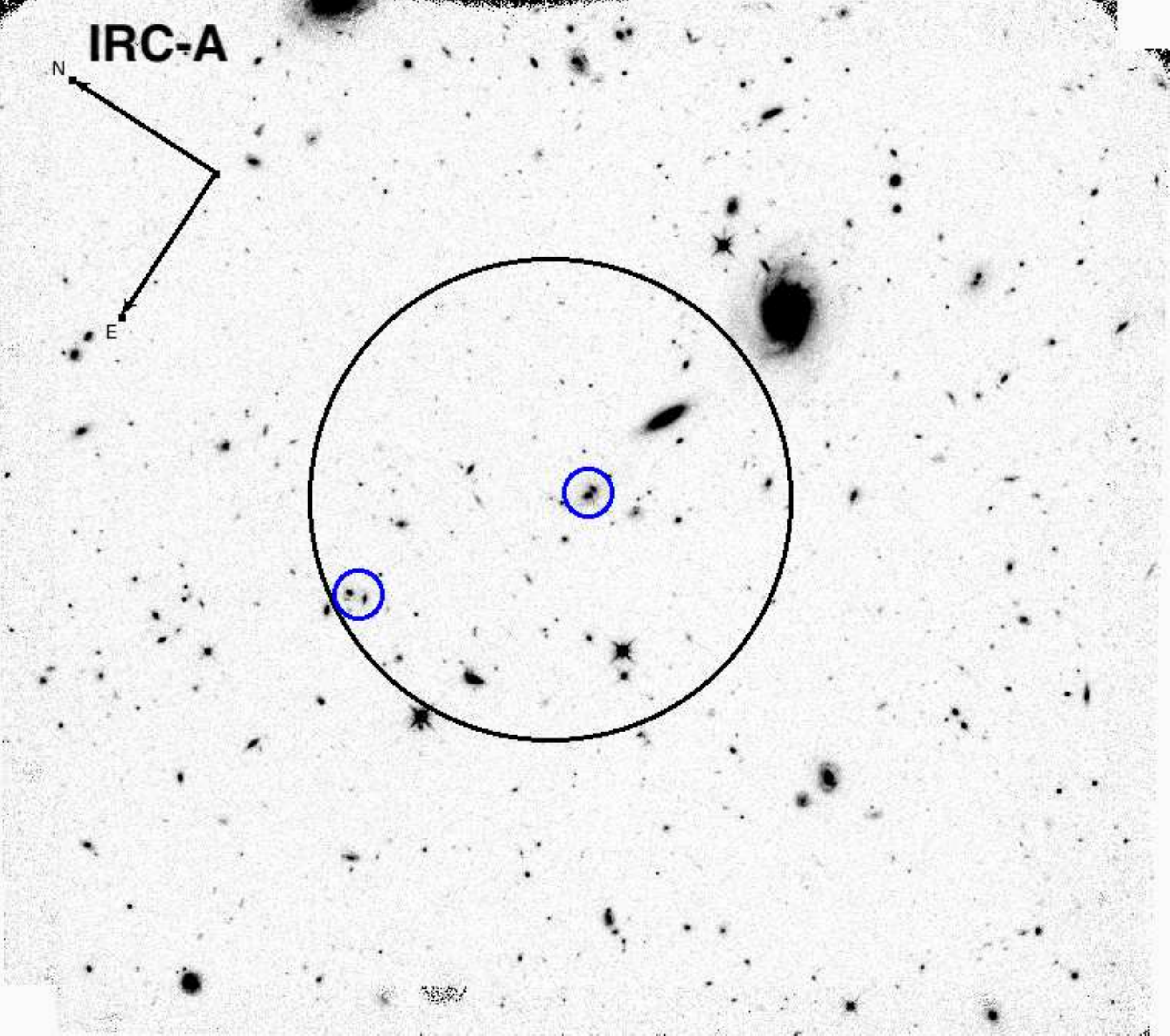}{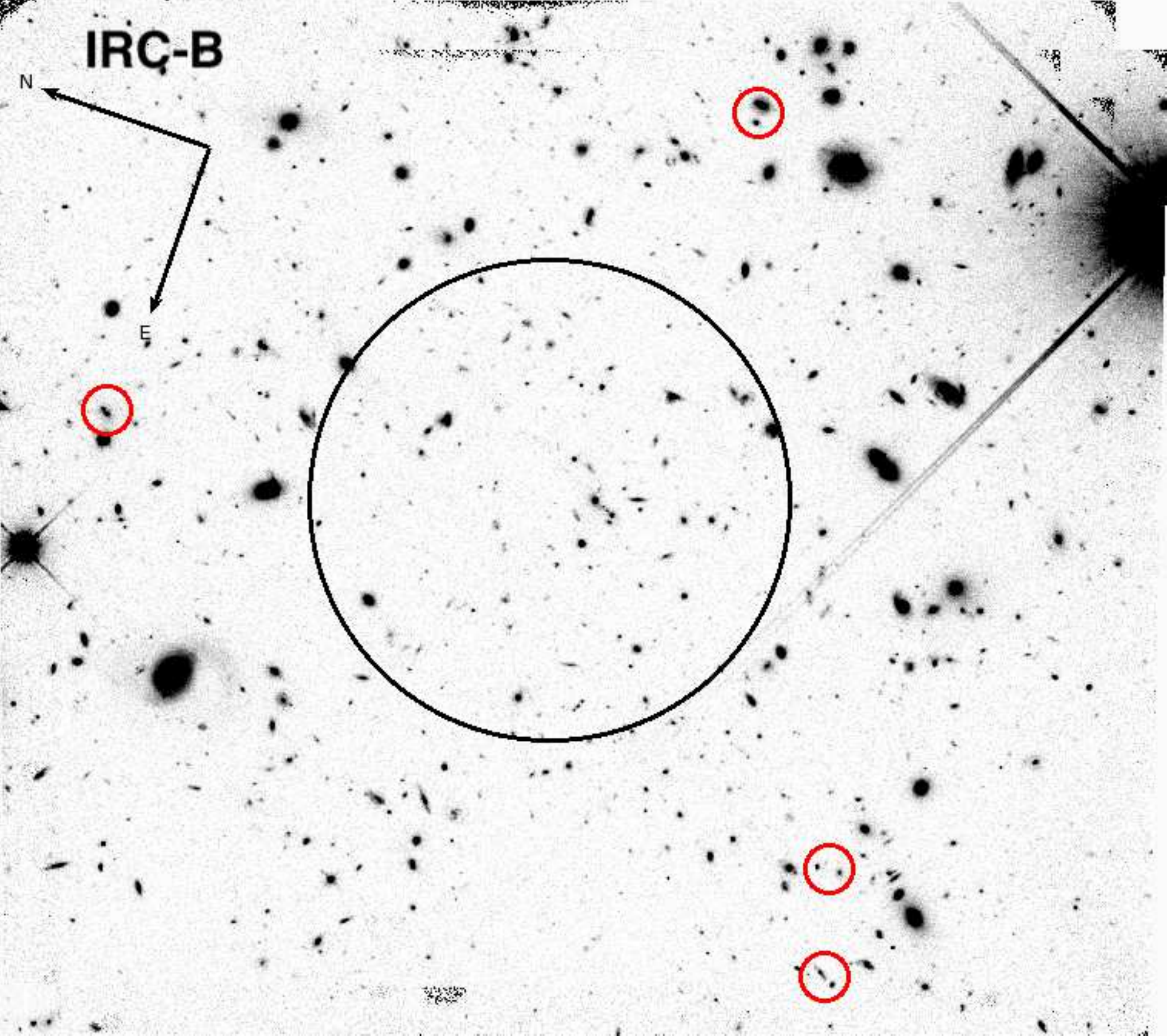}
\caption{Composite mosaics (F105W, F125W, F160W) for \emph{Left}: IRC0222A ($z\sim 1.6$); \emph{Right}: IRC0222B ($z\sim 1.9$); hereafter we refer to these candidate clusters as IRC-A and IRC-B respectively. Both fields measure $146\arcsec\times129\arcsec$. We have included a compass in the upper left corner to show the direction of N/E. The black circle is centered on the candidate cluster's coordinates, i.e. the peak of the detected overdensity (see \S \ref{sec: XMMLSS}), with a radius of $30\arcsec$, corresponding to the typical virial size of a cluster. The colored regions indicate the locations of cluster merger candidates selected via photometric or grism redshifts (see \S \ref{sec:III}) \label{rgbs}}
\end{figure*}

\subsubsection{WFC3 Imaging}

To analyze the WFC3 photometry, we follow the method described in \cite{Skelton2014}.  We apply a number of corrections to improve the data's quality and produce the final data products: masking satellite trails, persistence correction, sky-subtraction, flat-field re-application, initial astrometric alignment, and additional cosmic ray and bad pixel rejection. 

We began with \verb|Astrodrizzle|, using the default parameters and changing the bit value to 8192, \verb|final_wht_type|$=$\verb|IVM|, and \verb|final_pixfrac|$=$\verb|0.8| \citep{Skelton2014}, with an initial run to remove cosmic rays and bad pixels. This allowed for improved alignment within \verb|Tweakreg|. The initial drizzling also enabled the use of the F160W image as a reference image to improve alignment in \verb|Tweakreg|. The initial run produced ``crclean'' images, indicating they were then free of most of their cosmic rays. 

After the cosmic rays were removed, we used \verb|Tweakreg| to align all crclean images within a given filter to a common World Coordinate System (WCS) using the F160W image as a reference for alignment. This means that the F125W and F105W images were aligned according to the coordinate information contained in the F160W header.  Once aligned, the headers of all the crclean images were updated with the new coordinate information and we propagated the new coordinate information back to the original {\tt \_flt.fits} images using \verb|Tweakback|.

Using the newly updated {\tt \_flt.fits} images, we reran \verb|Astrodrizzle| to combine all the images within each filter. For this final run, we used the same parameters described earlier with the addition of final\_scale$=$0.03, to set the final pixel scale at $0.03\arcsec$ pixel. The resulting mosaics for the F105W, F125W, and F160W filters comprised an area of $4864\times 4308$ pixels or $146''\times129''$. Figure \ref{rgbs} shows the combined mosaics with the center of the cluster indicated by the black circle. The smaller, colored circles indicate identified merger candidates and are described in \S3.3. 

\subsubsection{Photometry}

For both cluster fields we use the HST/WFC3 F160W image for source detection. All ground-based optical and near-infrared images (u-Ks bands) are resampled to the pixel scale of HST imaging (0.03\arcsec/pixel). We utilize the SCAMP software \citep{Bertin2006a} to correct for small astrometric distortions, such as offsets and rotations, among the final mosaics.

\begin{figure}
\plotone{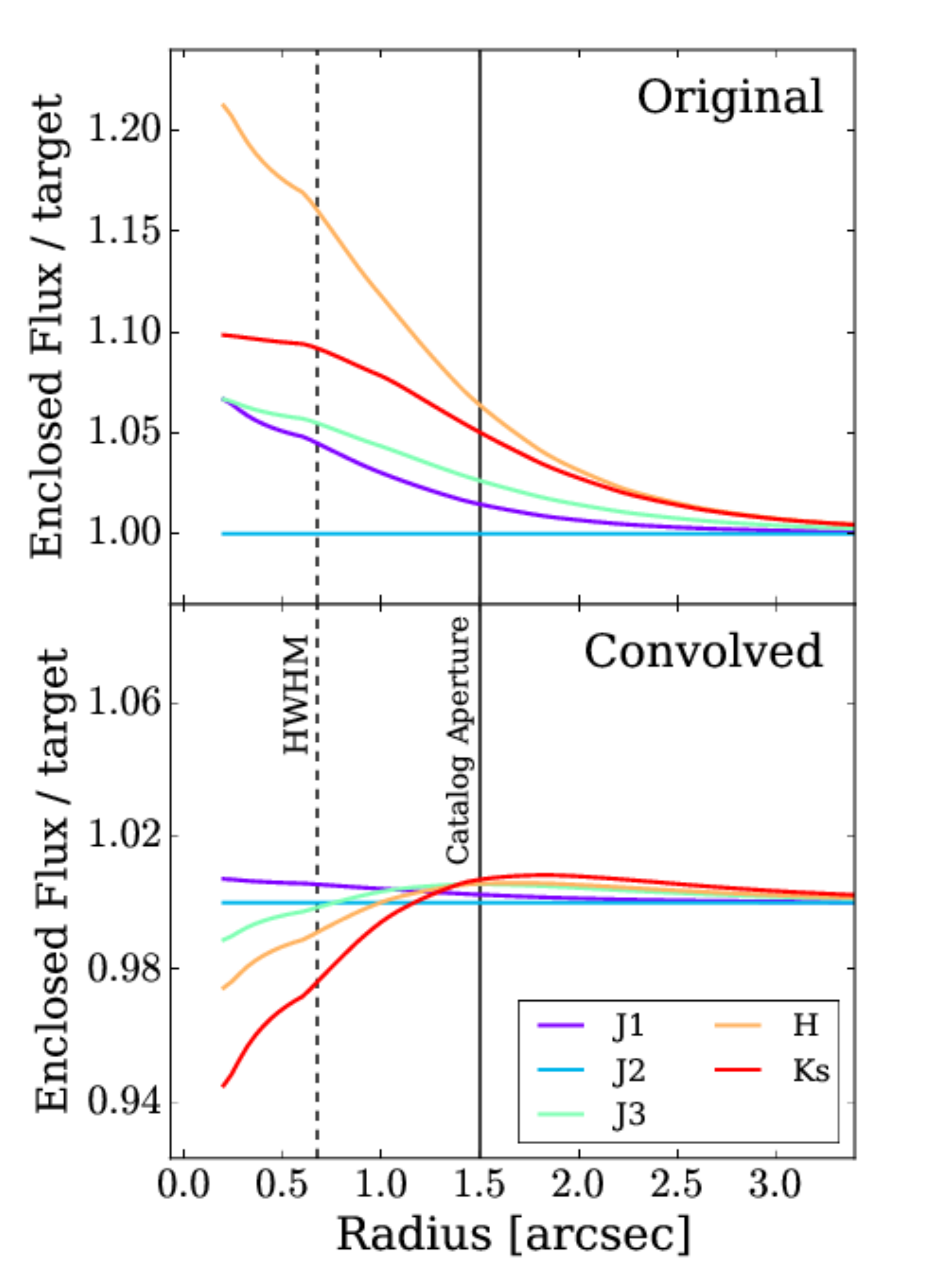}
\caption{Both panels plot the growth curve of each NEWFIRM image divided by the growth curve of the "target" PSF (i.e. the image with the largest FWHM) as a function of radius. The top panel shows these curves before PSF matching and the bottom shows it after matching. The dashed vertical line shows the size of the "target" PSF (half-width-half-max: HWHM) and the solid vertical line shows the aperture radius used for the photometric catalog. \label{psf}}
\end{figure}

We generate point spread functions (PSFs) for each image by stacking a selection of point sources in each image. These PSFs are used to construct convolution kernels (using {\tt scikit-image} in Python: \cite{vanderWalt2014}) in order to smooth all optical-NIR images to the largest PSF. Once images have been PSF-matched, photometry is extracted in fixed circular apertures ($r=1.7$\arcsec) by running Source Extractor \citep{Bertin1996} in dual-image mode using the F160W image for source detection. Figure \ref{psf} shows the accuracy of our PSF matching process. 

Due to the significantly larger PSFs of {\it Spitzer}/IRAC images ($\sim 2$\arcsec ) we use the {\tt T-PHOT} software \citep{Merlin2015a} to measure photometry. {\tt T-PHOT} is designed to extract photometry from images that suffer from significant blending by making use of a higher-resolution image as a prior. Positions and morphologies of objects from the high-resolution image (in our case the F160W image) are taken from a segmentation map produced by SExtractor. Then, a low-resolution model of each object is created using a user-provided convolution kernel. These models are then fit to the low-resolution IRAC image in a simultaneous fashion, allowing the profiles of neighboring objects to overlap. Once a global solution is found {\tt T-PHOT} reports the best-fit total flux for each object.

From these flux measurements we were able to obtain photometric redshifts and rest-frame colors using EAZY \citep{Brammer2008}. Stellar population properties (e.g. stellar masses) were estimated using the Fitting and Assessment of Synthetic Templates Software(FAST:\cite{Kriek2009}). For this portion of the analysis we adopt \cite{Bruzual2003} template library assuming a \cite{Chabrier2003} stellar initial mass function, constant solar metallicity, delayed exponential star-formation histories ($\rm SFR \propto t \times e^{-t/\tau}$), and a \cite{Calzetti2000} dust extinction curve.

\begin{figure*}[ht!]
\plottwo{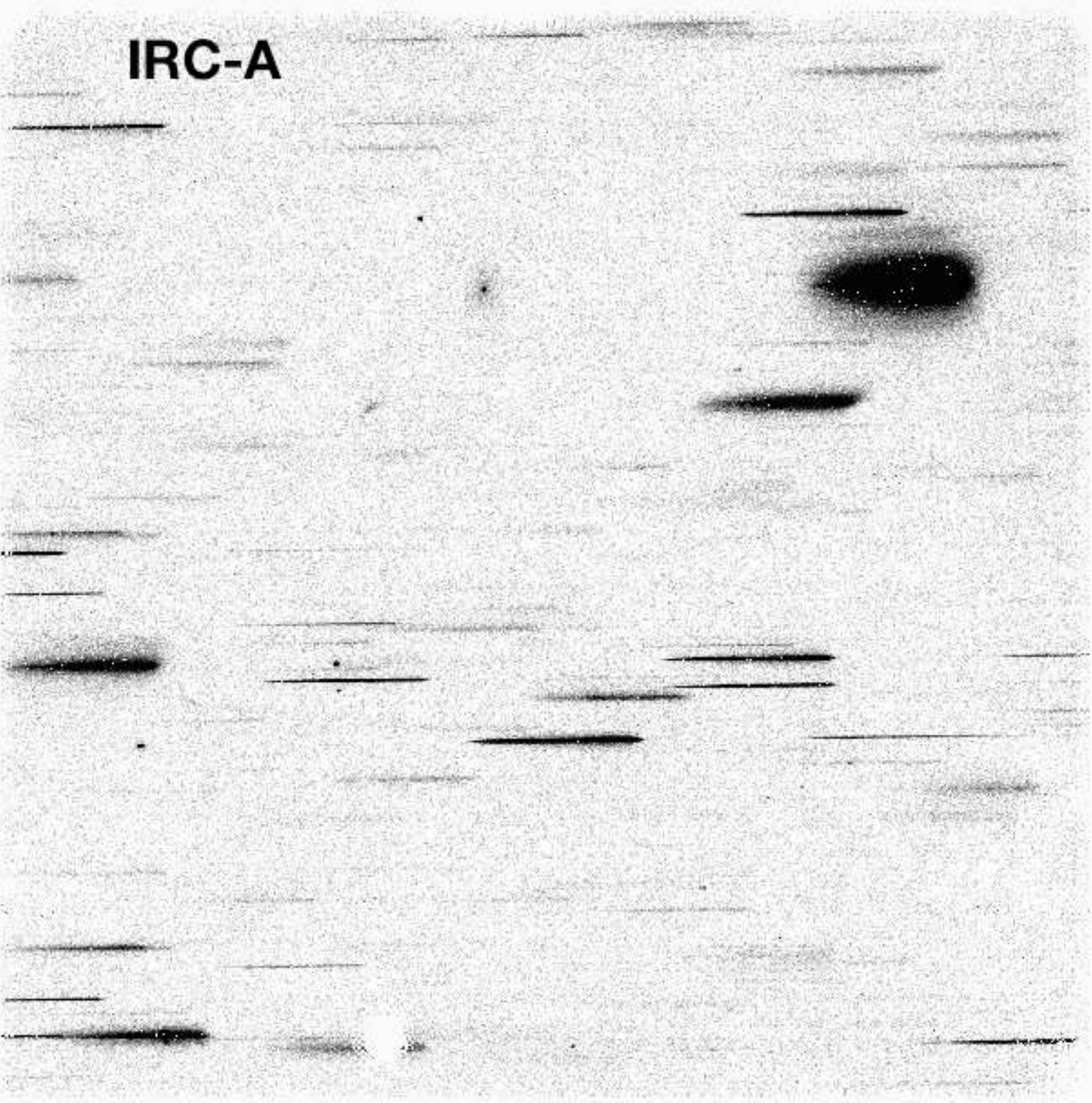}{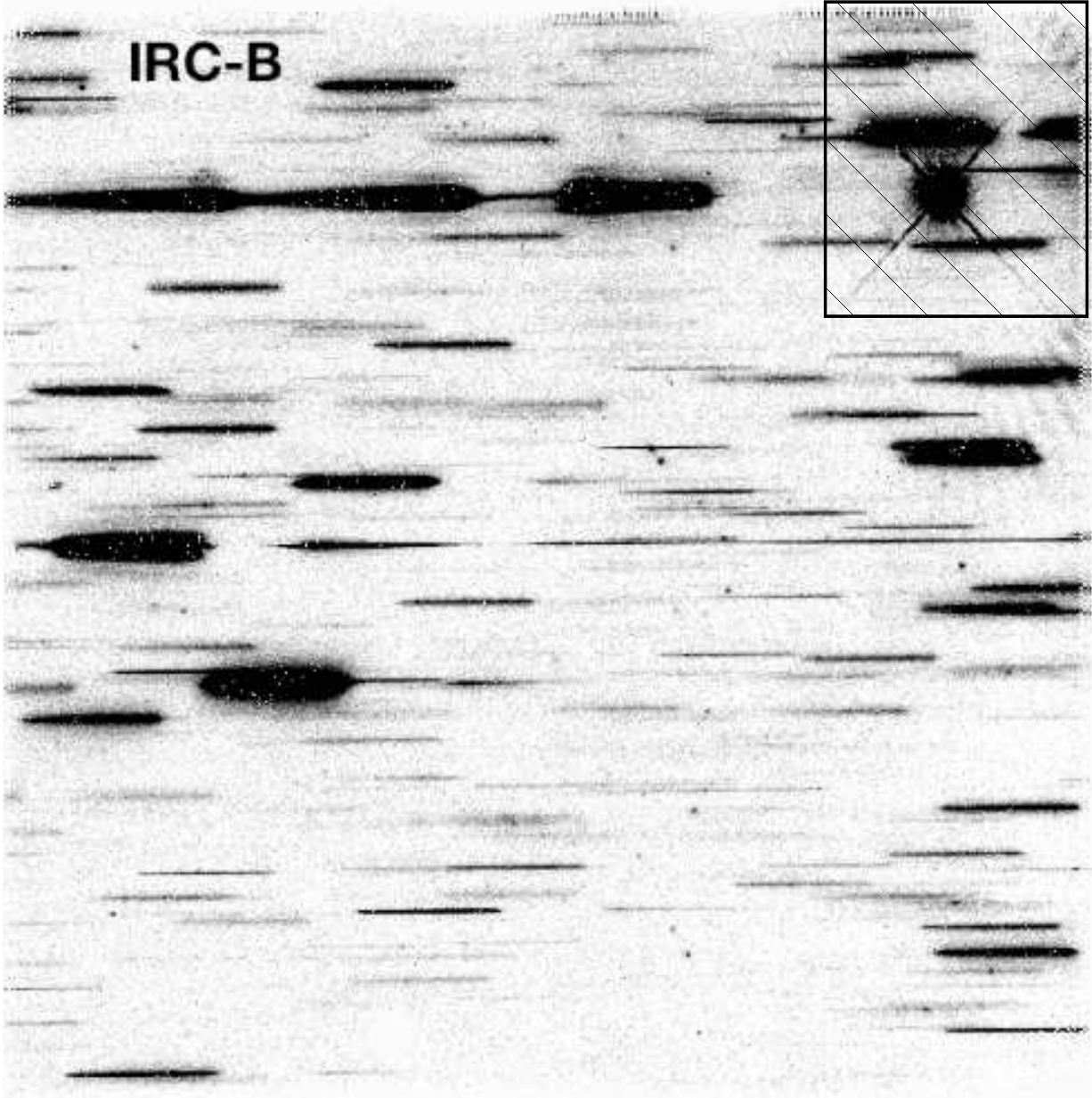}
\caption{The full interlaced grism image for \emph{Left}: IRC-A ($1.0 < z < 1.9$); \emph{Right}: IRC-B ($1.6 < z < 2.4$). Both fields measure $149\arcsec \times 125\arcsec$ . The hatched region in IRC-B indicates the region that was masked out in order to reduce contamination due to the star for the selection of our merger candidates. \label{gris}}
\end{figure*}

\subsection{Grism Observations}

We use the 3D-HST pipeline \citep{VanDokkum2011,Brammer2012a,Skelton2014,Momcheva2015} to reduce the grism observations for both cluster fields. For information on the reduction process, we refer the reader to \cite{Momcheva2015}. The observations were taken with G102 and G141 grisms on the WFC3 camera. Grisms are placed in the path of the incoming light to obtain spectra of all objects in the field of view simultaneously. 

For galaxy cluster candidate IRC-A at $z\sim1.6$, we use the G102 grism in combination with the F105W filter for direct imaging.  The G102 grism covers a wavelength range of 800--1150nm with a dispersion of 2.45nm/pixel.  For IRC-B at $z\sim2$, we use the G141 grism and F140W filter. The G141 grism covers a range of 1075--1700nm, with a dispersion of 4.65nm/pixel. The wavelength ranges of both grisms enable us to see emissions of \Ha, \Hb, \Oiii, and \Oii, elements essential in the galaxy formation process.  Figure \ref{gris} shows the full interlaced grism images for both cluster candidates, each measuring $2488\times 2092$ pixels, or $149\arcsec \times 125\arcsec$ ($0.06\arcsec/$pixel scale).

\subsubsection{WFC3 G141 \& G102 Redshifts}\label{sec:flags}

Redshifts are measured by fitting Spectral Energy Distributions (SEDs) to the grism spectroscopy \citep[see][]{Brammer2012a,Momcheva2015} and identifying spectral features including stellar continuum and emission lines. The grism spectra are matched to objects detected in the direct imaging. To determine the quality of the fits, we began with visual inspection of each grism image. Grism imaging and spectroscopy for all candidate cluster members is shown in Figures \ref{gris_pan_A1}-\ref{gris_pan_B3} in the Appendix. Our primary concern is determining whether the redshift identification is clearly affected by any error in the spectrum. To quantify the accuracy of both the data quality and the redshift reliability, we assign flags to denote the level of robustness of the measurements.

The data quality (\dq) flags are used to note whether the spectra were affected by known failure modes. Common failures that affect the data's quality include: incomplete masking of 0th-order spectra that can mimic emission lines, residuals from the spectra of very bright stars that may not be subtracted properly, and instances where corrupted photometric measurements lead to errors in the spectral fit. Objects are assigned a \dq\ value of either ``1'' or ``0'' corresponding to ``good'' or ``bad'', respectively. In all the figures, tables, and calculations presented here, we use only \dq\ flag values of 1, unless otherwise noted.

Redshift quality (\qz) flags are assigned as values of ``-1''  to ``3'' to indicate reliability of the grism measurement.  \qz\ flags are assigned based on visual inspections of the 1D and 2D spectra, comparison to the model data, and comparison to the photometric redshift data.  Objects flagged with a redshift quality (\qz) of ``3'' indicate a robust grism measurement, while those with \qz\ of ``2'' are less robust, but still reasonably reliable. \qz\ flag values of ``1'' or ``0'' specify the lowest reliabilities and a value of ``$-1$'' indicates an object is a star. 

All of the visual inspections and flag assignments described above were completed by CW, KT, LA, and IS.  

\subsection{GMOS}

Observations of the NCS fields were taken on GMOS-South in November 2012. The data were processed using the Gemini IRAF data reduction pipeline for GMOS, which included flat-fielding, wavelength calibration using arc lamps, and sky subtraction. At the time of our observations, the GMOS-S detectors had low quantum efficiency at $\lambda\gtrsim7000$\AA\ which limited our abilities to detect targeted lines for galaxies at $z>1.5$.  

In the NCS field, we measured the redshifts of 6  objects at $z=0.17, 0.32,0.33, 0.64$, and $0.80$ using \Ha/[NII] and \Oiii/\Hb. Four objects at $z>1$ had single, unidentified emission lines. No $z>1$ objects displayed multiple emission lines with which we could identify a grism redshift. Due to the non-detections of $z>1.5$ galaxies and the confirmation of several low-redshift galaxies, we infer that the targeted cluster galaxies are not at $z<1.5$, i.e. they are likely to be cluster members at $z>1.5$.

\begin{deluxetable*}{cccccccc}
\tablecaption{GMOS Observations \label{gmos}
}
\tablehead{\colhead{Field} & \colhead{Mask} & \colhead{RA} & \colhead{Dec} & \colhead{Mask PA} & \colhead{Observation Date} & \colhead{Central Wavelength [\AA]} & \colhead{Exposure Time [hr]}}
\startdata
UDS & Mask1 & 02:18:23.568 & -05:10:29.05 & 154 & 2012-11-14 & 8550 & 2.67 \\
UDS & Mask2 & 02:18:23.568 & -05:10:29.05 & 154 & 2012-11-14 & 8550 &  2.67 \\
NCS & Mask7 & 02:22:18.024 & -04:21:34.90 & 249.5 & 2012-11-15 & 8550 &  1.33 \\
NCS & Mask6 & 02:22:03.105 & -04:11:22.97 & 180 & 2012-11-15 & 8550 &  1.33 \\
NCS & Mask5 & 02:23:20.389 & -04:28:48.37 & 120 & 2012-11-15 & 8550 &  1.33 \\
UDS & Mask1 & 02:18:23.568 & -05:10:29.05 & 154 & 2012-11-16 & 8550 &  1.33 \\
UDS & Mask2 & 02:18:23.568 & -05:10:29.05 & 154 & 2012-11-16 & 8550 &  2.67 \\
UDS & Mask4  & 02:18:19.892 & -05:10:32.48 & 289 & 2012-11-16 & 8550 & 0.67 \\
\enddata
\end{deluxetable*}
 
\subsection{Field Sample}

Our field population is selected from the FourStar Galaxy Evolution Survey (ZFOURGE\footnote{\url{https://zfourge.tamu.edu}}) \citep{Straatman2016} in the COSMOS \citep{Scoville2007} and CDFS \citep{Giacconi2002} fields. The ZFOURGE coverage of both COSMOS and CDFS is approximately $10'\times10'$ and $>10$ times larger than the cluster field of view.  ZFOURGE provides precise photometric redshifts \citep[$\sigma_{\rm z}\sim2$\%;][]{Nanayakkara2016a} for galaxies spanning a wide range in redshift. In this paper, we use publicly available photometric catalogs from the ZFOURGE survey prepared by \cite{Straatman2016}. Grism redshifts for COSMOS and CDFS come from the available 3D-HST catalogs \citep{VanDokkum2011,Brammer2012a,Momcheva2015}.

The COSMOS field contains at least one known cluster within our redshift range at $z=2.095$ with 57 spectroscopically confirmed members \citep{Yuan2014}. To assess the best method for removing any potential impact by the cluster on the galaxy pair fraction, we first calculated the pair fraction using the full catalog of objects, including the known cluster members. We then implemented two methods to remove cluster members: (1) Rejecting objects with redshifts within 3$\sigma$ of the median redshift ($z=2.095$), where $\sigma = 0.00578$ is the standard deviation of redshift within confirmed cluster members, and (2) directly removing objects flagged as known cluster members by \cite{Yuan2014} in the ZFIRE catalogs \citep{Nanayakkara2016a}. 

We calculate the median pair fraction in the COSMOS field, including the cluster galaxies, for F160W $\leq 23$ to be $10.3^{+1.6}_{-1.3}\%$. Using method (1), we see no significant change in the pair fraction, calculated as $10.0^{+1.6}_{-1.3}\%$. With method (2), we find $10.2^{+1.6}_{-1.3}\%$. Though exclusion of the cluster members in the calculation of the pair fraction shows no significant effect, we choose to use method (2) to mask known members of the cluster located at $z=2.095$ in COSMOS. We use only those objects who were not flagged as cluster members in Figures \ref{svp} - \ref{uvj} and Table \ref{fracTable}. Explanations of how pairs were selected can be found in \S\ref{sec:mergID} and Table \ref{selects}. 

\subsection{Matched Object Catalogs}

We integrate the grism results with our photometric catalogs using the Astropy's \verb|match_to_catalog_sky()| routine that matches the coordinates of the objects with grism data to the objects detected in our SExtractor catalogs. We then generated a line-matched catalog containing the object's ID, grism ID, grism redshift, \dq,  and \qz\ flag values. Many of the objects had been previously studied in the NEWFIRM Cluster Survey (NCS), a ground based photometric survey. These are identified by their ``NCS ID'' in the grism output catalogs. 

For each of the two cluster fields, we generate four line-matched catalogs: (1) a photometric catalog containing flux measurements from multiple different filters; (2) a catalog output from EAZY containing the photometric redshifts; (3) a catalog output from FAST containing the star formation histories, masses, ages, and stellar attenuations; and (4) a grism catalog containing the grism redshifts, the data quality (\dq) and redshift quality (\qz) flags (see \S\ref{sec:flags}), as well as a merger flag value (see \S \ref{sec:mergID}) . Our final line matched catalogs consist of 499 objects in our IRC-A field of view and 661 objects in the IRC-B field of view. Relevant information from these tables for candidate cluster members are shown in Table \ref{TabA} and \ref{TabB} (see \S \ref{sec:III})

\subsection{Redshift Quality Control}\label{sec:zquality}
\begin{figure*}
\plottwo{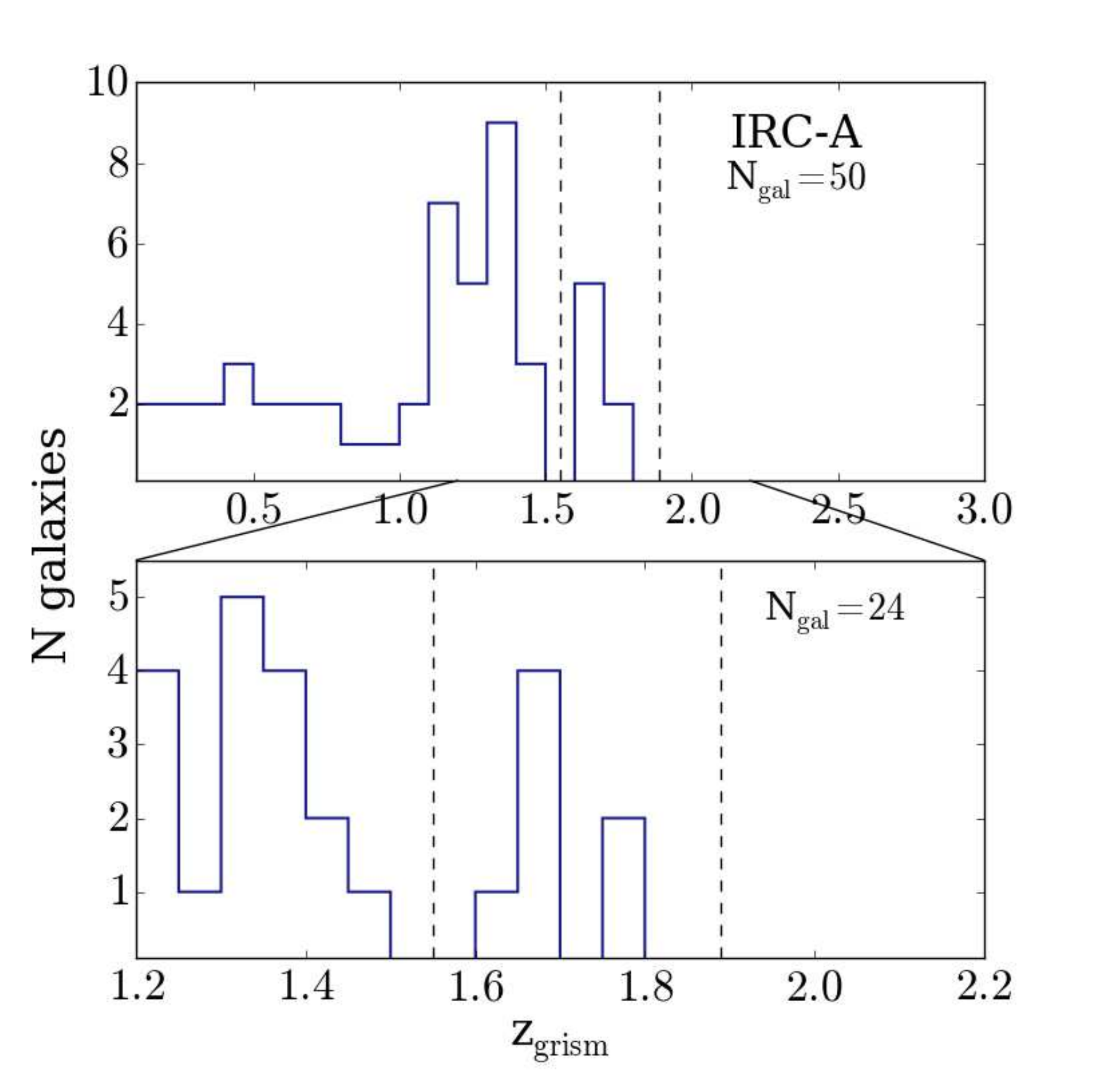}{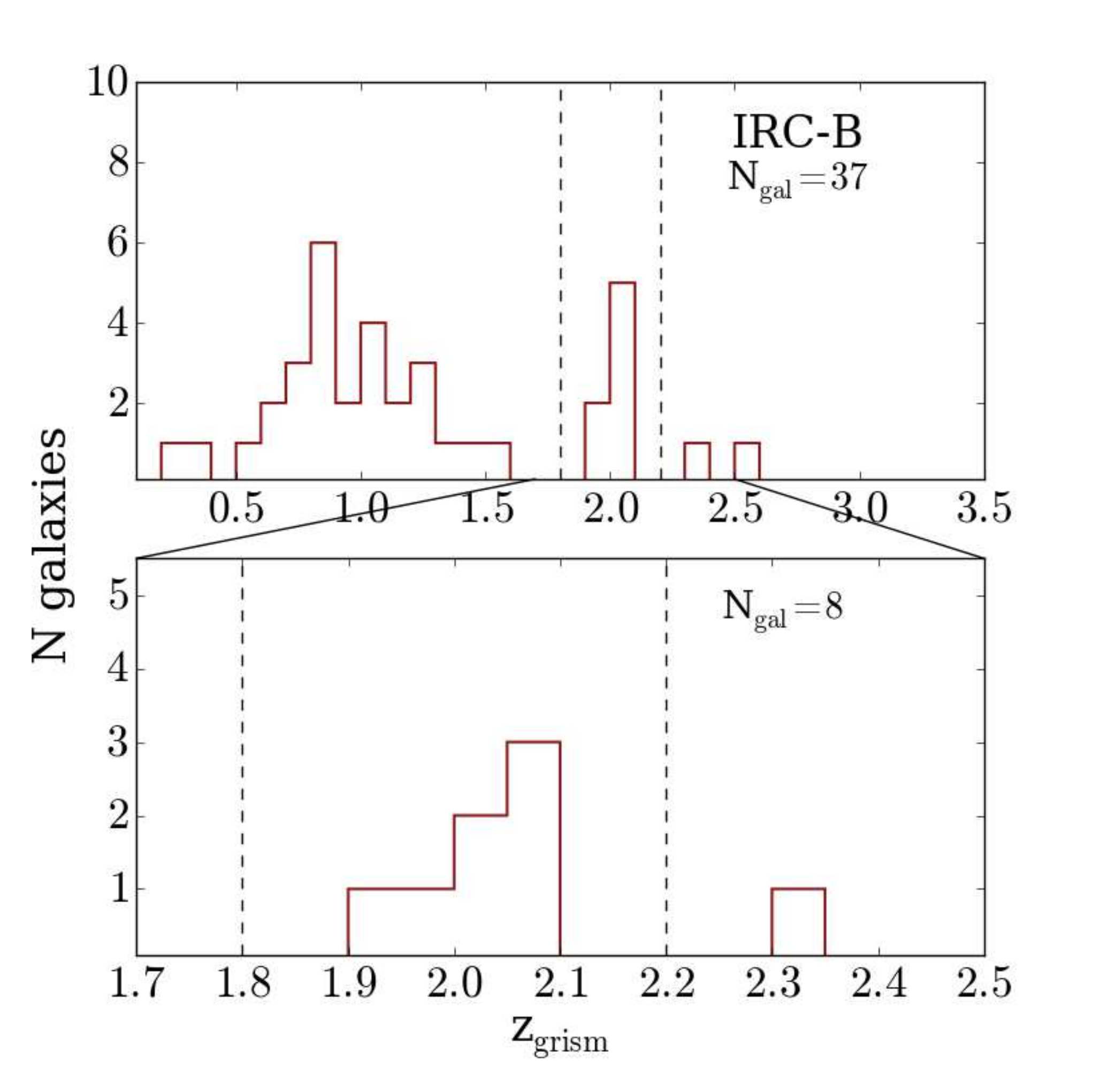}
\caption{Histograms of IRC-A (\emph{left}) and IRC-B (\emph{right}) showing the distribution of recorded grism redshifts within the entire field of view (top), and zoomed in to emphasize the projected redshift range of the candidate cluster (IRC-A: [1.55-1.9] ; IRC-B: [1.8-2.2]), indicated by the vertical dashed lines (bottom). All objects shown have data quality flag (\dq) values of ``1'' and redshift quality flag (\qz) values of ``3'' (see \S \ref{sec:zquality}).  \label{hist}}
\end{figure*}

Figure \ref{hist} shows the distribution of grism redshifts within each field of view.  All objects shown in Fig.~\ref{hist} have \dq\ flag values of ``1'' and \qz\ flag values of ``3'' (see \S\ref{sec:flags}). The histograms show a galaxy overdensity in IRC-A at redshift of $z \sim 1.6$ while IRC-B has galaxy overdensities at redshifts of $z \sim 0.9$ and $z \sim 2.1$. We define a redshift range for each candidate galaxy cluster: IRC-A ($1.55< z <1.9$) and IRC-B ($1.8< z <2.2$). 

\begin{figure*}
\plottwo{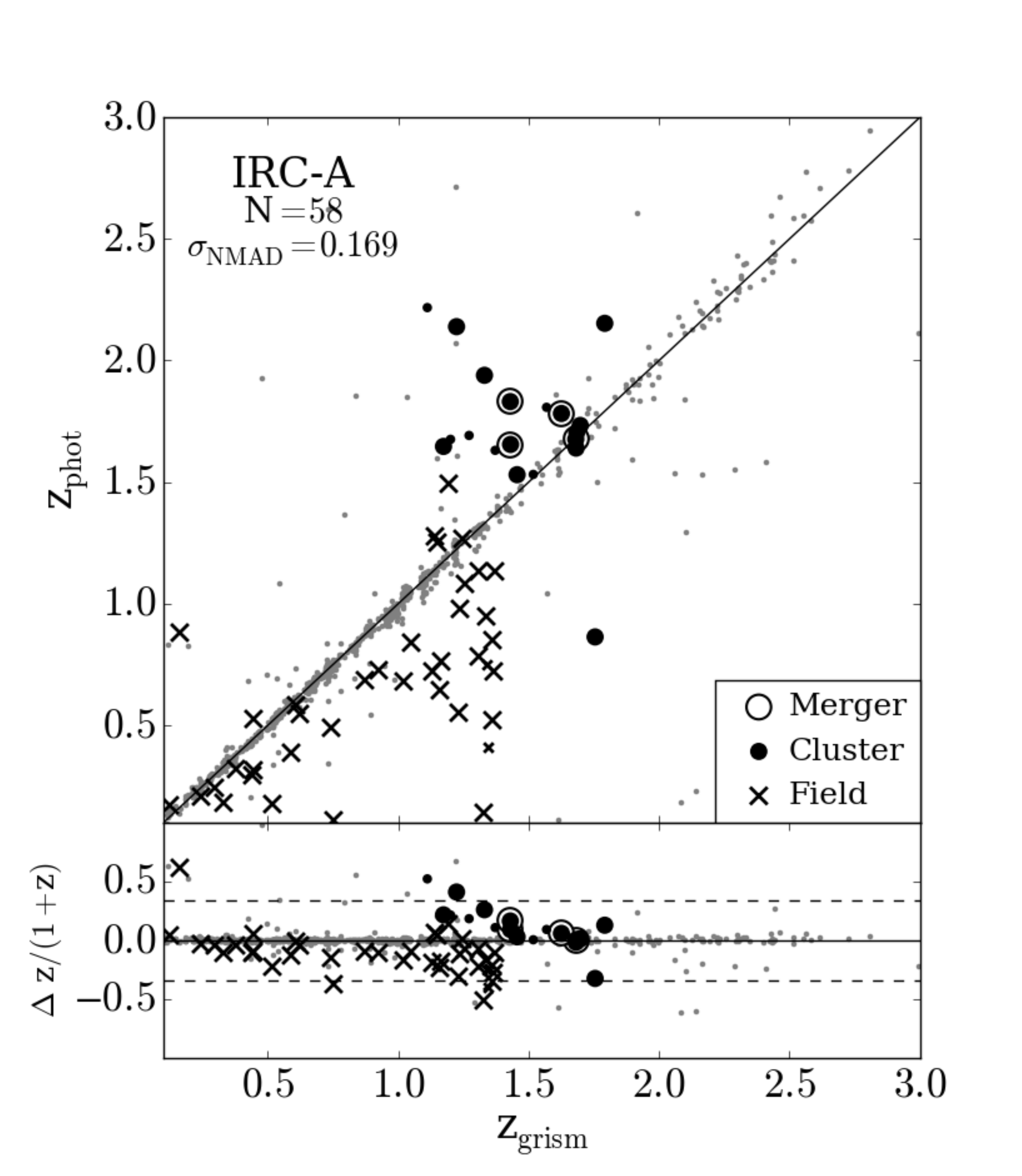}{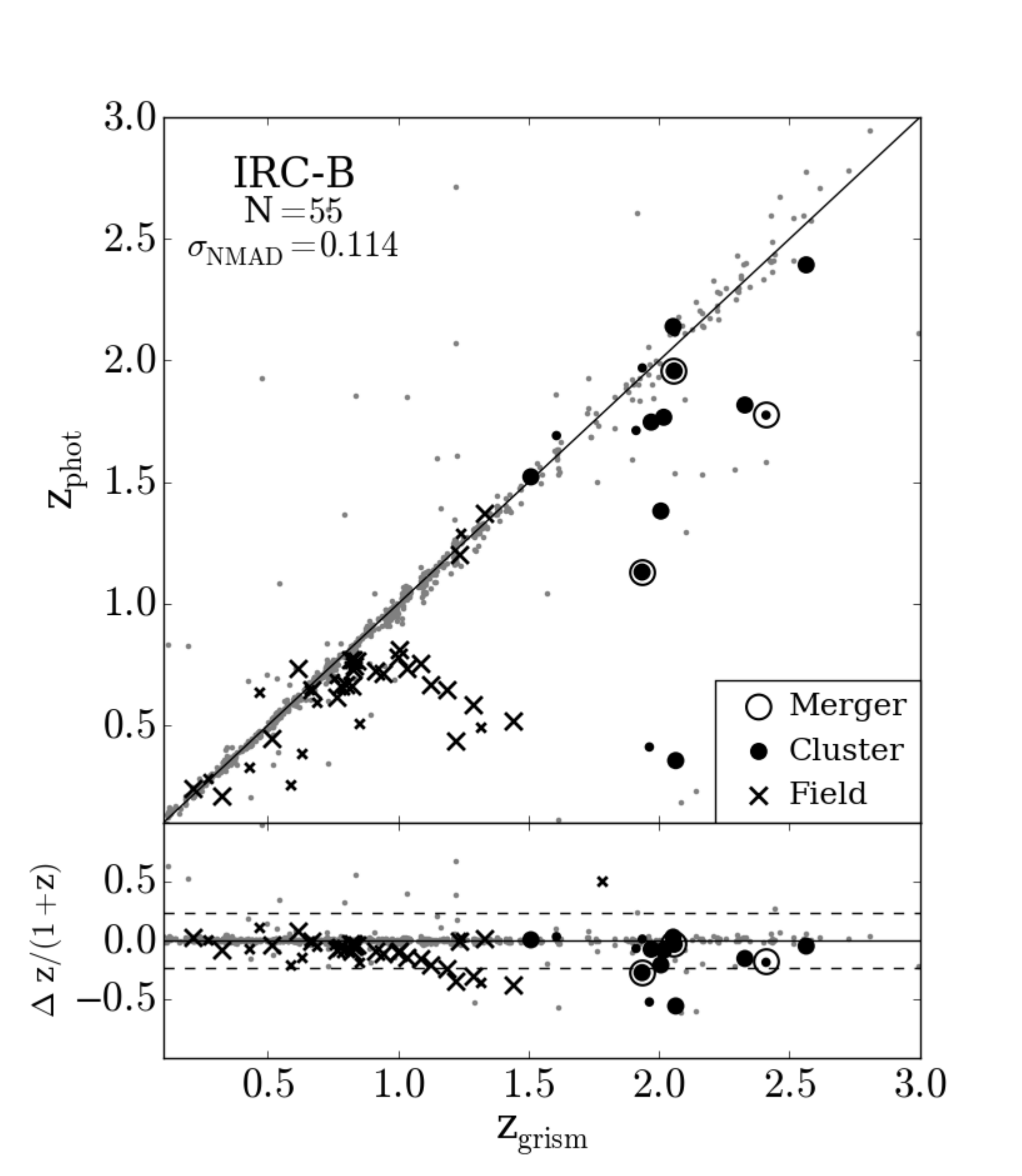}
\caption{Comparison between photometric and grism redshifts in IRC-A(\emph{left}) and IRC-B (\emph{right}). \emph{Top:} grism vs. photometric redshift for objects with recorded grism data. \emph{Bottom:} $\Delta z/(1+z)$ as a function of grism redshift with $\pm 2\sigma_{NMAD}$ indicated by the dashed lines. All objects shown have F160W $\leq 23$, data quality flag values \dq\ = 1, and redshift quality flag values \qz\ = 2 or 3. Points are sized according to their \qz value, with \qz\ = 2 represented by smaller points and \qz\ = 3 represented by larger points. Filled circles indicate galaxies with grism redshifts within each candidate clusters' projected redshift range (see \S\ref{sec:zquality}) or photometric redshifts within $1.5\leq z_{phot} \leq 2.5$ (see \S \ref{sec:III} and Table \ref{selects}). Galaxies detected in the WFC3 imaging with grism redshifts outside the projected range are indicated by crosses. Galaxies identified as merger candidates via selection method III (see \S \ref{sec:III} and Table \ref{selects}) are represented by open circles. The gray points represent our combined COSMOS and CDFS field population. \label{svp}}
\end{figure*}

As part of our analysis, we use multiple methods to identify  galaxy merger candidates (see Table \ref{selects}).  We consider only galaxies brighter than AB magnitude of 23.0 in F160W for both the cluster and field samples. Because we do not have grism redshifts for all galaxies with F160W$<23.0$ mag, we define a photometric redshift window by anchoring to the existing grism redshifts. Figure \ref{svp} shows the photometric verses grism redshift comparison for objects in our master catalogs with quality flag values of \dq\ = 1 and \qz $\geq 2$. Candidate cluster members are selected based on their grism or photometric redshifts and are indicated by filled circles (see \S \ref{sec:III}). Field galaxies in our master catalogs are indicated by x's.

We calculate the redshift difference as 
\begin{equation}
\frac{\Delta z}{(1+z)} = \frac{z_{phot}-z_{grism}}{1+ z_{grism}}
\end{equation}
We calculate the NMAD scatter ($\sigma_{NMAD}$) as:
\begin{equation}
\sigma_{NMAD} = median(|\Delta z - median(\Delta z)|) \times 1.48
\end{equation}
We find an NMAD scatter of $\sigma_{NMAD} = 0.169$ and $\sigma_{NMAD} = 0.114$ for IRC-A and IRC-B, respectively. Galaxies with photometric redshifts within $\pm 2\sigma_{NMAD}$ are used in one of our merger candidate selection methods described in \S\ref{sec:II}. 

\section{Results}

\subsection{Identifying Galaxy Merger Candidates}\label{sec:mergID}

In our analysis, we compare results from three selection methods for identifying galaxy merger candidates.  All selection methods require F160W$\leq 23.0$, angular separations $\lesssim3\arcsec$, and selection of galaxies based on grism or photometric redshifts. Note that our magnitude cut for our candidate clusters is slightly brighter than adopted in other studies (e.g. \cite{Lotz2008} uses $I_{814} \leq 24.0$ and \cite{Delahaye2017} uses $m_{160} < 23.25$). 

The pair fractions measured for each of the methods described below (summarized in Table \ref{selects}) are presented in Table \ref{fracTable} in the Appendix. Uncertainties are calculated assuming binomial statistics for $68\%$ confidence intervals, following \cite{Cameron2011A}.  

\subsubsection{Selection Method I}
 
The first selection method selects galaxy pairs with photometric redshifts between $1.5\leq z_{phot} \leq 2.5$; this corresponds to the broadest redshift range for candidate cluster members.  This method allows for comparison of our measured pair fraction to previous studies who rely only on photometric redshifts \citep{Lopez-Sanjuan2009}.

Of the 499 objects in our IRC-A catalog, 27 were selected based on their photometric redshift to be cluster members. Within this sub-sample we identified 4 cluster merger candidates, giving a pair fraction of \AIfrac. In IRC-B, 29 cluster galaxies were selected based on their photometric redshift from the parent sample consisting of 661 objects. From the sub-sample of 29 galaxies, 6 were selected as cluster merger candidates, giving a pair fraction of \BIfrac.

\subsubsection{Selection Method II \label{sec:II}}
Selection method II selects galaxy pairs with grism redshifts within the candidate clusters' redshift range ($\Azrange$ in IRC-A and  $\Bzrange$ in IRC-B)  or photometric redshifts within 2\signm of the candidate clusters' redshifts (see \S\ref{sec:zquality}):  $1.21 \leq z_{phot} \leq 2.24$ for IRC-A and $1.57 \leq z_{phot} \leq 2.43$ for IRC-B. This method was chosen to mimic the selection techniques in \cite{Delahaye2017}.

In IRC-A, 34 cluster galaxies were selected using method II from the parent population of 499 catalog objects. Within this sub-sample, 4 galaxies were identified as cluster merger candidates, giving a pair fraction of \AIIfrac. In IRC-B, 6 cluster merger candidates were selected from a sub-sample of 32 cluster galaxies (selected using method II), resulting in a pair fraction of \BIIfrac. 

\subsubsection{Selection Method III\label{sec:III}}
Selection method III selects galaxy pairs with grism redshifts within the candidate cluster's range ($\Azrange$ in IRC-A and  $\Bzrange$ in IRC-B) or photometric redshifts within $1.5 \leq z_{phot} \leq 2.5$. Galaxies that are selected based on their grism redshifts must also have a data quality flag \dq = 1 and redshift quality flag \qz $\geq 2$. This ensures that we are using the most reliable grism measurements while also excluding objects that have been flagged as stars (\qz\ = -1). 

\begin{figure*}[h!]
\plottwo{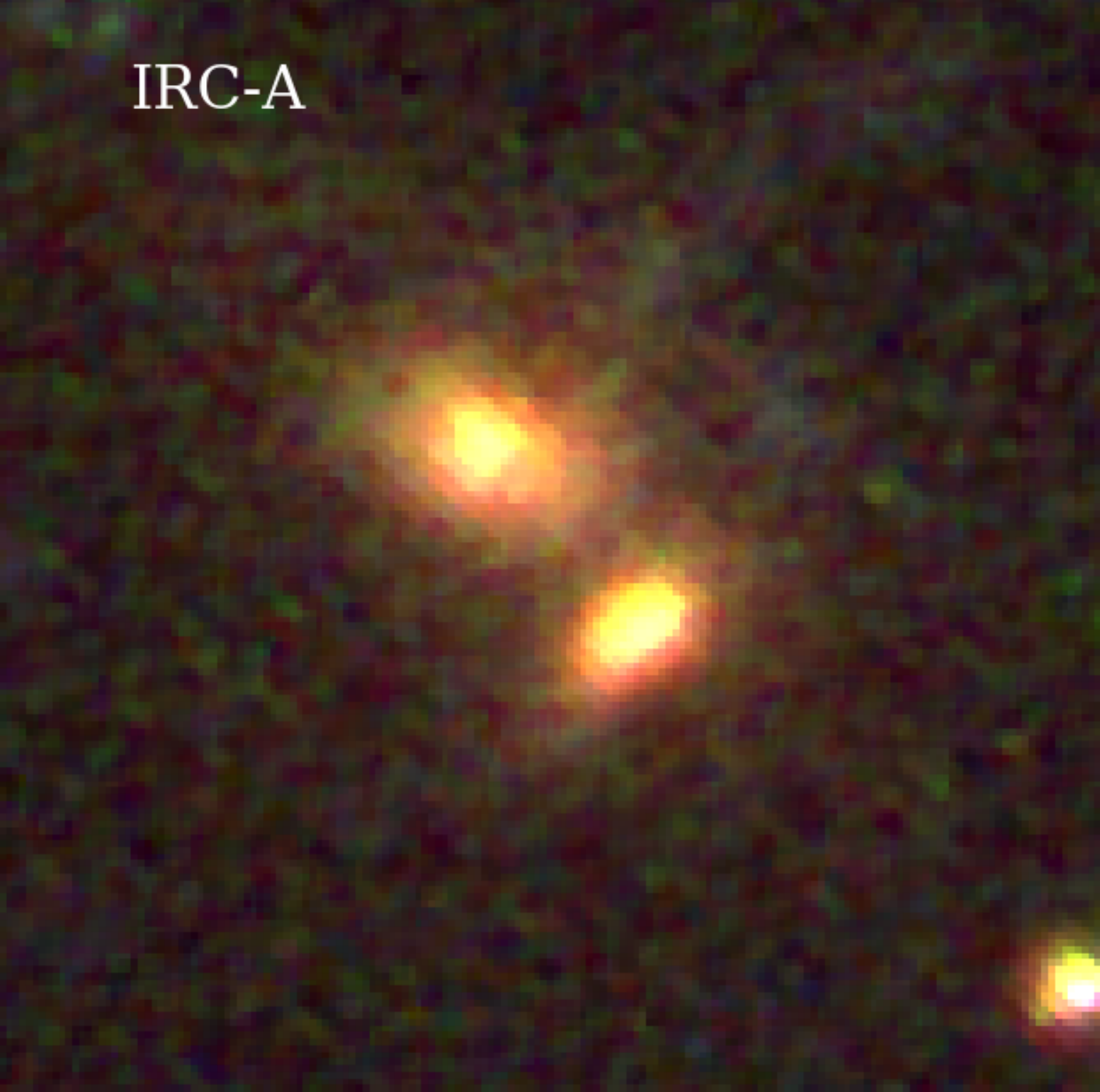}{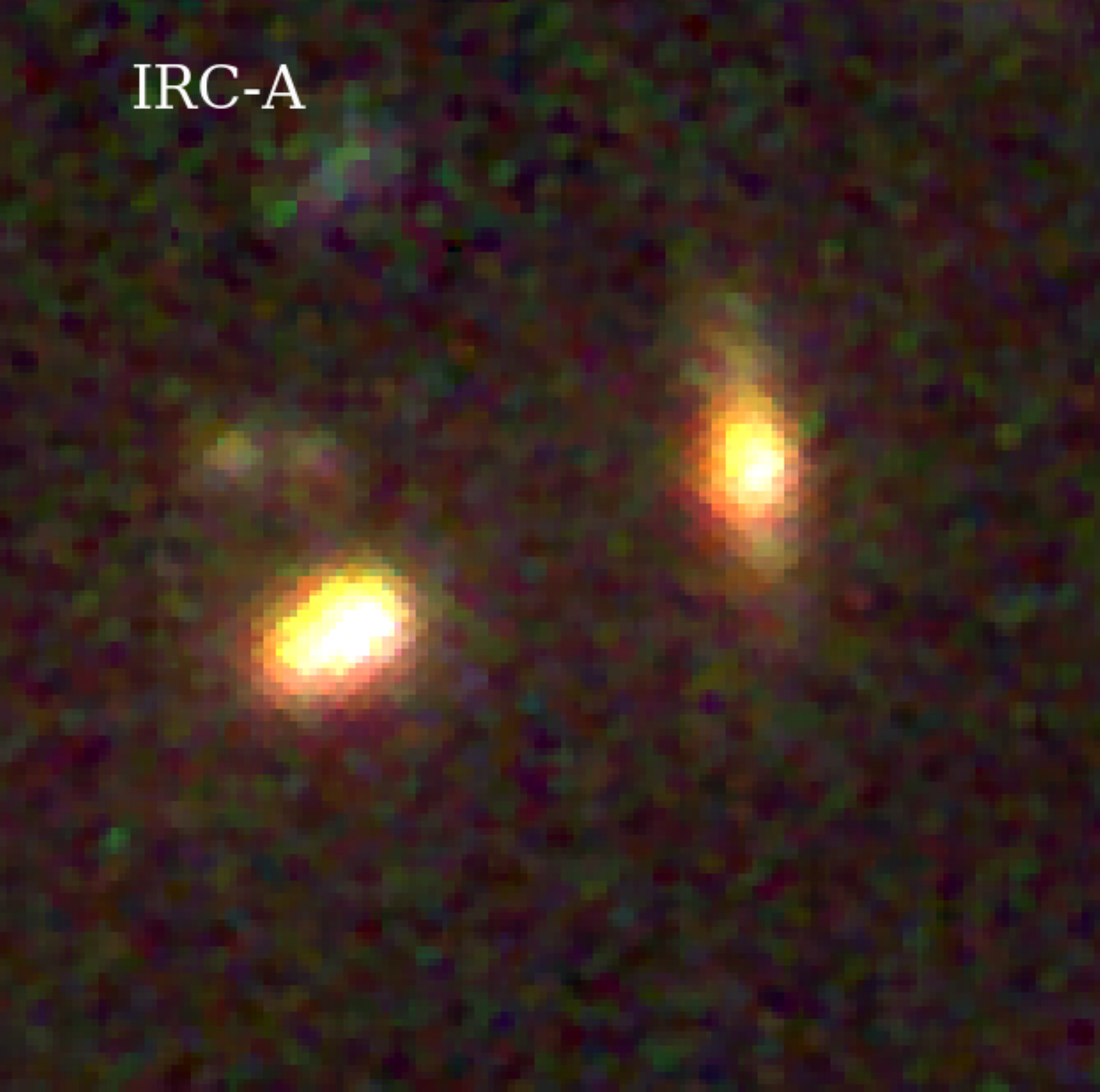}
\caption{Thumbnails of the mergers candidates in IRC-A selected using Selection Method III (\S \ref{sec:III} and used in the calculation of the merger fractions shown in Table \ref{fracTable}. Thumbnails were created by stacking each of the F160W, F125W, and F105W filters to create an RGB image with a 0.03\arcsec/pix scale, resulting in a 5\arcsec$\times$5\arcsec snapshot of the merger candidates. \emph{Left:} Object IDs 235 \& 241, located 5\arcsec from the cluster center. \emph{Right:} Object IDs 327 \& 329, located 27\arcsec from the cluster center. \label{panA}}
\end{figure*}

In IRC-A, 27 cluster galaxies were selected using method III. Within this sub-sample, we identified 4 cluster merger candidates, giving a pair fraction of \AIIIfrac. Selection method III resulted in a sub-sample of 33 cluster galaxies in IRC-B. Within this sub-sample we identified 8 cluster merger candidates, giving a pair fraction of \BIIIfrac. The cluster merger candidates identified using method III are highlighted in the composite mosaics of IRC-A and IRC-B in Figure \ref{rgbs}. Thumbnails of the cluster merger candidates identified via method III are shown in Figures \ref{panA} and \ref{panB}

We choose to focus on selection method III for the remainder of our analysis because it combines the photometric redshifts with our grism data. Tables \ref{TabA} and \ref{TabB} list the photometry and grism redshift information for the sub-sample of cluster galaxies selected via method III for our candidate clusters IRC-A and IRC-B. We also include the grism imaging for the cluster galaxies, selected by method III, in the Appendix in Figures \ref{gris_pan_A2} and \ref{gris_pan_B3}. Note that some of the candidate cluster members are selected based on their photometric redshifts and therefore may not have available grism data. The grism images presented in the Appendix encompass all objects, with recorded grism measurements, that are selected via method III without the added constraint on the \dq and \qz flags for the grism redshifts. This is to show why we chose to include the constraint on \dq and \qz because, for IRC-A in particular, there are some instances where objects that had been selected as candidate cluster members based on their grism redshifts were previously flagged as stars (\qz = -1) and therefore should not be included in the calculation of the pair fraction.

\subsubsection{Field Merger Fraction}

For comparison to the field, we select merger candidates in COSMOS and CDFS using method I (selection based only on photometric redshift) and a modified version of III (selection based on photometric or grism redshift) where the grism redshift range was chosen to be $1.55\leq z_{grism} \leq 2.2$, to cover the range of each of the candidate clusters. Note that the COSMOS field identifies a number of substructures between $1<z<3$ \citep{Scoville2013} that may drive the high pair fraction measured in both the corrected (see next section) and uncorrected pair fractions shown in Table \ref{fracTable}. 

Selection method I results in a sub-sample of 800 galaxies in our combined COSMOS + CDFS field population. Within this sub-sample, we identified 60 merger candidates, giving a field pair fraction of \fieldIfracs. Method III results in a sub-sample of 888 galaxies, of which 62 were identified as merger candidates, giving a field pair fraction of \fieldIIIfracs (see Table \ref{contamTab}).

\begin{table*}
\begin{center}
\caption{Summary of selection processes used to calculate the galaxy-galaxy merger fractions in Table \ref{fracTable}. See detailed explanation in \S \ref{sec:mergID}. \label{selects}}
\begin{tabular}{ccccc}
&&&& \\
\hline\hline
Selection & IRC-A& IRC-B& Field\tablenotemark{a}\\ 
Process	&	&	&	\\ \hline
I& 				$1.5\leq z_{phot} \leq 2.5$&	$1.5\leq z_{phot} \leq 2.5$&	 $1.5\leq z_{phot} \leq 2.5$\\ \hline
\multirow{3}{*}{II}&	$1.55\leq z_{grism}\leq 1.9$ & $1.8\leq z_{grism}\leq2.2$ & 	\multirow{3}{*}{...}\\
			&		or			&			or			&				\\
			&	$1.211\leq z_{phot}\leq 2.239$&  $1.571\leq z_{phot} \leq 2.429$&			  \\ \hline
\multirow{4}{*}{III}&	$1.55\leq z_{grism}\leq 1.9$ &  $1.8\leq z_{grism}\leq 2.2$ &	$1.55\leq z_{grism}\leq 2.2$ \\
            &       \dq\ = 1 and \qz\ $\geq$ 2   &       \dq\ = 1 and \qz\ $\geq$ 2   &               \\
			&		or			   &			or			&			or			\\
			&	$1.5\leq z_{phot} \leq 2.5$		& $1.5\leq z_{phot} \leq 2.5$		&	$1.5\leq z_{phot}\leq 2.5$\\  \hline
\end{tabular}
\end{center}
\tablenotetext{a}{Combined COSMOS and CDFS field population.}
\end{table*}

\subsection{Correcting for Potential Contamination \label{sec:contam}}

The pair fractions calculated in Table \ref{fracTable} are most likely an overestimate of the true pair fraction due to false projections. To correct for this, we calculate corrected pair fractions by randomizing the positions of selected galaxies, measuring the resulting pair fraction, and subtracting this from the original, uncorrected, fraction. 

Candidate cluster galaxies are selected for grism redshifts within each cluster's redshift range and F160W magnitudes $\leq 23.0$ (see selection process (III) in \S \ref{sec:III} and Table \ref{selects}). Field galaxies are selected in a similar manner, with the grism redshift range defined to encompass the full range of both candidate clusters (see Table \ref{selects}). 

For both candidate cluster and field galaxies, we randomize the RA and DEC positions of each selected galaxy and measured the pair fraction of companions found within $3\arcsec$ separation. This process is repeated 1000 times. We measure the average fraction of false pairs in IRC-A and IRC-B to be $3.8^{+7.1}_{-1.4}\%$ and $6.6^{+6.5}_{-2.5}\%$, respectively. For our COSMOS and CDFS field galaxies, we measure the average fraction of false pairs, for both fields combined, to be $2.0^{+0.6}_{-0.4}\%$. Uncertainties are calculated assuming binomial statistics for $68\%$ confidence intervals.

Corrected pair fractions for both candidate clusters, the COSMOS and CDFS field galaxies, and the combined field population are shown in Table \ref{contamTab}.

\begin{table*}
\begin{center}
\caption{Corrected galaxy-galaxy merger fractions (see \ref{sec:contam}) for IRC-A, IRC-B, COSMOS, and CDFS for different selection processes (see Table \ref{selects}) for $F160W \leq 23.0$. Uncertainties are calculated assuming binomial statistics for $68\%$ confidence intervals. \label{contamTab}}
\begin{tabular}{ccccc}
&&&&\\
\hline \hline
               & Selection  &	\multirow{2}{*}{$f_{merg, contam}$\tablenotemark{b}}& \multirow{2}{*}{$f_{merg,cor}$\tablenotemark{c}}    \\
               &	Process\tablenotemark{a}&	&	\\ \hline
IRC-A			&III	&	$3.8^{+7.1}_{-1.4}\%$&	$11_{-3.2}^{+8.2}\%$\\ 
IRC-B			&III	&	$6.6^{+6.5}_{-2.5}\%$&	$18_{-4.5}^{+7.8}\%$\\ 
COSMOS\tablenotemark{d}			&III	&	$2.9^{+1.0}_{-0.6}\%$&	$7.2^{+1.3}_{-1.0}\%$\\ 
CDFS			&III	&	$1.3^{+0.6}_{-0.5}\%$&	$2.4^{+0.9}_{-0.6}\%$\\ 
Field Population\tablenotemark{e}		&III	& 	$2.0^{+0.6}_{-0.4}\%$&	$5.0^{+1.1}_{-0.8}\%$\\ \hline

\end{tabular}
\end{center}
\tablenotetext{a}{See Table \ref{selects}}
\tablenotetext{b}{False pair contamination rate. See \S \ref{sec:contam}}
\tablenotetext{c}{Corrected pair fraction }
\tablenotetext{d}{Masked to exclude confirmed members of $z=1.62$ cluster \citep{Yuan2014}.}
\tablenotetext{e}{Combined COSMOS and CDFS populations}
\end{table*}

\begin{figure}
\plotone{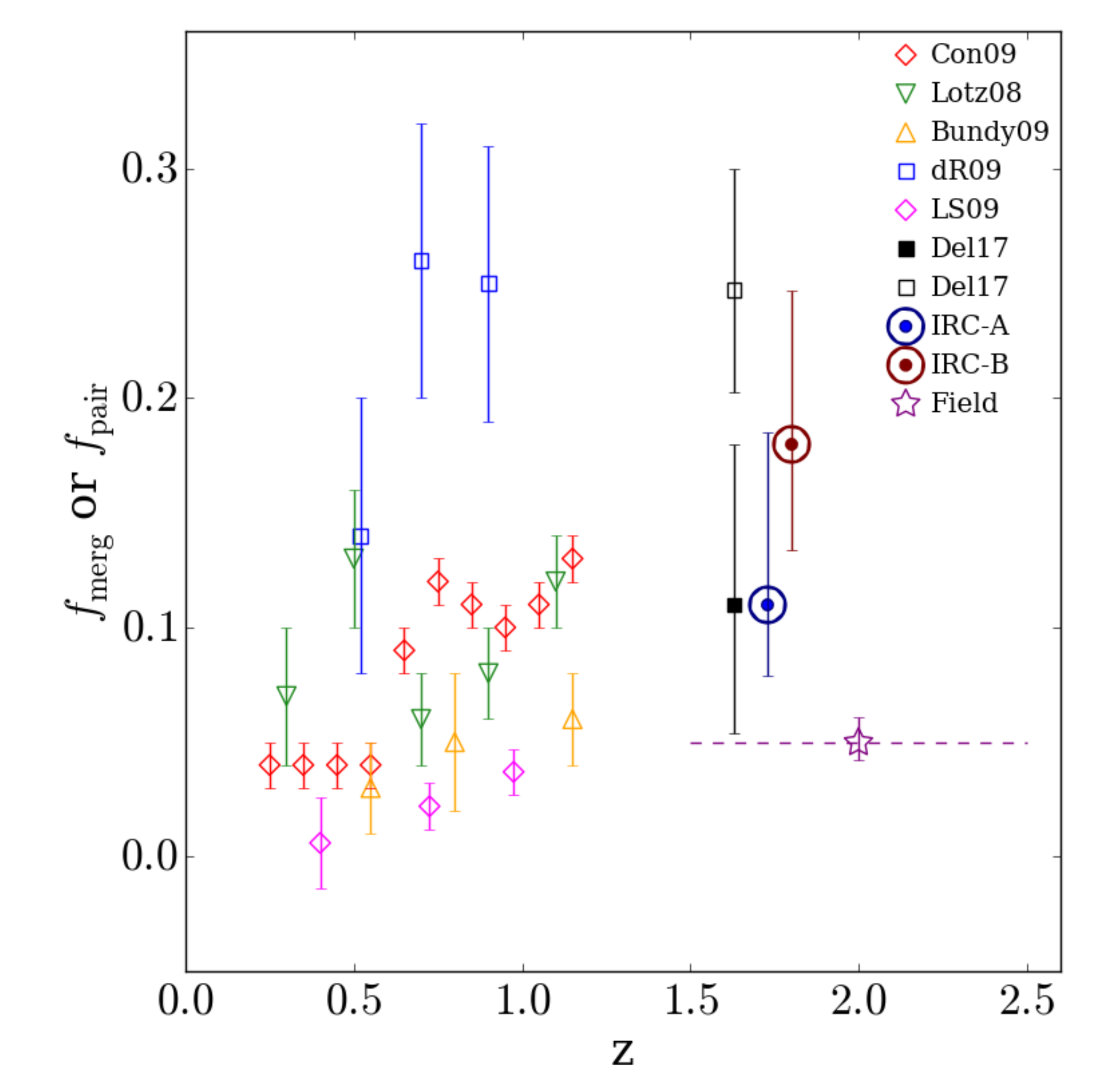}
\caption{Merger fractions as a function of redshift for objects selected by stellar mass ($M_{*} \geq 10^{10} M_{\odot}$), for separation distances of $\leq3$". The points for IRC-A, IRC-B and the field average represent the corrected pair fractions listed in Table \ref{contamTab} for selection process (III) (see \S \ref{sec:III} and Table \ref{selects}). Unfilled markers represent the field pair fractions measurements from the studies referenced in \S \ref{sec:comp} as well as the combined field measurement presented in this work. Filled markers represent the cluster pair fraction measurements from \cite{Delahaye2017} and this work. The dashed horizontal line for the field average represents the redshift range the field pair fraction spans. Note that for our candidate cluster and field populations the same selection method was used in each, allowing for more accurate comparison. 
\label{litfracs}}
\end{figure}

\section{Discussion}
\subsection{A Higher Galaxy Merger Fraction in Clusters}

Figure \ref{litfracs} presents the corrected pair fractions (see \S \ref{sec:contam}) for both candidate clusters and the field average of our COSMOS and CDFS populations. An earlier version of Figure \ref{litfracs} was presented in \cite{Lotz2011} and shows the merger or pair fractions from 5 previous field studies. These are discussed in more detail in the next section.

Compared to the pair statistics and morphological studies shown in Fig. \ref{litfracs}, in particular \cite{Bundy2009} and \cite{DeRavel2009} who use methods very similar to those presented here, we see a higher pair fraction within the two candidate clusters than what is measured in the field populations. We also find that the merger fraction within the candidate clusters is consistently higher than that of the field population. 

Correcting for the false pair contamination we calculate the corrected pair fractions (shown in Table \ref{contamTab}) of IRC-A and IRC-B to be \Afraccor and \Bfraccor, respectively. The corrected pair fraction for our combined field population is \fieldfraccor. With the corrected pair fractions, we still see a higher fraction in the candidate clusters than the field population.

\subsection{Observed and Rest-Frame Colors}

\begin{figure*}
\plottwo{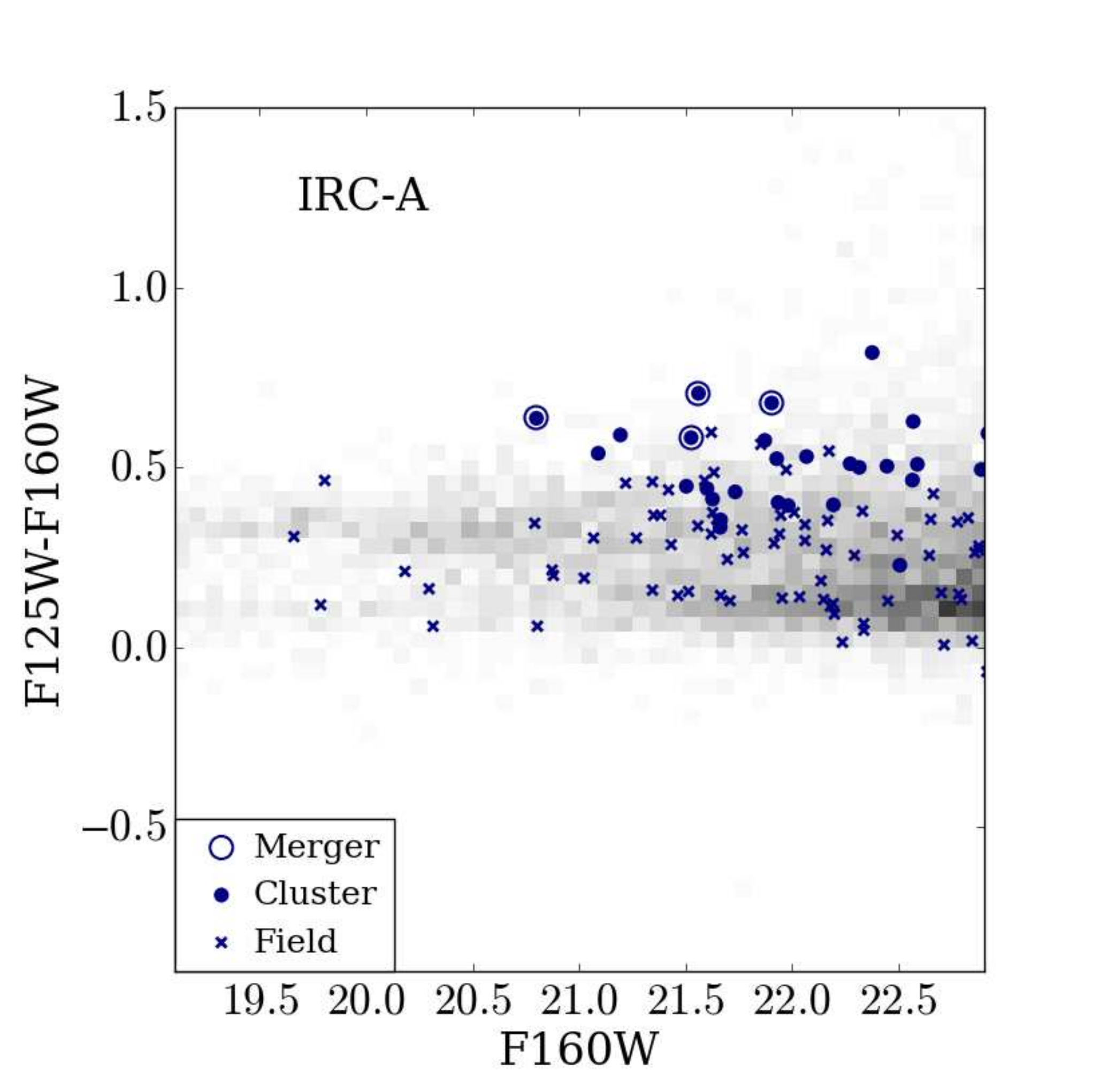}{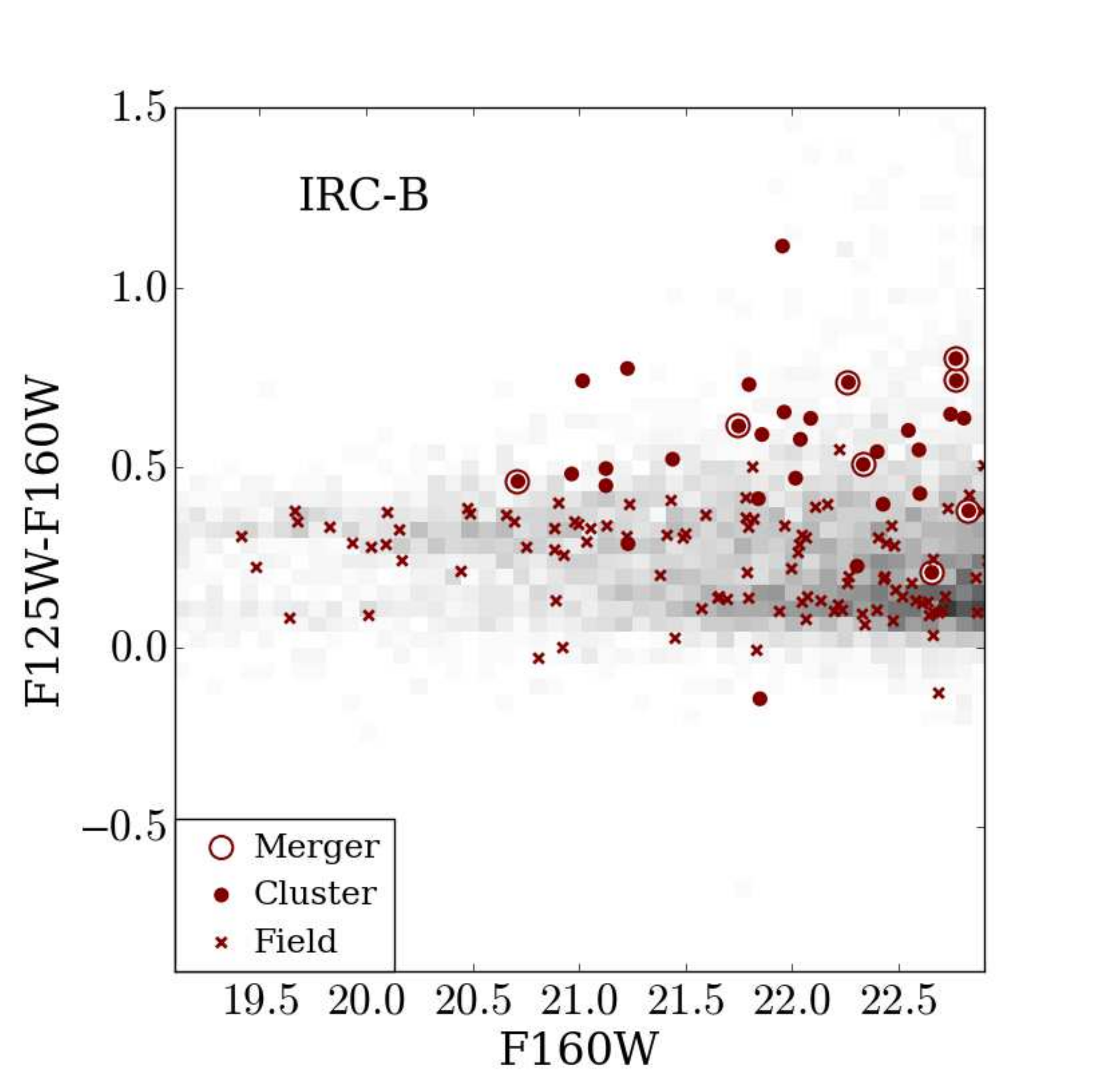}
\caption{Color Magnitude Diagram of IRC-A (\emph{left}) and IRC-B (\emph{right}). The filled circles represent candidate cluster members selected using method III, i.e. selection based on grism or photometric redshift (see \S \ref{sec:III}). Open circles indicate candidate merger candidates, identified via method III. Field galaxies from our master catalogs are indicated by crosses. 
The gray spatial histogram represents our combined COSMOS and CDFS field population. All galaxies shown have F160W$\leq 23.$ \label{cmd}}
\end{figure*}

Using the WFC3 imaging, we show the observed F125W and F160W color magnitude diagrams for IRC-A and IRC-B in Fig. \ref{cmd}. The objects in each cluster field are separated into candidate cluster members and mergers.  We include the COSMOS field sample, over all redshift, for comparison. COSMOS and photometry is taken from the zFOURGE photometric catalogs \citep{Straatman2016}. Candidate cluster members tend have redder (F125W-F160W) colors compared to the general field population.

\begin{figure*}
\plottwo{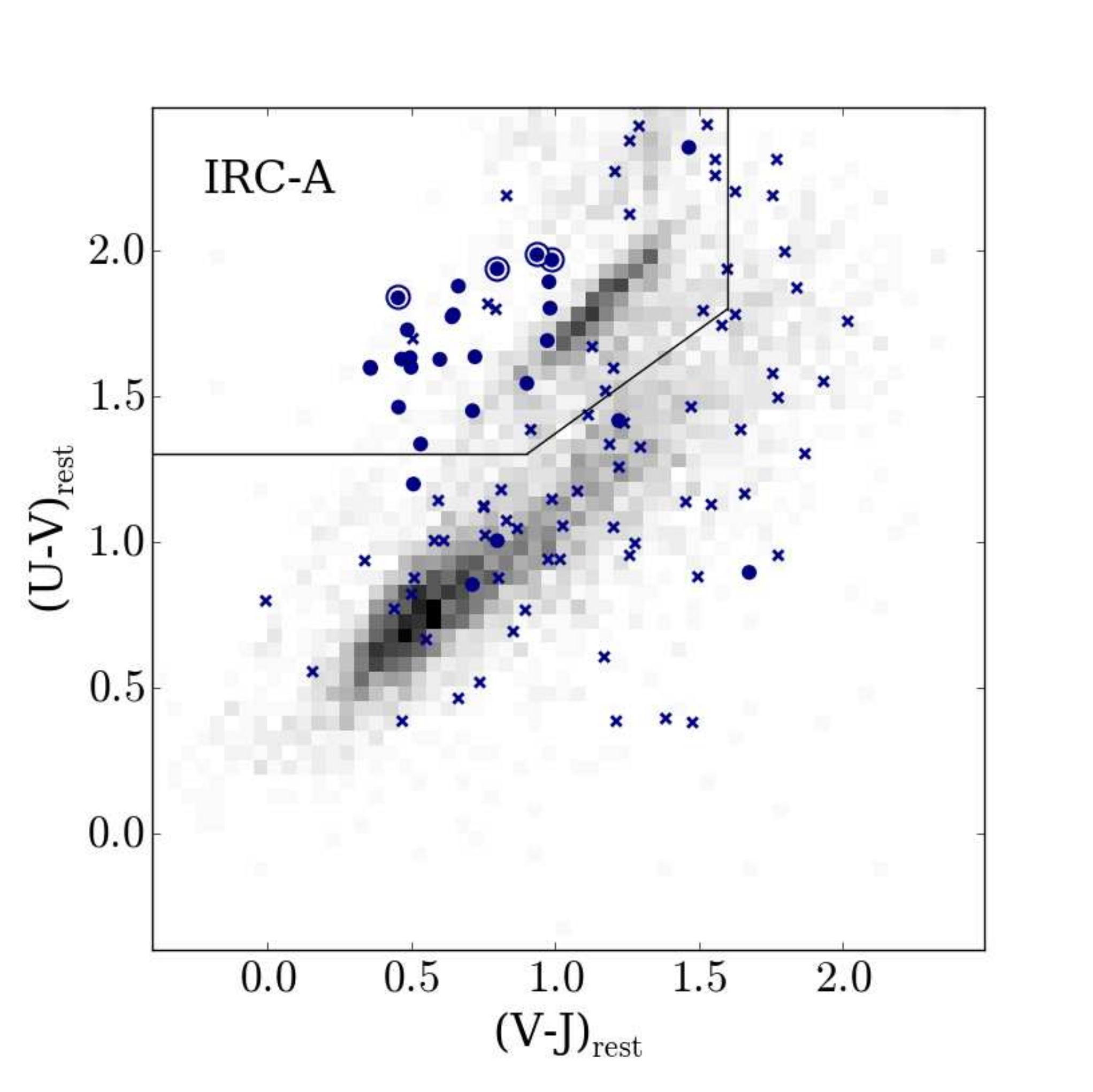}{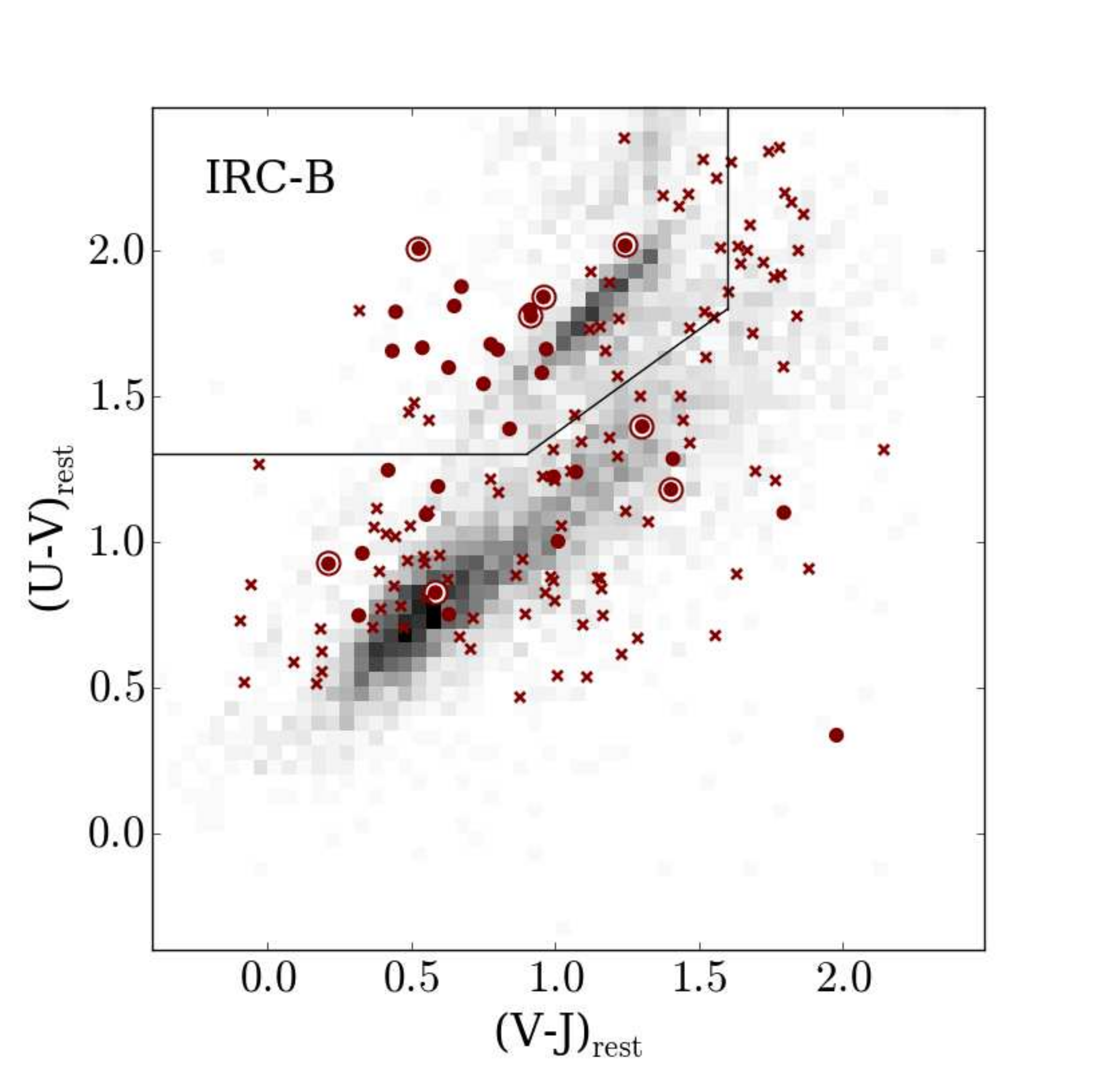}
\caption{Rest-frame $U-V$ vs. $V-J$ colors. Objects located within the black boxed region are generally considered to be quiescent while those located outside the boxed region are generally star-forming. Refer to the caption in Figure \ref{cmd} for explanation of symbols. \label{uvj}}
\end{figure*}

To assess the rest-frame colors we also generate UVJ diagrams, shown in Figure \ref{uvj}. When considering candidate cluster member, we find that $\sim 82\%$ in IRC-A and $\sim 52\%$ in IRC-B are quiescent. We also see that of the candidate cluster members identified as mergers, $100\%$ in IRC-A are quiescent while only $50\%$ in IRC-B are quiescent.

Simulations indicate that galaxy-galaxy merging enhances star formation \citep{Toomre1972,DiMatteo2007,Mihos1994}. \cite{Coogan2018} and \cite{Hine2016} see enhancement of star formation due to merging in high-redshift cluster environments ($z=1.99$ and $z=1.3$, respectively), in good agreement with predictions. However, \cite{Lotz2013} sees a dominance of quiescent galaxy merger candidates in a $z=1.62$ proto-cluster in comparison to the field population. Our findings in IRC-A, where we find that 100\% of the candidate cluster members identified as mergers are quiescent is in good agreement with the findings of \cite{Lotz2013}. 

We find that in IRC-A, the merger candidates are redder than the field population and show no evidence of on-going star formation,in good agreement with the findings in \cite{Lotz2013}. However in IRC-B only about half of the merger candidates are quiescent. Our findings imply that, depending on the dynamical state of the system, a cluster can be dominated by red mergers. However, given our limited sample, more clusters are needed.

\subsection{Spatial Distribution of Merger Candidates}

If galaxies in over-dense environments quench their star formation earlier than their field counterparts, galaxy-galaxy mergers in cluster cores may be redder than in the field \citep{VanDokkum1999,Tran2005}. Whether this is true at $z>1$, when galaxy clusters are even rarer and dynamically younger, is unknown. 

\begin{figure*}
\plottwo{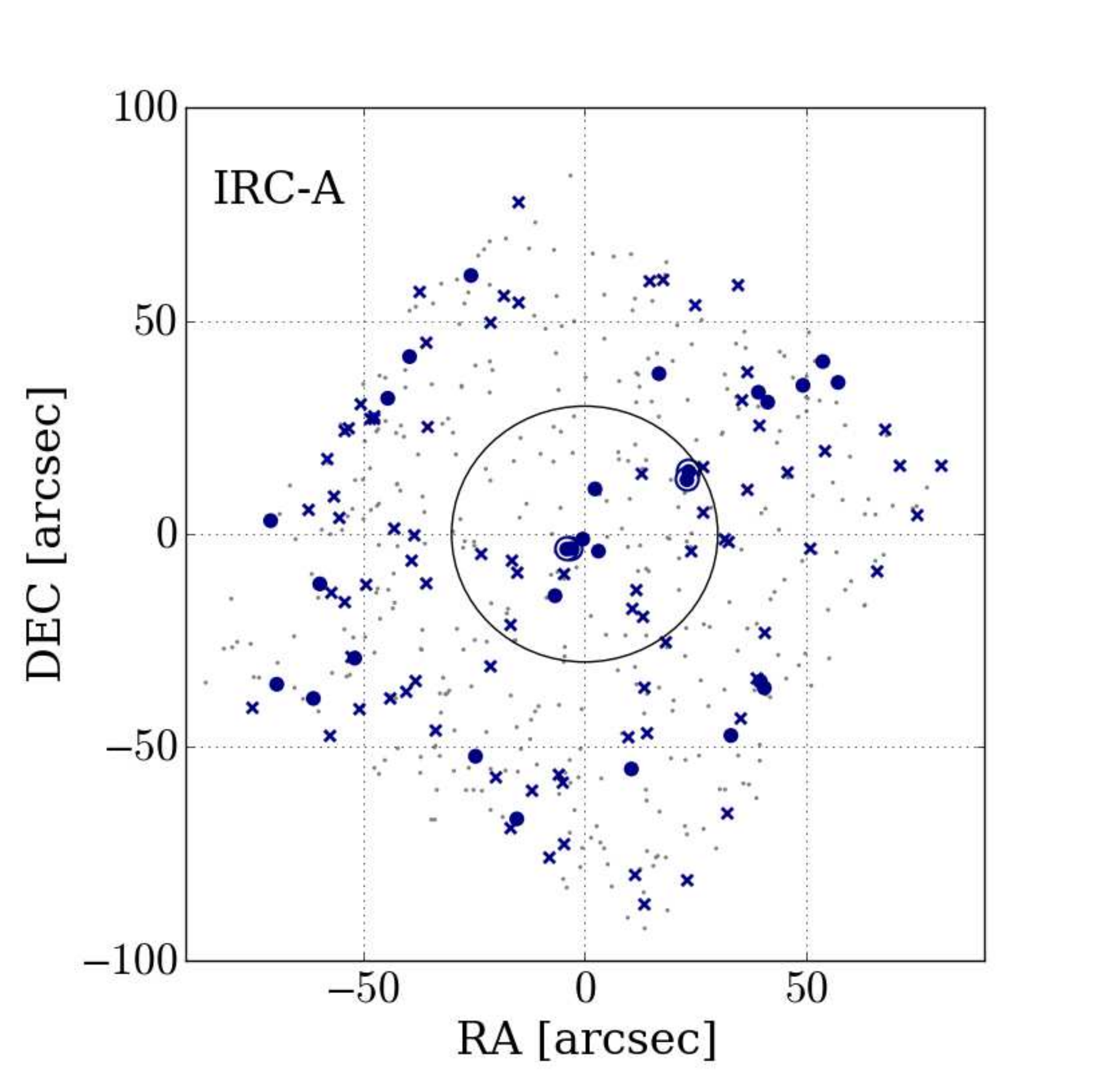}{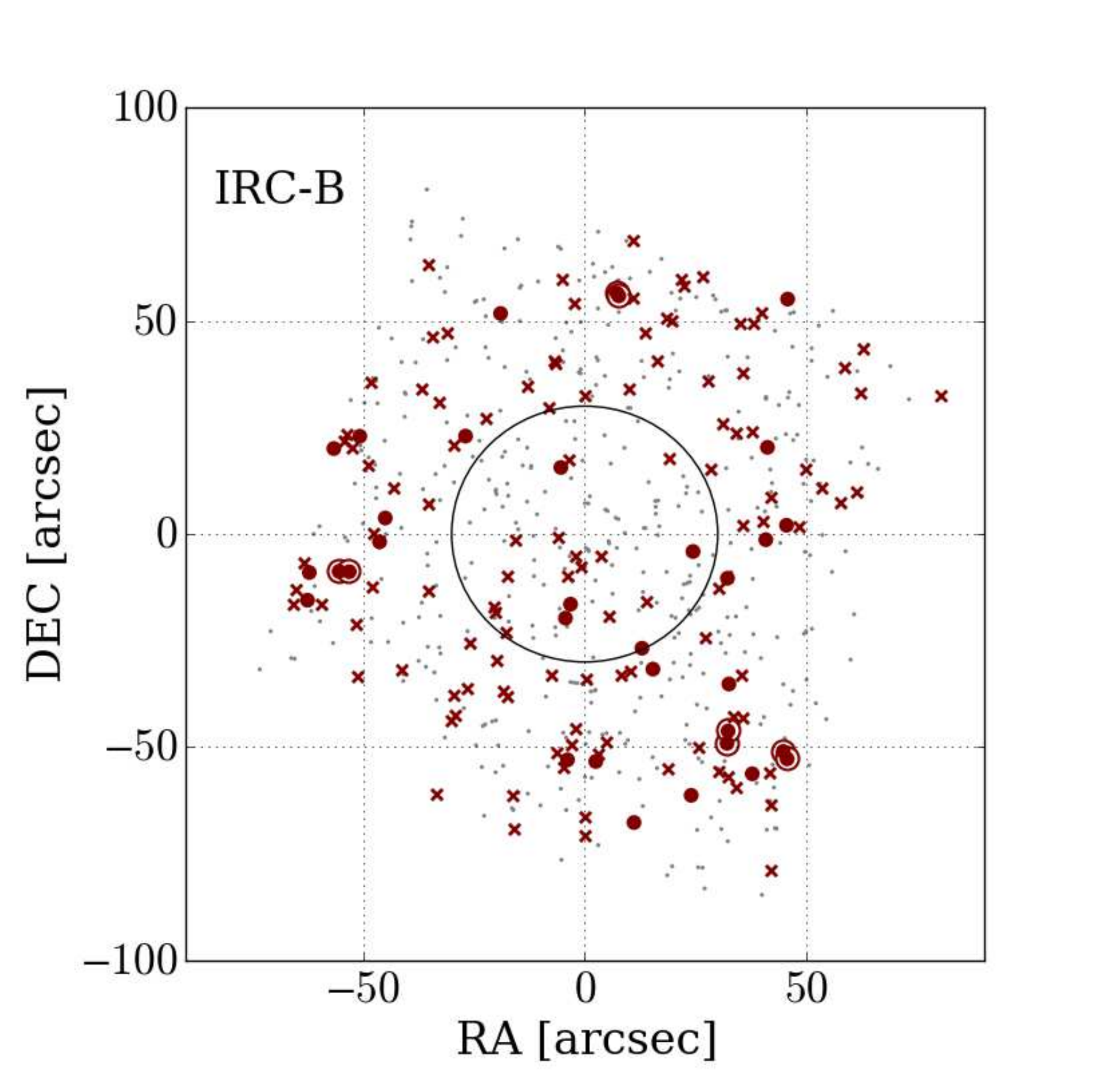}
\caption{Spatial distribution of galaxies in the candidate clusters IRC-A and IRC-B. Filled circles indicate candidate cluster members. Objects identified as merger candidates are represented by open circles. Grey points represent field galaxies detected in each WFC3 field of view. The black circle is centered on the candidate cluster's coordinates, with a radius of $30\arcsec$. Refer to the caption in Figure \ref{cmd} for explanation of symbols. \label{spatial}}
\end{figure*}

Figure \ref{spatial} shows the spatial distributions of the candidate cluster members and those identified as mergers for both IRC-A and IRC-B, with the cluster's center indicated by a $30\arcsec$ radius circle. We hypothesize that IRC-A at $z\sim1.6$ is a more evolved cluster, with candidate members identified as mergers (see Table \ref{TabA}) located within the cluster's core. In comparison, IRC-B is a less evolved cluster ($z\sim2$) where all of its merger candidates are located beyond a projected cluster radius of $30\arcsec$. In IRC-B, the merger candidates are most likely embedded in subclumps that will merge with the cluster's core \citep{VanDokkum1999}. We thus infer from our study that quenching of star formation in the cluster members occurs before galaxy-galaxy merging. However, more clusters are needed at this epoch to further strengthen our results.

\subsection{Are these robust galaxy clusters?}

\subsubsection{Analogs to Spectroscopically Confirmed Galaxy Clusters at $z\sim2$}

IRC-A was identified using the same method from \cite{Papovich2010} where a cluster at $z=1.62$ was first identified in the UDS field with the 1.6$\mu$m bump technique (see \S \ref{sec: XMMLSS}). For the UDS cluster, follow-up ground-based spectroscopy confirmed the galaxy cluster's redshift \citep{Tran2015}.  Likewise, IRC-A was first identified using the SWIRE observations that identified a galaxy overdensity at $>10\sigma$ relative to the field.  Follow-up medium-band imaging from NEWFIRM to track spectral features such as the D4000 break confirm a cluster redshift of $z\sim1.6$ which is consistent with our grism results.  

IRC-B also was initially identified using the $1.6\mu$m technique as an overdensity with a surface density $>10\sigma$ above the field.  From our NEWFIRM imaging, we determine a photometric redshift of $z\sim2$.  We note that our approach mirrors that of \citep{Spitler2012} who identified a galaxy cluster at $z\sim2$ in the the COSMOS field from similar near-infrared imaging taken with the Four-Star camera on Magellan.  The COSMOS cluster redshift was then confirmed with spectroscopy from Keck/MOSFIRE to be $z_{\rm spec}=2.1$  \cite{Yuan2014}.

\subsubsection{Testing for Random Projections}

To test if the galaxies we associate with our cluster candidates are chance projections, we use the ZFOURGE catalogs of field galaxies in CDFS and COSMOS (see \S3.4).  We randomly place apertures the size of our WFC3 footprint and select galaxies that satisfy our selection method III (see \S \ref{sec:mergID} and Table \ref{selects}).  Using 100,000 realizations, we find that on average, only $\sim2$\% of the field galaxies satisfy our selection criteria with method III.  In comparison, the fraction of galaxies selected using method III in both the IRC-A and IRC-B WFC3 catalogs is $\sim 5\%$. Thus we are confident that both cluster candidates are true overdensities at $z\sim2$.  We refer the reader to Appendix \S \ref{sec: prob} and Table \ref{prob} for further details of our randomization test.

\subsubsection{Predictions from IllustrisTNG}

Cosmological simulations are extremely useful in predicting the number density of an object with respect to the survey volume and redshift coverage. We use $\sim (300\, {\rm Mpc})^3$ volume simulations from IllustrisTNG \cite[TNG300;][]{Pillepich2017,  Nelson2017} to estimate the number of galaxy clusters in the XMM-LSS field-of-view (FOV) between $1.3<z<2.0$. The number density of galaxy clusters with halo masses greater than $10^{13.5}$ M$_{\odot}$ at redshift $z=1.3$ in TNG300 is 4510\,Gpc$^{-3}$. The comoving volume of our XMM-LSS observations is $0.08\,{\rm Gpc}^3$, which corresponds to about 360 galaxy clusters within the FOV assuming the same number density of galaxy clusters in the universe as TNG300. Thus we are confident that the existence and discovery of two galaxy clusters, IRC-A and IRC-B, in the XMM-LSS observations is highly probable.

\subsection{Comparison to Previous Studies \label{sec:comp}}
The ability to compare the measurements we report here to previous studies is complicated by the fact that different studies use different methods of selecting merger candidates. Here we discuss the similarities and differences in merger selection criteria between the studies presented in Fig. \ref{litfracs} and our work. 

The studies referenced in Fig. \ref{litfracs} all use a stellar mass cut at log$M_* > 10 M_{\odot}$. Although we apply no mass constraint, all of the merger candidates used in the calculation of our pair fractions have masses log$M_* > 10 M_{\odot}$. Note that the pair fractions presented in Fig. \ref{litfracs} for \cite{Lotz2008} (Lotz08),\cite{Conselice2009} (Con09), \cite{Lopez-Sanjuan2009} (LS09), \cite{Bundy2009} (Bundy09), \cite{DeRavel2009} (dR09) and \cite{Delahaye2017} (Del17) as well as the point for our combined COSMOS and CDFS field population all represent measurements in field environments. These are indicated by unfilled markers. The measurements presented in Fig. \ref{litfracs} for Del17 and our candidate clusters, IRC-A and IRC-B, represent pair fractions in cluster environments. Cluster measurements are indicated by filled markers. 

We refer the reader to \cite{Lotz2011} for a more in-depth explanation of each study referenced in Fig.\ref{litfracs}. Here we provide brief descriptions of each of these studies.

\subsubsection{Field Merger Candidates Identified via Morphology}

\begin{itemize}
\item Lotz08 selects merger candidates via $G-M_{20}$ between redshift $0.2<z<1.2$ in AEGIS \citep{Davis2006} by defining a parent sample selected based on luminosity (with the assumption of pure luminosity evolution). Their merger candidates consist of close pairs and morphologically disturbed galaxies selected using either spectroscopic or photometric redshifts (about $60 \%$ of the parent sample has spectroscopic redshifts while the reminaing $40\%$ have photometric redshifts). They find that the merger fraction remains roughly constant for this redshift range.

\item Con09 measured merger fractions by selecting merger candidates based on the calculated assymmetry from COSMOS, GEMS \citep{Rix2004}, GOODS \citep{Giavalisco2004}, and AEGIS for $0.2<z<1.2$. Redshift selection is done using photometric or spectroscopic, but it is unclear what percentage of their sample is based on spectroscopic vs photometric selection. 

\item LS09 measured the merger fraction in SPITZER/IRAC selected objects of GOODS-S using morphological asymmetry and distortions for $z\leq 1.3$. Merger candidates are selected from either luminosity or stellar mass selected samples and use only photometric redshifts. 

\item Del17 measured the merger fraction in 4 clusters ($1.59<z<1.71$) as well as the UDS field. Merger candidates were selected using both spectroscopic and photometric (when no spectroscopic measurements available) redshifts as well as stellar mass and magnitude limits. UDS merger candidates were selected in the range $1.55<z<1.75$. Candidates were identified via either close pair statistics or visual disturbance. 
\end{itemize}

\subsubsection{Field Merger Candidates Identified via Projected Separation}

Bundy09 and dR09 both provide mass selected pair fractions with 4:1 mass ratios and projected separations of $5\ kpc\ h^{-1} < R_{proj} < 20\ kpc\ h^{-1}$ and $R_{proj} < 100\ kpc\ h^{-1}$, respectively. Note that our selection criteria of a separation less than 3'' corresponds to a projected separation of $R_{proj} < 25\ kpc\ h^{-1}$. 

\begin{itemize}
\item dr09 measured the close pair fraction of galaxies in the VMOS VLT Deep Survey \citep{LeFevre2005} to $z\sim 1$. Merger candidates are selected based on spectroscopic redshift from either luminosity or stellar mass selected samples.  The merger fractions of $\sim12-25$\% are among the highest of all the comparison studies.  However, they use $R_{proj} < 100\ kpc\ h^{-1}$ compared to our criterion of $R_{proj} < 25\ kpc\ h^{-1}$.

\item Bundy09 measured the close pair fraction of galaxies in GOODS. Merger candidates are selected using either spectroscopic or photometric (when no spectroscopic measurement available) redshift from K$_{s}$-selected catalogs. The merger fraction of $\lesssim5$\% is consistent with our field measurement.
\end{itemize}

\subsection{Our Study in Context}
Our study is set apart by the combination of the area covered, the quality of the photometric redshifts, and the high resolution HST imaging. In particular, the inclusion of the ZFOURGE catalogs provides a more statistically robust measurement of the field merger fraction relative to other studies at $z\sim2$.

We measure the field merger fraction at $z\sim 2$ to be roughly half what \cite{Lotz2008} reports. However, this is likely due to the different selection methods for identify merger candidates:  our study relies solely on close pair identification while \cite{Lotz2008} uses morphologically disturbed galaxies and includes systems fainter than our magnitude limit of $F160W\leq23.0$.

Del17 uses merger candidate selection methods that are most similar to our selection method (III) (see \S \ref{sec:mergID} and Table \ref{selects}):  they also select  merger candidates based on magnitude and both photometric and spectroscopic redshifts. Our measurement for the pair fraction in IRC-A ($z\sim 1.6$) is consistent with the total cluster pair fraction presented in Del17. In contrast, our measurement of the pair fraction in the field population is roughly five times lower than Del17.

We note that all of our selection methods to identify mergers for both the candidate galaxy clusters and the combined field population yield similar results.  Thus we are confident that the relative fraction of mergers in the clusters is higher than that of the field at $z\sim2$.

\section{Summary}


We measure the galaxy merger fractions in two cluster candidates initially identified as overdensities using the $1.6\mu$m bump technique (\S \ref{sec: XMMLSS}). Follow up with near-IR medium-band imaging from NEWFIRM provides more precise photometric redshifts ($dz\sim0.05$) that indicates these two galaxy cluster candidates to be $z\sim1.6$ and $z\sim2$.  Additional Gemini/MOS observations and the grism observations presented in this study are consistent with both cluster candidates being at $z\sim2$.  

Using a combination of high-resolution HST imaging with grism spectroscopy, we identify galaxy merger candidates in IRC-A at $1.55\leq z_{grism}\leq 1.9$ and in IRC-B at $1.8\leq z_{grism}\leq 2.2$. By implementing a magnitude cut at $F160W \leq 23.0$, a pair separation of $\leq 3$'', and various photometric and/or grism redshift ranges, we measure the fraction of galaxy-galaxy mergers in both candidate clusters.  We compare our results to a control fraction measured in the COSMOS and CDFS fields selected using the same methods. The different selection methods applied to both the candidate clusters and field populations yield similar results for the measured pair fractions. 

To determine whether our merger fractions were affected by chance projections, we performed a randomization test on both the candidate cluster and field populations. This consists of randomizing the RA and DEC positions of each galaxy selected using method (III) (see \S \ref{sec:mergID} and Table \ref{selects}) and measuring the fraction of pairs found to be within $3\arcsec$ separation. This process is repeated 1000 times and the average pair fraction from all iterations is subtracted from the original pair fraction (listed in Table \ref{fracTable}). Correcting for false projections, we find corrected merger fractions of \Afraccor in IRC-A and \Bfraccor in IRC-B. For our COSMOS and CDFS populations we measure a corrected merger fraction of $7.2_{-1.0}^{+1.3}\%$ and $2.4_{-0.6}^{+0.9}\%$, respectively, giving us an average corrected merger fraction for the combined field population of \fieldfraccor. Overall, we find that the galaxy-galaxy merger fraction in the candidate clusters is enhanced compared to the field population, even when correcting for chance projections. Our results of an increase in the galaxy pair fraction with increasing redshift is consistent with overall field studies \citep{Lotz2008,Conselice2009,Lopez-Sanjuan2009,Bundy2009,DeRavel2009}. 

Although some studies \citep{Delahaye2017} report little to no increase in the pair fraction in clusters with respect to the field,  we note that our merger fractions in the clusters ($\sim11-13$\%) are comparable to their measurements.  We hypothesize that the discrepancy is due to differences in measurements of the  field merger fraction at $z\sim2$.  Our combination of precision photometric redshifts from ZFOURGE over a larger cosmic volume relative to existing studies finds a field merger fraction of $\sim5$\% at this epoch.  Our results are in good agreement with studies that report higher pair fractions in cluster environments in comparison to the field population \citep{Lotz2013,Coogan2018,Hine2016}. 

We find no conclusive evidence of enhanced star formation in merging galaxies within the candidate clusters with respect to the field. However, our findings imply that the dominance of quiescent mergers could be attributed to the dynamical state of the cluster. In IRC-A, at $z\sim 1.6$, all candidate cluster members identified as mergers are quiescent (as identified using the UVJ diagram in Figure \ref{uvj}) and, from the spatial distributions in Figure \ref{spatial}, are located within the cluster's core. We therefore hypothesize that IRC-A is likely a more evolved cluster. IRC-B, at $z\sim 2$, is a less evolved cluster, with only $\sim 52\%$ of candidate cluster members identified as mergers being quiescent, and all located outside the core. The differences in the spatial distributions and star forming populations of IRC-A and IRC-B supports the picture where evolution in clusters begins to separate at $z\sim 2$, postulated by other studies \citep{Brodwin2013,Tran2015,Kawinwanichakij2017,Papovich2018}.

The differences in the various methods of identifying merger candidates throughout previous studies \citep{Lopez-Sanjuan2009,Man2012,Man2016,Conselice2009,Williams2011,Tran2008,DeRavel2009,Lotz2008,Lotz2013,Bundy2009,Delahaye2017} presents a difficult challenge for direct comparison of merger fractions. The advantage to our study is our use of similar selection processes for both candidate clusters and the COSMOS field population, therefore our relative comparison of the candidate clusters' pair fractions with respect to the field is robust. However, given our limited statistics, more observations are need.

\acknowledgements

C. Watson and K. Tran are grateful for support by the National Science Foundation under Grant Number 1410728.
Support for program HST-GO-12896 was provided by NASA through a grant from the Space Telescope Science Institute, which is operated by the Association of Universities for Research in Astronomy, Inc., under NASA contract NAS 5-26555. This work is based on observations taken by the 3D-HST Treasury Program (GO 12177 and 12328) with the NASA/ESA HST, which is operated by the Association of Universities for Research in Astronomy, Inc., under NASA contract NAS5-26555. This work is based partly on observations obtained at the Kitt Peak National Observatory.  The National Optical Astronomy Observatory (NOAO) is operated by the Association of Universities for Research in Astronomy (AURA) under cooperative agreement with the National Science Foundation. The NEWFIRM Medium-Band Survey was supported by the National Science Foundation grant AST-0807974. The combination of NEWFIRM Medium-Band Survey and Spitzer data was supported by NASA?s Astrophysics Data Analysis Program, under grant number NNX11AB08B.

\bibliography{Paper2017.bib}

\begin{thebibliography}{}
\expandafter\ifx\csname natexlab\endcsname\relax\def\natexlab#1{#1}\fi

\bibitem[{Baugh {et~al.}(1996)Baugh, Cole, \& Frenk}]{Baugh1996}
Baugh, C.~M., Cole, S., \& Frenk, C.~S. 1996, Monthly Notices of the Royal
  Astronomical Society, 283, 1361

\bibitem[{Bertin(2006)}]{Bertin2006a}
Bertin, E. 2006, Astronomical Data Analysis Software and Systems XV, 351, 112

\bibitem[{Bertin \& Arnouts(1996)}]{Bertin1996}
Bertin, E., \& Arnouts, S. 1996, Astronomy {\&} Astrophysics, 117, 393

\bibitem[{Bezanson {et~al.}(2016)Bezanson, Wake, Brammer, van Dokkum, Franx,
  Labb{\'{e}}, Leja, Momcheva, Nelson, Quadri, Skelton, Weiner, \&
  Whitaker}]{Bezanson2015}
Bezanson, R., Wake, D.~a., Brammer, G.~B., {et~al.} 2016, The Astrophysical
  Journal, 822, 27

\bibitem[{Brammer {et~al.}(2008)Brammer, van Dokkum, \& Coppi}]{Brammer2008}
Brammer, G.~B., van Dokkum, P.~G., \& Coppi, P. 2008, The Astrophysical
  Journal, 686, 1503

\bibitem[{Brammer {et~al.}(2012)Brammer, van Dokkum, Franx, Fumagalli, Patel,
  Rix, Skelton, Kriek, Nelson, Schmidt, Bezanson, da~Cunha, Erb, Fan,
  {F{\"{o}}rster Schreiber}, Illingworth, Labb{\'{e}}, Leja, Lundgren, Magee,
  Marchesini, McCarthy, Momcheva, Muzzin, Quadri, Steidel, Tal, Wake, Whitaker,
  \& Williams}]{Brammer2012a}
Brammer, G.~B., van Dokkum, P.~G., Franx, M., {et~al.} 2012, The Astrophysical
  Journal Supplement Series, 200, 13

\bibitem[{Brodwin {et~al.}(2013)Brodwin, Stanford, Gonzalez, Zeimann, Snyder,
  Mancone, Pope, Eisenhardt, Stern, Alberts, Ashby, Brown, Galametz, Gettings,
  Jannuzi, Miller, Moustakas, \& Moustakas}]{Brodwin2013}
Brodwin, M., Stanford, S.~A., Gonzalez, A.~H., {et~al.} 2013, The Astrophysical
  Journal, 779, 15

\bibitem[{Bruzual \& Charlot(2003)}]{Bruzual2003}
Bruzual, G., \& Charlot, S. 2003, Monthly Notices of the Royal Astronomical
  Society, 344, 1000

\bibitem[{Bundy {et~al.}(2009)Bundy, Fukugita, Ellis, Targett, Belli, \&
  Kodama}]{Bundy2009}
Bundy, K., Fukugita, M., Ellis, R.~S., {et~al.} 2009, The Astrophysical
  Journal, 697, 1369

\bibitem[{Calzetti {et~al.}(2000)Calzetti, Armus, Bohlin, Kinney, Koornneef, \&
  Storchi‐Bergmann}]{Calzetti2000}
Calzetti, D., Armus, L., Bohlin, R.~C., {et~al.} 2000, The Astrophysical
  Journal, 533, 682

\bibitem[{Cameron(2011)}]{Cameron2011A}
Cameron, E. 2011, Publications of the Astronomical Society of Australia, 28,
  128

\bibitem[{Chabrier(2003)}]{Chabrier2003}
Chabrier, G. 2003, The Publications of the Astronomical Society of the Pacific,
  115, 763

\bibitem[{Conselice {et~al.}(2009)Conselice, Yang, \& Bluck}]{Conselice2009}
Conselice, C.~J., Yang, C., \& Bluck, A. F.~L. 2009, Monthly Notices of the
  Royal Astronomical Society, 394, 1956

\bibitem[{Coogan {et~al.}(2018)Coogan, Daddi, Sargent, Strazzullo, Valentino,
  Gobat, Magdis, Bethermin, Pannella, Onodera, Liu, Cimatti, Dannerbauer,
  Carollo, Renzini, \& Tremou}]{Coogan2018}
Coogan, R.~T., Daddi, E., Sargent, M.~T., {et~al.} 2018, Monthly Notices of the
  Royal Astronomical Society, 479, 703

\bibitem[{Davis {et~al.}(2006)Davis, Guhathakurta, Konidaris, Newman, Ashby,
  Biggs, Barmby, Bundy, Chapman, Coil, Conselice, Cooper, Croton, Eisenhardt,
  Ellis, Faber, Fang, Fazio, Georgakakis, Gerke, Goss, Gwyn, Harker, Hopkins,
  Huang, Ivison, Kassin, Kirby, Koekemoer, Koo, Laird, Floc'h, Lin, Lotz,
  Marshall, Martin, Metevier, Moustakas, Nandra, Noeske, Papovich, Phillips,
  Rich, Rieke, Rigopoulou, Salim, Schiminovich, Simard, Smail, Small, Weiner,
  Willmer, Willner, Wilson, Wright, \& Yan}]{Davis2006}
Davis, M., Guhathakurta, P., Konidaris, N., {et~al.} 2006, The Astrophysical
  Journal, 1

\bibitem[{de~Ravel {et~al.}(2009)de~Ravel, {Le Fevre}, Tresse, Bottini,
  Garilli, {Le Brun}, Maccagni, Scaramella, Scodeggio, Vettolani, Zanichelli,
  Adami, Arnouts, Bardelli, Bolzonella, Cappi, Charlot, Ciliegi, Contini,
  Foucaud, Franzetti, Gavignaud, Guzzo, Ilbert, Iovino, Lamareille, McCracken,
  Marano, Marinoni, Mazure, Meneux, Merighi, Paltani, Pello, Pollo, Pozzetti,
  Radovich, Vergani, Zamorani, Zucca, Bondi, Bongiorno, Brinchmann, Cucciati,
  {de La Torre}, Gregorini, Memeo, Perez-Montero, Mellier, Merluzzi, \&
  Temporin}]{DeRavel2009}
de~Ravel, L., {Le Fevre}, O., Tresse, L., {et~al.} 2009, Astronomy and
  Astrophysics, 498, 379

\bibitem[{Delahaye {et~al.}(2017)Delahaye, Webb, Nantais, DeGroot, Wilson,
  Muzzin, Yee, Foltz, Noble, Demarco, Tudorica, Cooper, Lidman, Perlmutter,
  Hayden, Boone, \& Surace}]{Delahaye2017}
Delahaye, A.~G., Webb, T. M.~A., Nantais, J., {et~al.} 2017, The Astrophysical
  Journal, 843, 9

\bibitem[{{Di Matteo} {et~al.}(2007){Di Matteo}, Combes, Melchior, \&
  Semelin}]{DiMatteo2007}
{Di Matteo}, P., Combes, F., Melchior, A.-L., \& Semelin, B. 2007, Astronomy
  {\&} Astrophysics, 468, 61

\bibitem[{Fakhouri \& Ma(2008)}]{Fakhouri2008}
Fakhouri, O., \& Ma, C.~P. 2008, Monthly Notices of the Royal Astronomical
  Society, 386, 577

\bibitem[{Giacconi {et~al.}(2002)Giacconi, Zirm, Wang, Rosati, Nonino, Tozzi,
  Gilli, Mainieri, Hasinger, Kewley, Bergeron, Borgani, Gilmozzi, Grogin,
  Koekemoer, Schreier, Zheng, \& Norman}]{Giacconi2002}
Giacconi, R., Zirm, A., Wang, J., {et~al.} 2002, The Astrophysical Journal
  Supplement Series, 139, 369

\bibitem[{Giavalisco {et~al.}(2004)Giavalisco, Ferguson, Koekemoer, Dickinson,
  Alexander, Bauer, Bergeron, Biagetti, Brandt, Casertano, Cesarsky,
  Chatzichristou, Conselice, Cristiani, Costa, Dahlen, Mello, Eisenhardt,
  Erben, Fall, Fassnacht, Fosbury, Fruchter, Gardner, Grogin, Hook,
  Hornschemeier, Idzi, Jogee, Kretchmer, Laidler, Lee, Livio, Lucas, Madau,
  Mobasher, Moustakas, Nonino, Padovani, Papovich, Park, Ravindranath, Renzini,
  Richardson, Riess, Rosati, Schirmer, Schreier, Somerville, Spinrad, Stern,
  Stiavelli, \& Strolger}]{Giavalisco2004}
Giavalisco, M., Ferguson, H.~C., Koekemoer, A.~M., {et~al.} 2004, The
  Astrophysical Journal, 600, L93

\bibitem[{Hine {et~al.}(2016)Hine, Geach, Alexander, Lehmer, Chapman, \&
  Matsuda}]{Hine2016}
Hine, N.~K., Geach, J.~E., Alexander, D.~M., {et~al.} 2016, Monthly Notices of
  the Royal Astronomical Society, 455, 2363

\bibitem[{Kauffmann(1996)}]{Kauffmann1996}
Kauffmann, G. 1996, Monthly Notices of the Royal Astronomical Society, 281, 487

\bibitem[{Kawinwanichakij {et~al.}(2017)Kawinwanichakij, Papovich, Quadri,
  Glazebrook, Kacprzak, Allen, Bell, Croton, Dekel, Ferguson, Forrest, Grogin,
  Guo, Kocevski, Koekemoer, Labb{\'{e}}, Lucas, Nanayakkara, Spitler,
  Straatman, Tran, Tomczak, \& van Dokkum}]{Kawinwanichakij2017}
Kawinwanichakij, L., Papovich, C., Quadri, R.~F., {et~al.} 2017, The
  Astrophysical Journal, 847, 134

\bibitem[{Kriek {et~al.}(2009)Kriek, van Dokkum, Labbe, Franx, Illingworth,
  Marchesini, \& Quadri}]{Kriek2009}
Kriek, M., van Dokkum, P.~G., Labbe, I., {et~al.} 2009, The Astrophysical
  Journal, 700, 221

\bibitem[{{Le F{\`{e}}vre} {et~al.}(2005){Le F{\`{e}}vre}, Vettolani, Maccagni,
  Picat, Adami, Arnaboldi, Bardelli, Bondi, Bottini, Bolzonella, Busarello,
  Cappi, Ciliegi, Contini, Charlot, Foucaud, Franzetti, Garilli, Gavignaud,
  Guzzo, Ilbert, Iovino, Brun, Marano, Marinoni, McCracken, Mathez, Mazure,
  Mellier, Meneux, Merluzzi, Paltani, Pell{\`{o}}, Pollo, Pozzetti, Radovich,
  Rizzo, Scaramella, Scodeggio, Tresse, Zamorani, Zanichelli, \&
  Zucca}]{LeFevre2005}
{Le F{\`{e}}vre}, O., Vettolani, G., Maccagni, D., {et~al.} 2005, Astronomy
  {\&} Astrophysics, 439, 845

\bibitem[{Lidman {et~al.}(2013)Lidman, Iacobuta, Bauer, Barrientos, Cerulo,
  Couch, Delaye, Demarco, Ellingson, Faloon, Gilbank, Huertas-Company, Mei,
  Meyers, Muzzin, Noble, Nantais, Rettura, Rosati, S??nchez-Janssen,
  Strazzullo, Webb, Wilson, Yan, \& Yee}]{Lidman2013}
Lidman, C., Iacobuta, G., Bauer, A.~E., {et~al.} 2013, Monthly Notices of the
  Royal Astronomical Society, 433, 825

\bibitem[{Lonsdale {et~al.}(2003)Lonsdale, Smith, Rowan-robinson, Surace,
  Shupe, \& Xu}]{Lonsdale2003}
Lonsdale, C.~J., Smith, H.~E., Rowan-robinson, M., {et~al.} 2003, Publications
  of the Astronomical Society of the Pacific, 115, 897

\bibitem[{Lopez-Sanjuan {et~al.}(2009)Lopez-Sanjuan, Balcells, Pablo, Barro,
  Garc, \& May}]{Lopez-Sanjuan2009}
Lopez-Sanjuan, C., Balcells, M., Pablo, G.~P., {et~al.} 2009, Astronomy and
  Astrophysics, 501, 505

\bibitem[{Lotz {et~al.}(2011)Lotz, Jonsson, Cox, Croton, Primack, Somerville,
  \& Stewart}]{Lotz2011}
Lotz, J.~M., Jonsson, P., Cox, T.~J., {et~al.} 2011, The Astrophysical Journal,
  742, 103

\bibitem[{Lotz {et~al.}(2008)Lotz, Davis, Faber, Guhathakurta, Gwyn, Huang,
  Koo, Lin, Newman, Noeske, Papovich, Willmer, Coil, Conselice, Cooper,
  Hopkins, Metevier, Primack, Rieke, \& Weiner}]{Lotz2008}
Lotz, J.~M., Davis, M., Faber, S.~M., {et~al.} 2008, The Astrophysical Journal,
  672, 177

\bibitem[{Lotz {et~al.}(2013)Lotz, Papovich, Faber, Ferguson, Grogin, Guo,
  Kocevski, Koekemoer, Lee, McIntosh, Momcheva, Rudnick, Saintonge, Tran,
  van~der Wel, \& Willmer}]{Lotz2013}
Lotz, J.~M., Papovich, C., Faber, S.~M., {et~al.} 2013, The Astrophysical
  Journal, 773, 154

\bibitem[{Man {et~al.}(2012)Man, Toft, Zirm, Wuyts, \& van~der Wel}]{Man2012}
Man, A., Toft, S., Zirm, A., Wuyts, S., \& van~der Wel, A. 2012, The
  Astrophysical Journal, 744, 85

\bibitem[{Man {et~al.}(2016)Man, Zirm, \& Toft}]{Man2016}
Man, A. W.~S., Zirm, A.~W., \& Toft, S. 2016, The Astrophysical Journal, 830,
  arXiv:arXiv:1410.3479v1

\bibitem[{McIntosh {et~al.}(2008)McIntosh, Guo, Hertzberg, Katz, Mo, {Van Den
  Bosch}, \& Yang}]{McIntosh2008}
McIntosh, D.~H., Guo, Y., Hertzberg, J., {et~al.} 2008, Monthly Notices of the
  Royal Astronomical Society, 388, 1537

\bibitem[{Merlin {et~al.}(2015)Merlin, Fontana, Ferguson, Dunlop, Elbaz,
  Bourne, Bruce, Buitrago, Castellano, Schreiber, Grazian, McLure, Okumura,
  Shu, Wang, Amorin, Boutsia, Cappelluti, Comastri, Derriere, Faber, \&
  Santini}]{Merlin2015a}
Merlin, E., Fontana, A., Ferguson, H.~C., {et~al.} 2015, Astronomy {\&}
  Astrophysics, 582, 21

\bibitem[{Mihos \& Hernquist(1996)}]{Mihos1994}
Mihos, J.~C., \& Hernquist, L. 1996, The Astrophysical Journal, 431, L9

\bibitem[{Momcheva {et~al.}(2016)Momcheva, Brammer, van Dokkum, Skelton,
  Whitaker, Nelson, Fumagalli, Maseda, Leja, Franx, Rix, Bezanson, {Da Cunha},
  Dickey, Schreiber, Illingworth, Kriek, Labb{\'{e}}, Lange, Lundgren, Magee,
  Marchesini, Oesch, Pacifici, Patel, Price, Tal, Wake, van~der Wel, \&
  Wuyts}]{Momcheva2015}
Momcheva, I.~G., Brammer, G.~B., van Dokkum, P.~G., {et~al.} 2016, The
  Astrophysical Journal, 225, 27

\bibitem[{Nanayakkara {et~al.}(2016)Nanayakkara, Glazebrook, Kacprzak, Yuan,
  Tran, Spitler, Kewley, Straatman, Cowley, Fisher, Labbe, Tomczak, Allen, \&
  Alcorn}]{Nanayakkara2016a}
Nanayakkara, T., Glazebrook, K., Kacprzak, G.~G., {et~al.} 2016, The
  Astrophysical Journal, 828, 1

\bibitem[{Nelson {et~al.}(2017)Nelson, Pillepich, Springel, Weinberger,
  Hernquist, Pakmor, Genel, Torrey, Vogelsberger, Kauffmann, Marinacci, \&
  Naiman}]{Nelson2017}
Nelson, D., Pillepich, A., Springel, V., {et~al.} 2017, Monthly Notices of the
  Royal Astronomical Society, 475, 624

\bibitem[{{Omer Jr.} {et~al.}(1965){Omer Jr.}, Page, \& Wilson}]{OmerJr.1965}
{Omer Jr.}, G.~C., Page, T.~L., \& Wilson, A.~G. 1965, The Astronomical
  Journal, 70

\bibitem[{Papovich(2008)}]{Papovich2008}
Papovich, C. 2008, The Astrophysical Journal, 676, 206

\bibitem[{Papovich {et~al.}(2010)Papovich, Momcheva, Willmer, Finkelstein,
  Finkelstein, Tran, Brodwin, Dunlop, Farrah, Khan, Lotz, McCarthy, McLure,
  Rieke, Rudnick, Sivanandam, Pacaud, \& Pierre}]{Papovich2010}
Papovich, C., Momcheva, I., Willmer, C. N.~A., {et~al.} 2010, The Astrophysical
  Journal, 1

\bibitem[{Papovich {et~al.}(2012)Papovich, Bassett, Lotz, {Van Der Wel}, Tran,
  Finkelstein, Bell, Conselice, Dekel, Dunlop, Guo, Faber, Farrah, Ferguson,
  Finkelstein, H{\"{a}}ussler, Kocevski, Koekemoer, Koo, McGrath, McLure,
  McIntosh, Momcheva, Newman, Rudnick, Weiner, Willmer, \&
  Wuyts}]{Papovich2012}
Papovich, C., Bassett, R., Lotz, J.~M., {et~al.} 2012, Astrophysical Journal,
  750, arXiv:1110.3794

\bibitem[{Papovich {et~al.}(2018)Papovich, Kawinwanichakij, Quadri, Glazebrook,
  Labbe, Tran, Forrest, Kacprzak, Spitler, Straatman, \&
  Tomczak}]{Papovich2018}
Papovich, C., Kawinwanichakij, L., Quadri, R., {et~al.} 2018, The Astrophysical
  Journal, 854, 30

\bibitem[{Peebles(1970)}]{Peebles1970}
Peebles, P. J.~E. 1970, The Astronomical Journal, 75, 13

\bibitem[{Pillepich {et~al.}(2017)Pillepich, Nelson, Hernquist, Springe,
  {R{\"{u}}diger Pakmor}, Torrey, Weinberger, Gene, Naiman, Marinacci, \&
  Vogelsberger}]{Pillepich2017}
Pillepich, A., Nelson, D., Hernquist, L., {et~al.} 2017, Monthly Notices of the
  Royal Astronomical Society, 475, 648

\bibitem[{Rix {et~al.}(2004)Rix, Barden, Beckwith, Bell, Borch, Caldwell,
  Haussler, Jahnke, Jogee, McIntosh, Meisenheimer, Peng, Sanchez, Somerville,
  Wisotzki, \& Wolf}]{Rix2004}
Rix, H., Barden, M., Beckwith, S. V.~W., {et~al.} 2004, The Astrophysical
  Journal Supplement Series, 152, 163

\bibitem[{Rodriguez-Gomez {et~al.}(2015)Rodriguez-Gomez, Genel, Vogelsberger,
  Sijacki, Pillepich, Sales, Torrey, Snyder, Nelson, Springel, Ma, \&
  Hernquist}]{Rodriguez-gomez2015}
Rodriguez-Gomez, V., Genel, S., Vogelsberger, M., {et~al.} 2015, Monthly
  Notices of the Royal Astronomical Society, 449, 49

\bibitem[{{Ryan, Jr.} {et~al.}(2008){Ryan, Jr.}, Cohen, Windhorst, Silk, Ryan,
  Cohen, Windhorst, \& Silk}]{RyanJr.2008}
{Ryan, Jr.}, R.~E., Cohen, S.~H., Windhorst, R.~A., {et~al.} 2008, The
  Astrophysical Journal, 678, 751

\bibitem[{Scoville(2007)}]{Scoville2007}
Scoville, N. 2007, Bulletin of the American Astronomical Society, 205, 1467

\bibitem[{Scoville {et~al.}(2013)Scoville, Arnouts, Aussel, Benson, Bongiorno,
  Bundy, Calvo, Capak, Carollo, Civano, Dunlop, Elvis, Faisst, Finoguenov, Fu,
  Giavalisco, Guo, Ilbert, Iovino, Kajisawa, Kartaltepe, Leauthaud, {Le
  F{\`{e}}vre}, Lefloch, Lilly, Liu, Manohar, Massey, Masters, McCracken,
  Mobasher, Peng, Renzini, Rhodes, Salvato, Sanders, Sarvestani, Scarlata,
  Schinnerer, Sheth, Shopbell, Smol{\v{c}}i{\'{c}}, Taniguchi, Taylor, White,
  \& Yan}]{Scoville2013}
Scoville, N., Arnouts, S., Aussel, H., {et~al.} 2013, Astrophysical Journal,
  Supplement Series, 206, arXiv:arXiv:1303.6689v1

\bibitem[{Skelton {et~al.}(2014)Skelton, Whitaker, Momcheva, Brammer, van
  Dokkum, Labbe, Franx, van~der Wel, Bezanson, {Da Cunha}, Fumagalli,
  Schreiber, Kriek, Leja, Lundgren, Magee, Marchesini, Maseda, Nelson, Oesch,
  Pacifici, Patel, Price, Rix, Tal, Wake, Wuyts, Labb{\'{e}}, Franx, van~der
  Wel, Bezanson, {Da Cunha}, Fumagalli, {F{\"{o}}rster Schreiber}, Kriek, Leja,
  Lundgren, Magee, Marchesini, Maseda, Nelson, Oesch, Pacifici, Patel, Price,
  Rix, Tal, Wake, \& Wuyts}]{Skelton2014}
Skelton, R.~E., Whitaker, K.~E., Momcheva, I.~G., {et~al.} 2014, The
  Astrophysical Journal Supplement Series, 214, 24

\bibitem[{Spitler {et~al.}(2012)Spitler, Labb{\'{e}}, Glazebrook, Persson,
  Monson, Papovich, Tran, Poole, Quadri, {Van Dokkum}, Kelson, Kacprzak,
  McCarthy, Murphy, Straatman, \& Tilvi}]{Spitler2012}
Spitler, L.~R., Labb{\'{e}}, I., Glazebrook, K., {et~al.} 2012, Astrophysical
  Journal Letters, 748, 1

\bibitem[{Straatman {et~al.}(2016)Straatman, Glazebrook, Eric, Spitler, Quadri,
  Labb, Papovich, Tran, Brammer, Cowley, Tomczak, Nanayakkara, Alcorn, Allen,
  Broussard, Dokkum, Forrest, Houdt, Kacprzak, Kawinwanichakij, Kelson, Lee,
  Mccarthy, Mehrtens, Monson, Murphy, Rees, Tilvi, \& Whitaker}]{Straatman2016}
Straatman, C., Glazebrook, K., Eric, S., {et~al.} 2016, The Astrophysical
  Journal, 830, 1

\bibitem[{Tanaka {et~al.}(2010)Tanaka, Finoguenov, \& Ueda}]{Tanaka2010}
Tanaka, M., Finoguenov, A., \& Ueda, Y. 2010, Astrophysical Journal Letters,
  716, 152

\bibitem[{Toomre \& Toomre(1972)}]{Toomre1972}
Toomre, A., \& Toomre, J. 1972, Astrophysical Journal, 178, 623

\bibitem[{Tran {et~al.}(2008)Tran, Moustakas, Gonzalez, Bai, Zaritsky, \&
  Kautsch}]{Tran2008}
Tran, K.-V.~H., Moustakas, J., Gonzalez, A.~H., {et~al.} 2008, The
  Astrophysical Journal Letters, 683, L17

\bibitem[{Tran {et~al.}(2005)Tran, {Van Dokkum}, Franx, Illingworth, Kelson, \&
  Schreiber}]{Tran2005}
Tran, K.-V.~H., {Van Dokkum}, P., Franx, M., {et~al.} 2005, ArXiv eprints, 627,
  L25

\bibitem[{Tran {et~al.}(2010)Tran, Papovich, Saintonge, Brodwin, Dunlop,
  Farrah, Finkelstein, Finkelstein, Lotz, McLure, Momcheva, \&
  Willmer}]{Tran2010}
Tran, K. V.~H., Papovich, C., Saintonge, A., {et~al.} 2010, Astrophysical
  Journal Letters, 719, 126

\bibitem[{Tran {et~al.}(2015)Tran, Nanayakkara, Yuan, Kacprzak, Glazebrook,
  Kewley, Momcheva, Papovich, Quadri, Rudnick, Saintonge, Spitler, Straatman,
  \& Tomczak}]{Tran2015}
Tran, K. V.~H., Nanayakkara, T., Yuan, T., {et~al.} 2015, Astrophysical
  Journal, 811, 28

\bibitem[{van~der Walt {et~al.}(2014)van~der Walt, Sch{\"{o}}nberger,
  Nunez-Iglesias, Boulogne, Warner, Yager, Gouillart, \& Yu}]{vanderWalt2014}
van~der Walt, S., Sch{\"{o}}nberger, J.~L., Nunez-Iglesias, J., {et~al.} 2014,
  ArXiv e-prints, 1407.6245, arXiv:arXiv:1407.6245v1

\bibitem[{van Dokkum {et~al.}(1999)van Dokkum, Franx, Fabricant, Kelson, \&
  Illingworth}]{VanDokkum1999}
van Dokkum, P.~G., Franx, M., Fabricant, D., Kelson, D.~D., \& Illingworth,
  G.~D. 1999, The Astrophysical Journal, 520, L95

\bibitem[{van Dokkum {et~al.}(2009)van Dokkum, Labb{\'{e}}, Marchesini, Quadri,
  Brammer, Whitaker, Kriek, Franx, Rudnick, Illingworth, Lee, \&
  Muzzin}]{VanDokkum2009}
van Dokkum, P.~G., Labb{\'{e}}, I., Marchesini, D., {et~al.} 2009, Publications
  of the Astronomical Society of the Pacific, 121, 2

\bibitem[{van Dokkum {et~al.}(2011)van Dokkum, Brammer, Fumagalli, Nelson,
  Franx, Rix, Kriek, Skelton, Patel, Schmidt, Bezanson, Bian, da~Cunha, Erb,
  Fan, Schreiber, Illingworth, Labbe, Lundgren, Magee, Marchesini, McCarthy,
  Muzzin, Quadri, Steidel, Tal, Wake, Whitaker, \& Williams}]{VanDokkum2011}
van Dokkum, P.~G., Brammer, G., Fumagalli, M., {et~al.} 2011, The Astrophysical
  Journal Letters, 743, L15

\bibitem[{Whitaker {et~al.}(2011)Whitaker, Labb{\'{e}}, {Van Dokkum}, Brammer,
  Kriek, Marchesini, Quadri, Franx, Muzzin, Williams, Bezanson, Illingworth,
  Lee, Lundgren, Nelson, Rudnick, Tal, \& Wake}]{Whitaker2011}
Whitaker, K.~E., Labb{\'{e}}, I., {Van Dokkum}, P.~G., {et~al.} 2011,
  Astrophysical Journal, 735, arXiv:1105.4609

\bibitem[{Williams {et~al.}(2011)Williams, Quadri, \& Franx}]{Williams2011}
Williams, R.~J., Quadri, R.~F., \& Franx, M. 2011, The Astrophysical Journal,
  738, L25

\bibitem[{Yuan {et~al.}(2014)Yuan, Nanayakkara, Kacprzak, Tran, Glazebrook,
  Kewley, Spitler, Poole, Labb{\'{e}}, Straatman, \& Tomczak}]{Yuan2014}
Yuan, T., Nanayakkara, T., Kacprzak, G.~G., {et~al.} 2014, Astrophysical
  Journal Letters, 795, 2

\bibitem[{Zwicky(1957)}]{Zwicky1957}
Zwicky, F. 1957, Zeitschrift f{\{}$\backslash$"{\{}u{\}}{\}}r Astrophysik, 44,
  64

\end{thebibliography}

\newpage

\begin{table*}[h!]
\begin{center}
\caption{Galaxy-galaxy merger fractions for IRC-A, IRC-B, COSMOS, and CDFS for different selection processes (see Table \ref{selects}) for $F160W \leq 23.0$. Uncertainties are calculated assuming binomial statistics for $68\%$ confidence intervals. \label{fracTable}}
\begin{tabular}{ccccc}
&&&&\\
\hline \hline
                        		&Selection&	\multirow{2}{*}{$N_{tot}$}& \multirow{2}{*}{$N_{merg,3"}$}& \multirow{2}{*}{$f_{merg,3"}$}   \\ 
                                &		Process	\tablenotemark{a}&	&	\\ \hline	
\multirow{3}{*}{IRC-A}& 							I&	 		27& 			4&			$15_{-4.4}^{+9.4}\%$ \\ 
				&							II&		 	34& 			4&			$12_{-3.5}^{+7.8}\%$\\ 
				&							III&		 	28&	 		4& 			$14_{-4.3}^{+9.1}\%$\\ \hline
\multirow{3}{*}{IRC-B}&							I&		 	29& 			6& 			$21_{-5.5}^{+9.3}\%$ \\ 
				&							II&	 		32& 			6&			$19_{-5.1}^{+8.7}\%$\\ 
				&							III&			 33& 			8& 			$24_{-5.8}^{+8.9}\%$\\ \hline
\multirow{2}{*}{COSMOS\tablenotemark{b}}& 			I&			453&			46&			$10_{-1.3}^{+1.6}\%$ \\ 
					&						III&			455&			46&			$10_{-1.2}^{+1.6}\%$\\ \hline
\multirow{2}{*}{CDFS}& 							I&			427&			14&			$3.3^{+1.1}_{-0.7}\%$ \\
				&							III&			433&			16&			$3.7^{+1.1}_{-0.7}\%$ \\ \hline
\multirow{2}{*}{Field Population\tablenotemark{c}}&		I&			800&			60&			$6.7^{+1.4}_{-1.0}\%$\\ 
					&						III&			888&			62&			$6.9^{+1.4}_{-1.0}\%$\\ \hline

\end{tabular}
\end{center}
\tablenotetext{a}{See Table \ref{selects}}
\tablenotetext{b}{Masked to exclude confirmed members of $z=1.62$ cluster \citep{Yuan2014}.}
\tablenotetext{c}{Combined COSMOS and CDFS populations}
\end{table*}

\begin{figure*}[h!]
\plottwo{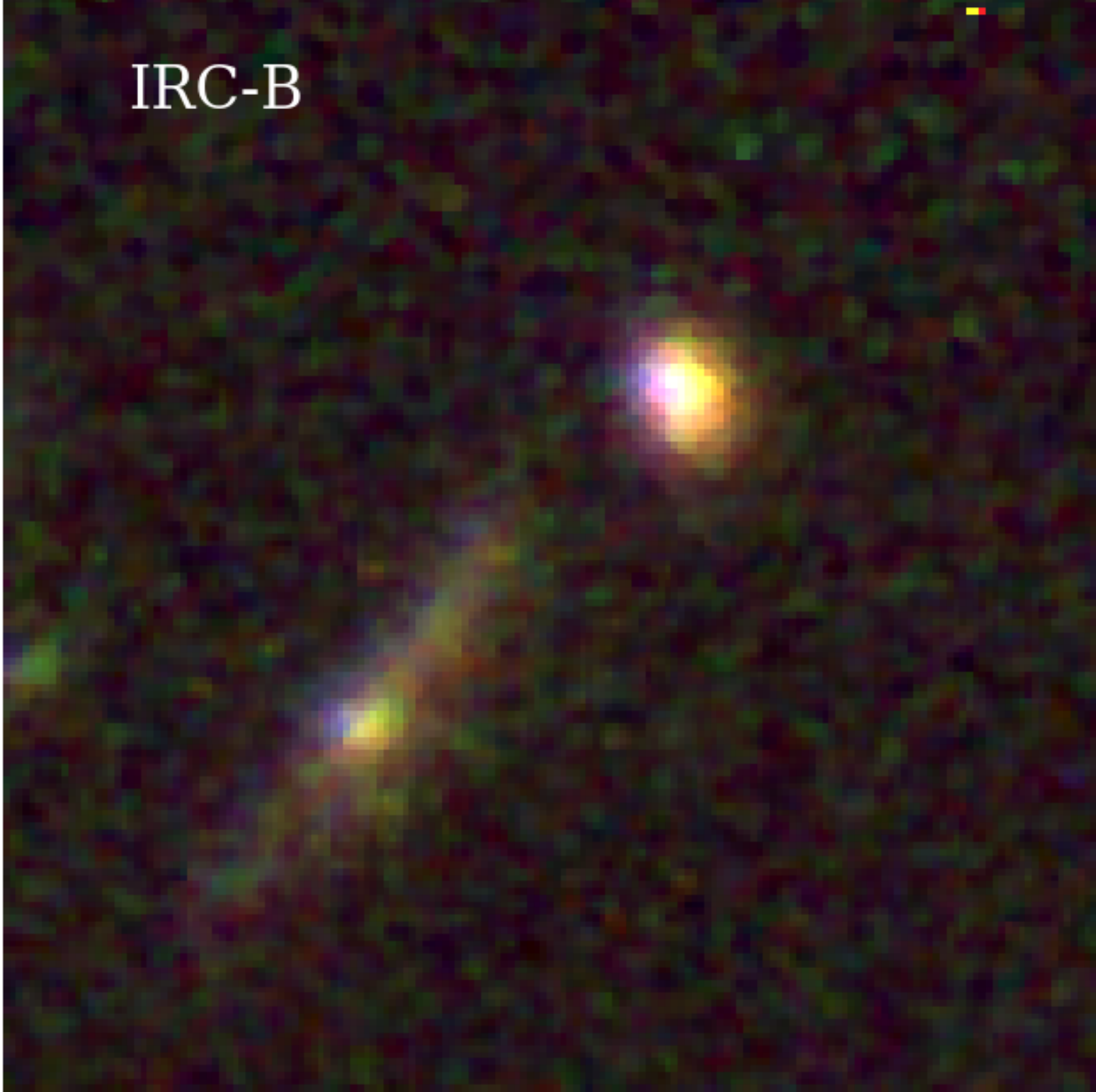}{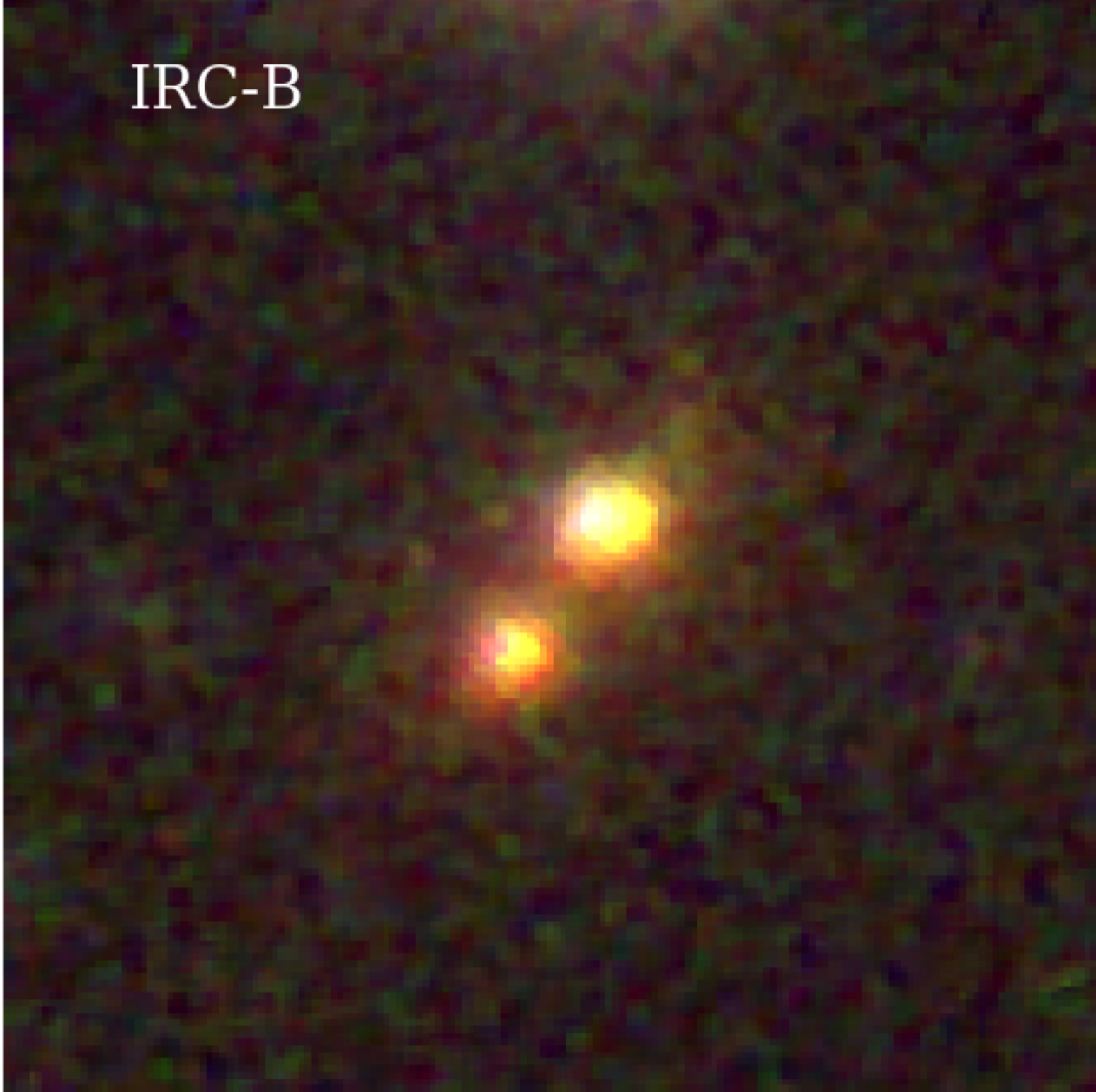}
\plottwo{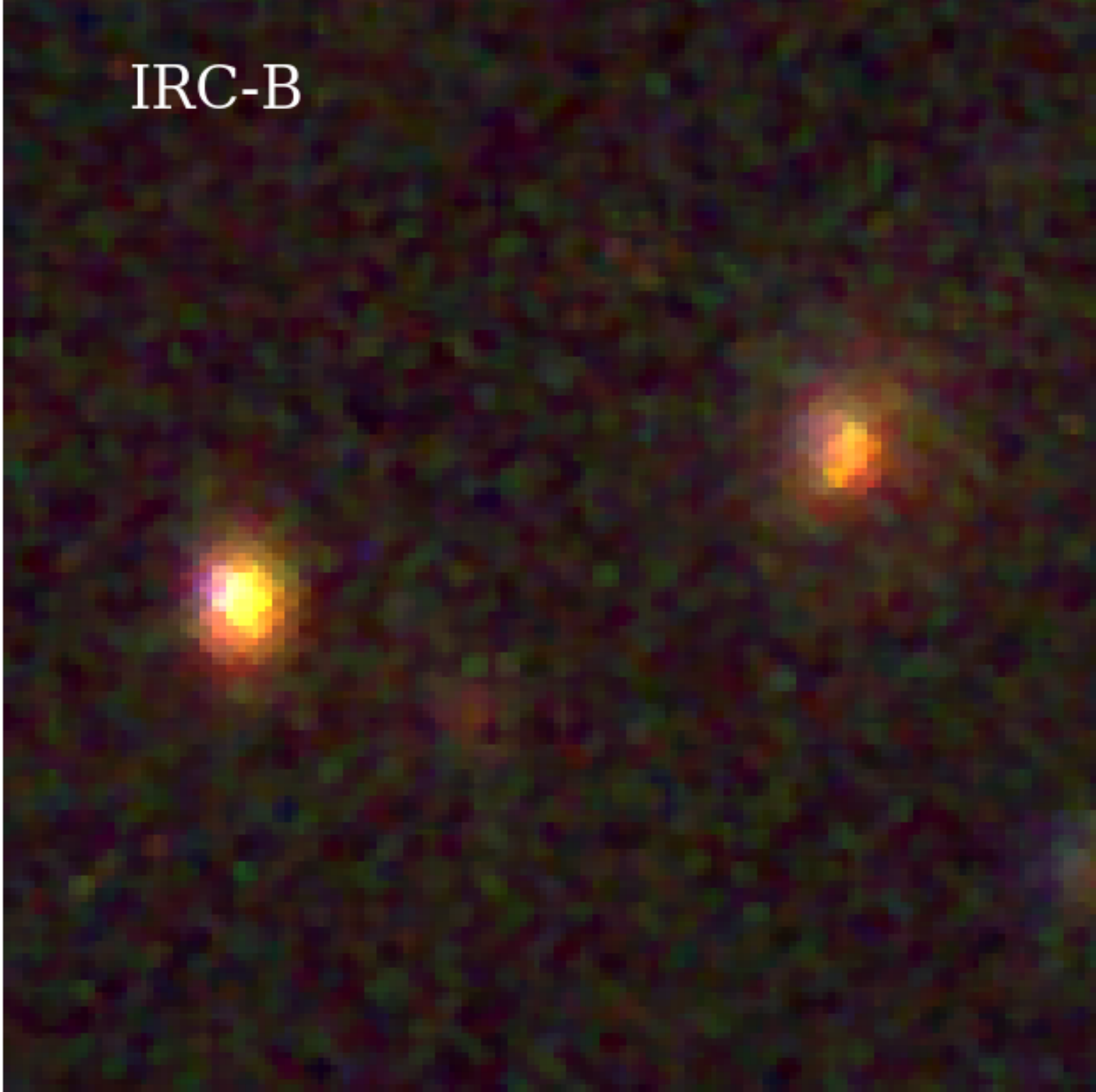}{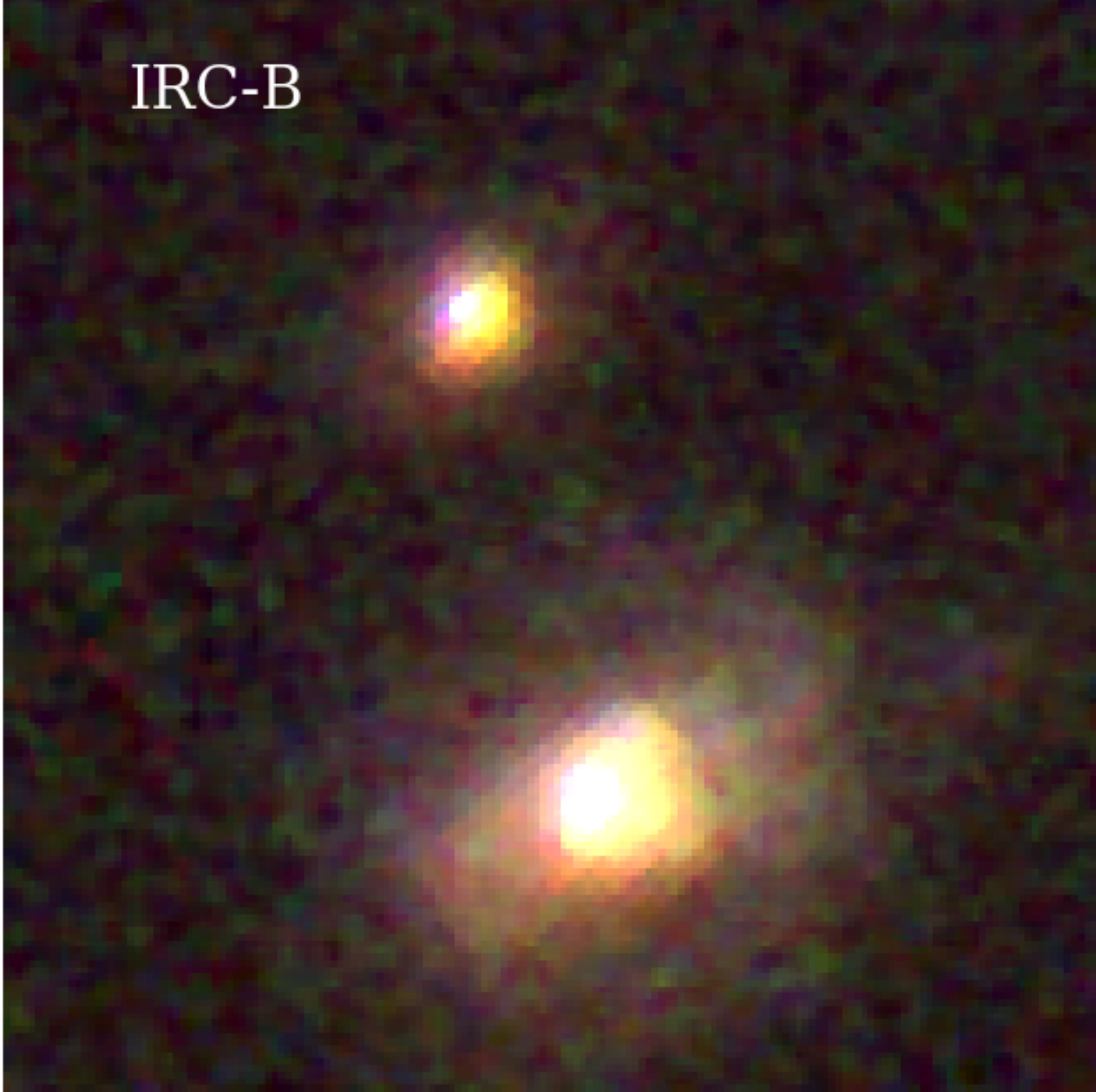}
\caption{Thumbnails of the mergers candidates in IRC-B. Thumbnails were created by stacking each of the F160W, F125W, and F105W filters to create an RGB image with a 0.03\arcsec/pix scale, resulting in a 5\arcsec$\times$5\arcsec snapshot of the merger candidates. \emph{Top Left:} Object IDs 105 \& 116, located 69\arcsec from the cluster center. \emph{Top Right:} Object IDs 646 \& 655, located 57\arcsec from the cluster center. \emph{Bottom Left:} Object IDs 126 \& 146, located 57\arcsec from the cluster center. \emph{Bottom Right:} Object IDs 300 \& 318, located 55'' from the cluster center. \label{panB}}
\vspace{80pt}
\end{figure*}

\newpage

\begin{deluxetable}{cccccccccccccc}[ht!]
\tablecaption{Photometry measurements and coordinate information for galaxies in IRC-A that satisfy selection method III (see \S \ref{sec:III}): grism redshifts within the cluster's redshift range ($1.5\leq z_{grism} \leq 1.9$) with \dq\ = 1 and \qz\ $\geq 2$ or photometric redshifts between $1.5\leq z_{phot} \leq 2.5$, and F160W magnitdue $\leq 23.0$. \label{TabA}}
\tablehead{
	\colhead{ID} & 
	\colhead{HST RA\tablenotemark{c}} & 
	\colhead{HST DEC\tablenotemark{c}} & 
	\colhead{$z_{grism}$} & 
	\colhead{$z_{phot}$} &
	 \colhead{Stellar Mass} & 
	 \colhead{F160W\tablenotemark{d}} & 
	 \colhead{F125W\tablenotemark{d}} & 
	 \colhead{U\tablenotemark{e}} & 
	 \colhead{V\tablenotemark{e}} &
	 \colhead{J\tablenotemark{e}} & 
	 \colhead{$DQ$\tablenotemark{f}} & 
	 \colhead{$Q_z$\tablenotemark{f}} &
	 \colhead{Merger Flag\tablenotemark{g}}
	  }
\startdata
31 & 2:22:02.479 & -4:13:12.80 & 0.750 & 1.692 & 10.42 & 22.31 & 22.82 & 24.00 & 22.40 & 21.90 & 0 & 0 & 0 \\
57 & 2:22:04.202 & -4:13:01.10 & 2.157 & 1.617 & 10.37 & 22.59 & 23.10 & 24.18 & 22.73 & 22.01 & 0 & 0 & 0 \\
70 & 2:22:01.859 & -4:12:58.11 & 1.382 & 1.818 & 10.60 & 21.93 & 22.34 & 23.26 & 21.92 & 21.39 & 0 & 0 & 0 \\
87 & 2:22:06.201 & -4:12:42.09 & ... & 1.670 & 10.02 & 22.51 & 22.73 & 23.44 & 22.55 & 20.87 & ... & ... & 1 \\
91 & 2:22:05.698 & -4:12:53.24 & 1.112 & 2.217 & 10.56 & 22.38 & 23.20 & 23.58 & 22.12 & 21.66 & 1 & 2 & 0 \\
116 & 2:21:59.422 & -4:12:44.55 & 1.580 & 1.573 & 10.35 & 22.27 & 22.78 & 24.29 & 22.52 & 21.88 & 0 & 0 & 0 \\
120 & 2:22:06.140 & -4:12:40.55 & 1.224 & 2.139 & 10.98 & 21.09 & 21.63 & 21.95 & 20.95 & 20.15 & 1 & 3 & 0 \\
131 & 2:21:58.872 & -4:12:41.23 & ... & 1.711 & 10.62 & 22.07 & 22.60 & 23.93 & 22.15 & 21.50 & ... & ... & 0 \\
142 & 2:22:00.043 & -4:12:35.10 & 0.619 & 1.625 & 11.13 & 21.19 & 21.78 & 23.26 & 21.37 & 20.39 & 0 & 0 & 0 \\
208 & 2:22:03.054 & -4:12:20.46 & 1.698 & 1.733 & 10.59 & 21.93 & 22.45 & 23.72 & 21.99 & 21.50 & 1 & 3 & 0 \\
218 & 2:21:59.519 & -4:12:17.69 & 1.456 & 1.531 & 10.56 & 22.19 & 22.59 & 23.85 & 22.43 & 21.21 & 1 & 3 & 0 \\
235 & 2:22:03.296 & -4:12:09.42 & 1.431 & 1.655 & 11.34 & 20.80 & 21.44 & 22.93 & 20.96 & 19.97 & 1 & 3 & 1 \\
241 & 2:22:03.229 & -4:12:09.52 & 1.430 & 1.831 & 11.04 & 21.56 & 22.27 & 23.46 & 21.52 & 20.73 & 1 & 3 & 1 \\
252 & 2:22:03.707 & -4:12:10.01 & 1.372 & 1.631 & 10.29 & 22.56 & 23.03 & 24.33 & 22.69 & 22.20 & 1 & 2 & 0 \\
261 & 2:22:03.470 & -4:12:07.20 & 1.272 & 1.692 & 10.49 & 22.92 & 23.51 & 24.82 & 23.02 & 22.04 & 1 & 2 & 0 \\
286 & 2:21:58.780 & -4:12:02.86 & 1.122 & 1.652 & 10.76 & 21.87 & 22.45 & 23.54 & 21.99 & 21.09 & 0 & 0 & 0 \\
321 & 2:22:03.655 & -4:11:55.41 & 1.174 & 1.647 & 10.62 & 21.60 & 22.04 & 23.31 & 21.71 & 21.36 & 1 & 3 & 0 \\
327 & 2:22:05.065 & -4:11:51.33 & 1.681 & 1.677 & 10.73 & 21.53 & 22.11 & 23.48 & 21.65 & 21.19 & 1 & 3 & 1 \\
329 & 2:22:05.037 & -4:11:53.27 & 1.626 & 1.782 & 10.96 & 21.90 & 22.58 & 23.91 & 21.93 & 20.99 & 1 & 3 & 1 \\
398 & 2:22:00.540 & -4:11:34.16 & 1.331 & 1.939 & 10.96 & 21.63 & 22.04 & 23.15 & 21.51 & 20.79 & 1 & 3 & 0 \\
399 & 2:22:06.250 & -4:11:35.06 & 1.682 & 1.640 & 10.60 & 21.73 & 22.16 & 23.47 & 21.84 & 21.24 & 1 & 3 & 0 \\
400 & 2:22:06.111 & -4:11:32.74 & 1.685 & 1.706 & 10.73 & 21.50 & 21.95 & 23.19 & 21.56 & 21.10 & 1 & 3 & 0 \\
411 & 2:22:06.781 & -4:11:31.12 & 1.569 & 1.808 & 10.61 & 22.57 & 23.20 & 24.25 & 22.56 & 21.59 & 1 & 2 & 0 \\
415 & 2:22:04.617 & -4:11:28.38 & 1.201 & 1.676 & 10.62 & 21.66 & 22.02 & 22.93 & 21.74 & 21.23 & 1 & 2 & 0 \\
418 & 2:22:07.309 & -4:11:30.44 & 1.518 & 1.532 & 10.09 & 22.89 & 23.38 & 25.02 & 23.15 & 22.48 & 1 & 2 & 0 \\
436 & 2:22:00.868 & -4:11:24.38 & 0.263 & 1.544 & 10.15 & 22.44 & 22.95 & 24.31 & 22.71 & 22.36 & 0 & 0 & 0 \\
437 & 2:22:07.079 & -4:11:25.53 & 1.792 & 2.153 & 10.66 & 21.98 & 22.38 & 22.71 & 21.85 & 21.14 & 1 & 3 & 0 \\
492 & 2:22:01.791 & -4:11:05.35 & 1.755 & 0.864 & 10.56 & 21.66 & 22.00 & 24.79 & 22.44 & 20.97 & 1 & 3 & 0 \\
\enddata
\tablenotetext{c}{HST coordinate information were obtained from the space-based Hubble data.}
\tablenotetext{d}{AB Magnitude}
\tablenotetext{e}{Rest Frame Magnitude}
\tablenotetext{f}{\dq and \qz values refer to flags assigned to note quality of data and reliability of grism redshift measurement (see \S \ref{sec:zquality}).}
\tablenotetext{g}{A merger flag value of ``1'' denotes galaxies identified as merger candidates used in the calculation of the pair fractions listed in Table \ref{fracTable}.}
\vspace{50pt}
\end{deluxetable}

\newpage

\begin{deluxetable}{cccccccccccccc}[ht!]
\tablecaption{Photometry measurements and coordinate information for galaxies in IRC-B that satisfy selection method III (see \S \ref{sec:III}): grism redshifts within the cluster's redshift range ($1.8\leq z_{grism} \leq 2.2$) with \dq\ = 1 and \qz\ $\geq 2$ or photometric redshifts between $1.5\leq z_{phot} \leq 2.5$, and F160W magnitdue $\leq 23.0$. \label{TabB}}
\tablehead{
	\colhead{ID} & 
	\colhead{HST RA\tablenotemark{c}} & 
	\colhead{HST DEC\tablenotemark{c}} & 
	\colhead{$z_{grism}$} & 
	\colhead{$z_{phot}$} &
	 \colhead{Stellar Mass} & 
	 \colhead{F160W\tablenotemark{d}} & 
	 \colhead{F125W\tablenotemark{d}} & 
	 \colhead{U\tablenotemark{e}} & 
	 \colhead{V\tablenotemark{e}} &
	 \colhead{J\tablenotemark{e}} & 
	 \colhead{$DQ$\tablenotemark{f}} & 
	 \colhead{$Q_z$\tablenotemark{f}} &
	 \colhead{Merger Flag\tablenotemark{g}}
	  }
\startdata
13 & 2:22:18.041 & -4:22:52.66 & 1.077 & 2.074 & 10.63 & 22.04 & 22.62 & 23.08 & 21.89 & 21.30 & 0 & 0 & 0 \\
49 & 2:22:18.900 & -4:22:46.28 & 2.064 & 0.356 & 7.95 & 22.55 & 23.15 & 24.74 & 24.40 & 22.43 & 1 & 3 & 0 \\
80 & 2:22:19.818 & -4:22:41.23 & 2.860 & 1.529 & 10.71 & 21.96 & 22.62 & 23.70 & 22.42 & 21.01 & 1 & 0 & 0 \\
81 & 2:22:17.041 & -4:22:37.95 & ... & 1.966 & 10.14 & 22.30 & 22.53 & 23.04 & 22.30 & 21.98 & ... & ... & 0 \\
91 & 2:22:17.467 & -4:22:38.33 & 0.140 & 1.748 & 10.56 & 22.60 & 23.03 & 24.01 & 22.62 & 21.78 & 0 & 0 & 1 \\
105 & 2:22:20.283 & -4:22:35.99 & ... & 1.791 & 10.01 & 22.66 & 22.86 & 23.57 & 22.65 & 22.44 & ... & ... & 1 \\
116 & 2:22:20.343 & -4:22:37.73 & 1.935 & 1.131 & 10.16 & 22.83 & 23.21 & 24.63 & 23.45 & 22.05 & 1 & 3 & 1 \\
126 & 2:22:19.439 & -4:22:34.08 & 2.058 & 1.957 & 10.67 & 22.77 & 23.51 & 24.47 & 22.63 & 21.67 & 1 & 3 & 1 \\
146 & 2:22:19.454 & -4:22:31.20 & 2.410 & 1.776 & 10.32 & 22.77 & 23.57 & 24.80 & 22.79 & 22.27 & 1 & 2 & 1 \\
203 & 2:22:18.325 & -4:22:16.68 & 2.054 & 2.140 & 11.23 & 21.23 & 22.00 & 22.55 & 21.00 & 20.25 & 1 & 3 & 0 \\
204 & 2:22:19.468 & -4:22:20.17 & 2.018 & 1.767 & 10.49 & 22.40 & 22.94 & 24.03 & 22.44 & 21.81 & 1 & 3 & 0 \\
237 & 2:22:18.161 & -4:22:11.78 & 1.970 & 1.747 & 10.63 & 22.60 & 23.14 & 24.42 & 22.63 & 21.72 & 1 & 3 & 0 \\
265 & 2:22:17.011 & -4:22:04.71 & 1.607 & 1.692 & 10.59 & 22.02 & 22.49 & 23.75 & 22.09 & 21.55 & 1 & 2 & 0 \\
278 & 2:22:13.133 & -4:22:00.51 & ... & 1.976 & 10.73 & 21.80 & 22.53 & 22.92 & 21.67 & 21.25 & ... & ... & 0 \\
283 & 2:22:17.087 & -4:22:01.44 & 1.962 & 1.855 & 10.74 & 22.09 & 22.72 & 23.92 & 22.04 & 21.37 & 1 & 1 & 0 \\
300 & 2:22:13.609 & -4:21:53.76 & 2.734 & 2.144 & 11.08 & 20.71 & 21.17 & 21.42 & 20.59 & 20.01 & 1 & 1 & 1 \\
315 & 2:22:13.161 & -4:21:54.02 & 1.936 & 1.970 & 10.80 & 21.23 & 21.52 & 22.13 & 21.17 & 20.84 & 1 & 2 & 0 \\
316 & 2:22:19.446 & -4:21:55.36 & 2.329 & 1.817 & 10.52 & 22.81 & 23.44 & 24.46 & 22.80 & 21.83 & 1 & 3 & 0 \\
318 & 2:22:13.762 & -4:21:53.83 & 2.721 & 1.510 & 10.54 & 22.33 & 22.84 & 24.07 & 22.67 & 21.37 & 1 & 1 & 1 \\
341 & 2:22:18.928 & -4:21:49.09 & 1.912 & 1.713 & 10.51 & 21.84 & 22.25 & 23.55 & 21.89 & 21.46 & 1 & 2 & 0 \\
342 & 2:22:14.217 & -4:21:46.82 & 2.007 & 1.381 & 10.83 & 20.96 & 21.45 & 22.62 & 21.38 & 20.31 & 1 & 3 & 0 \\
356 & 2:22:20.022 & -4:21:46.33 & 1.963 & 0.412 & 8.31 & 22.43 & 22.82 & 25.11 & 24.01 & 22.21 & 1 & 2 & 0 \\
374 & 2:22:20.333 & -4:21:42.94 & 2.061 & 2.117 & 11.04 & 21.13 & 21.62 & 22.09 & 21.00 & 20.45 & 1 & 2 & 0 \\
381 & 2:22:14.302 & -4:21:41.24 & 1.531 & 1.596 & 10.68 & 21.86 & 22.45 & 23.63 & 22.05 & 21.10 & 1 & 1 & 0 \\
443 & 2:22:16.942 & -4:21:29.38 & 1.509 & 1.522 & 10.56 & 21.13 & 21.58 & 22.43 & 21.43 & 20.42 & 1 & 3 & 0 \\
466 & 2:22:15.509 & -4:21:22.02 & 0.612 & 2.006 & 11.29 & 21.02 & 21.76 & 22.54 & 20.86 & 20.08 & 0 & 0 & 0 \\
473 & 2:22:20.047 & -4:21:24.66 & 2.564 & 2.393 & 10.97 & 21.95 & 23.07 & 23.20 & 21.41 & 20.97 & 1 & 3 & 0 \\
478 & 2:22:13.529 & -4:21:24.94 & ... & 1.557 & 10.58 & 21.85 & 21.71 & 23.60 & 21.94 & 21.15 & ... & ... & 0 \\
484 & 2:22:13.919 & -4:21:22.06 & ... & 2.303 & 10.94 & 21.44 & 21.96 & 22.07 & 21.31 & 20.68 & ... & ... & 0 \\
614 & 2:22:16.035 & -4:20:53.27 & 0.611 & 1.797 & 10.48 & 22.74 & 23.39 & 24.57 & 22.76 & 22.12 & 1 & 0 & 0 \\
624 & 2:22:20.350 & -4:20:49.89 & ... & 1.536 & 10.08 & 22.98 & 23.28 & 24.40 & 23.18 & 22.19 & ... & ... & 0 \\
646 & 2:22:17.787 & -4:20:48.56 & ... & 1.819 & 10.99 & 22.26 & 23.00 & 24.27 & 22.25 & 21.01 & ... & ... & 1 \\
655 & 2:22:17.813 & -4:20:49.14 & 2.002 & 1.797 & 11.00 & 21.75 & 22.36 & 23.54 & 21.77 & 20.85 & 0 & 0 & 1 \\
\enddata
\tablenotetext{c}{HST coordinate information were obtained from the space-based Hubble data.}
\tablenotetext{d}{AB Magnitude}
\tablenotetext{e}{Rest Frame Magnitude}
\tablenotetext{f}{\dq and \qz values refer to flags assigned to note quality of data and reliability of grism redshift measurement (see \S \ref{sec:zquality}).}
\tablenotetext{g}{A merger flag value of ``1'' denotes galaxies identified as merger candidates used in the calculation of the pair fractions listed in Table \ref{fracTable}.}
\end{deluxetable}

\newpage

\newpage 

\appendix 
\section{ Probability of randomly finding association of galaxies in IRC, COSMOS, and CDFS fields of view\label{sec: prob}}

We performed a randomization test to check whether or not we would be able to find an association of galaxies that satisfy the our selection method III (see \S \ref{sec:mergID} and Table \ref{selects}) in order to see whether the group of galaxies we associate with our candidate clusters is a chance occurrence. 

We choose 10'$\times$10' areas in our COSMOS and CDFS zFOURGE (with scales of 0.15\arcsec/pixel) fields to create a square area with well defined edges. We then randomly define a 2.5'$\times$2.5' search area to simulate the footprint of our HST fields of view. Within these 2.5'$\times$2.5' search areas we determine the number of galaxies that satisfy our selection criteria for COSMOS and CDFS, i.e. F160W $\leq 23.0$, $1.55 \leq z_{grism} \leq 2.2$, and $1.5 \leq z_{phot} \leq 2.5$. (see \S \ref{sec:mergID} and Table \ref{selects}). Table \ref{prob} shows the results of this randomization test for 100,000 iteration. Probability of finding an assosication of galaxies in redshift space that satisfy our cluster selection method is $\sim2\%$ in the field.

For comparison, applying the same methods to our candidate cluster catalogs, we find the fraction of galaxies satisfying this selection criteria is $\sim5\%$. This strengthens our case that IRC-A and IRC-B are robust cluster candidates.

\begin{table}[h!]
\begin{center}
\caption{Probability of finding galaxies within search area the size of HST field of view (2.5x2.5') within COSMOS and CDFS zFOURGE fields (10x10') that satisfy redshift and magnitude limits of selection method (III) (see \S \ref{sec:mergID} and Table \ref{selects}). \label{prob}}
\begin{tabular}{ccccc}
&&&&\\
\hline \hline
                            &	$N_{tot}$\tablenotemark{a} & $N_{cluster}$\tablenotemark{b}    &\\ \hline
IRC-A                       &  499                          & 28                         & $5.6_{-0.9}^{+1.2}\%$ \\ 
IRC-B                       & 661                           & 33                        &$5.0_{-0.7}^{+1.0}\%$ \\  

COSMOS\tablenotemark{c}     &772                            &22                         &$2.9^{+0.7}_{-0.5}\%$ \\ 

CDFS			            &1035	                        &17                         &$1.7^{+0.5}_{-0.3}\%$	\\ 

Field Population\tablenotemark{d}	&1807	                &39                         &$2.2^{+0.4}_{-0.3}\%$\\ \hline

\end{tabular}
\end{center}
\tablenotetext{a}{Mean number of galaxies found within search area.}
\tablenotetext{b}{Mean number of galaxies found within search area that satisfy selection method (III)}
\tablenotetext{c}{Masked to exclude confirmed members of $z=2.1$ cluster.}
\tablenotetext{d}{Combined COSMOS and CDFS populations}
\end{table}

\section{Grism Imaging and Spectroscopy for IRC-A and IRC-B candidate cluster members}

\begin{figure} 
\vspace{-10pt}
\figurenum{B1}
\begin{center}$
\begin{array}{ccc}
\includegraphics[width = 0.33 \textwidth]{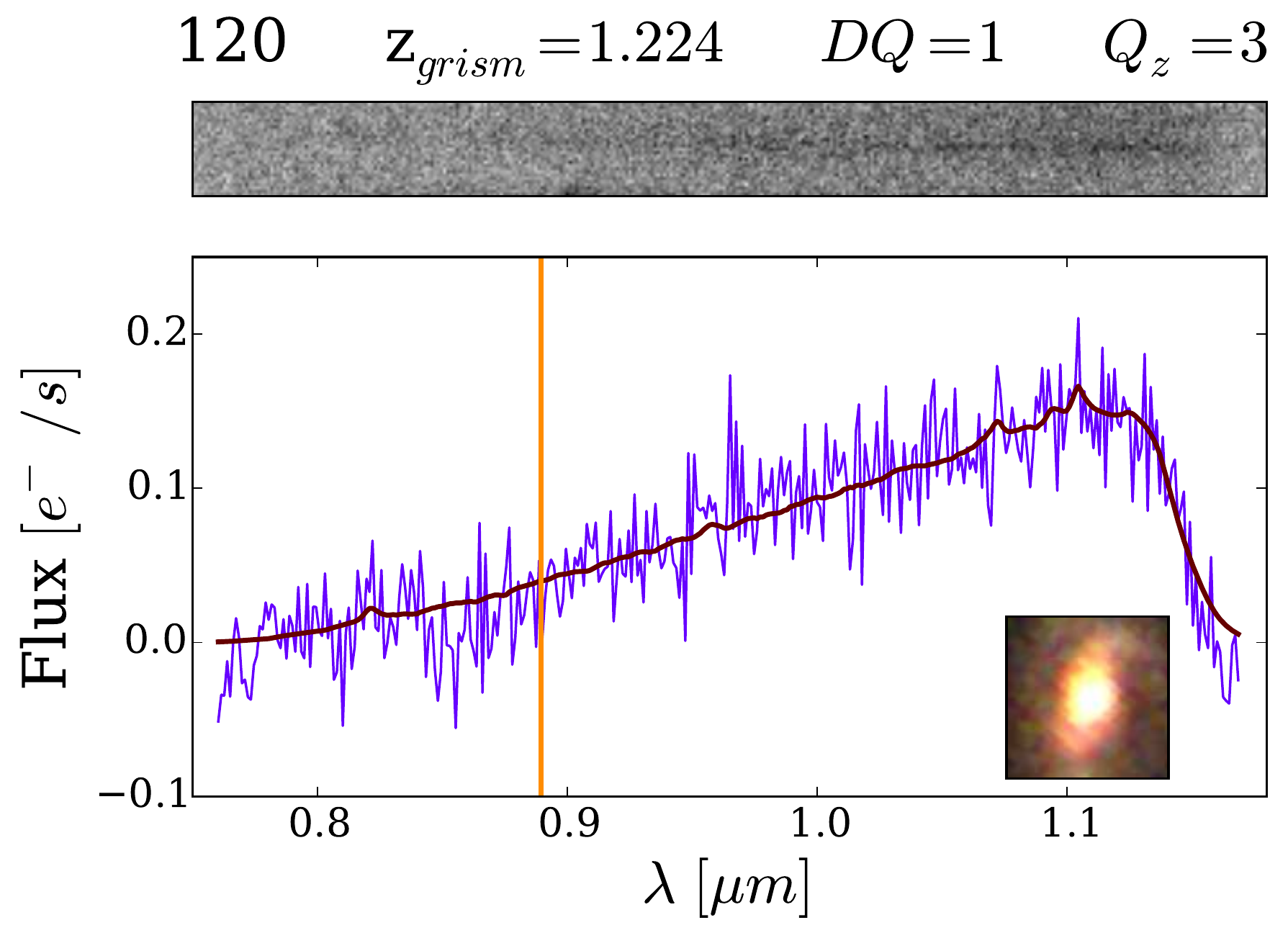} & 
\includegraphics[width = 0.33 \textwidth]{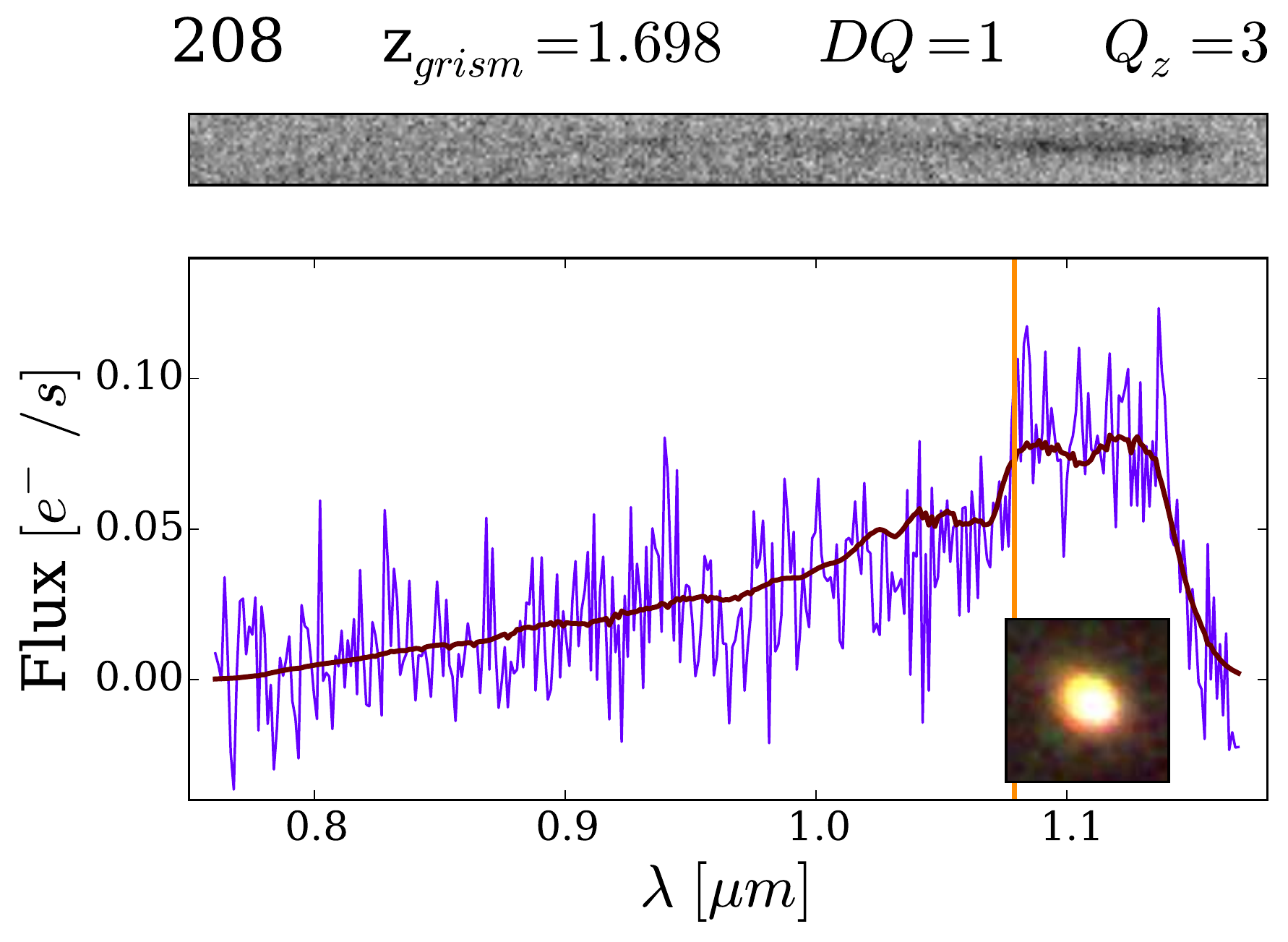} & 
\includegraphics[width = 0.33 \textwidth]{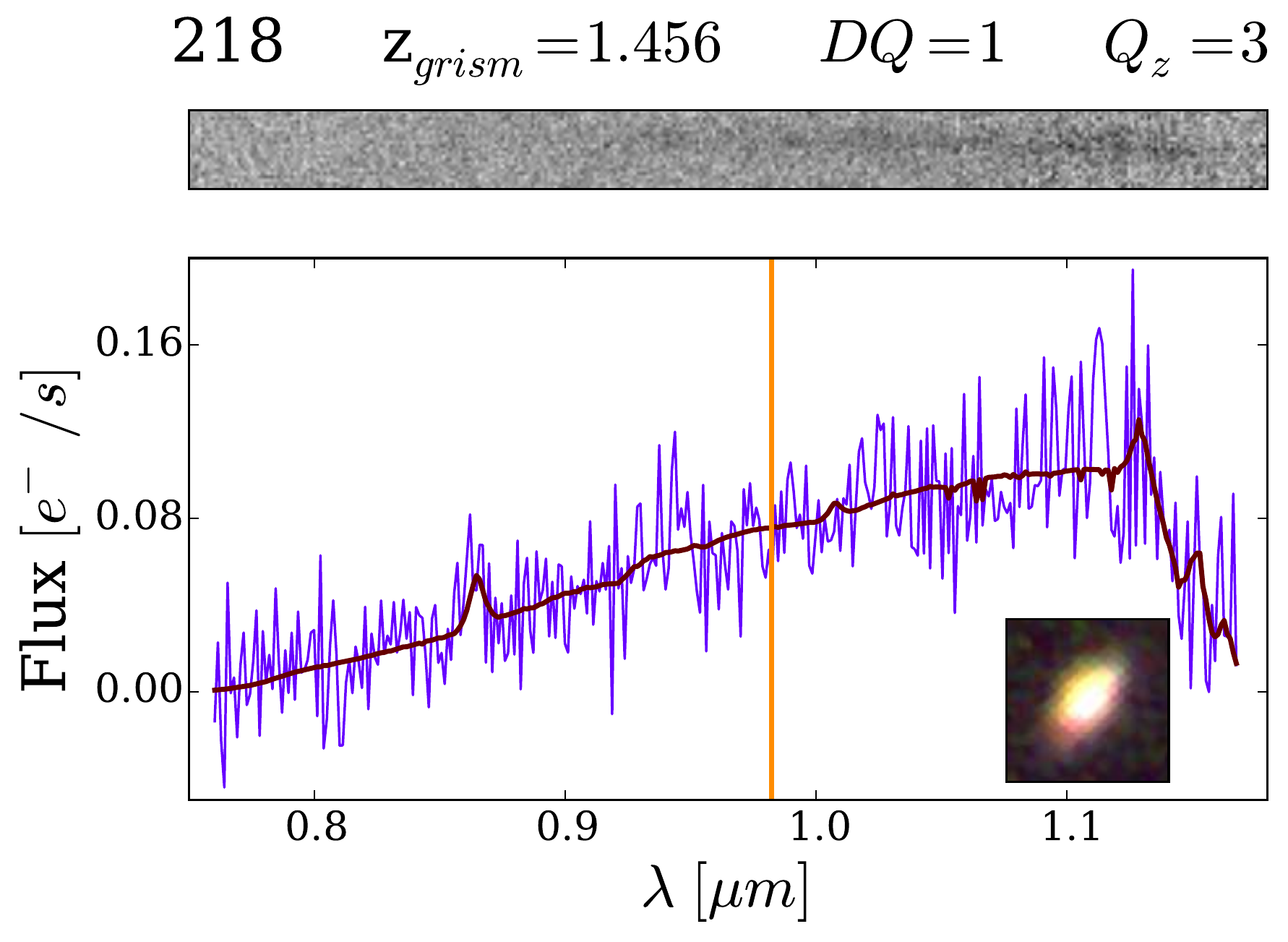} \\
\includegraphics[width = 0.33 \textwidth]{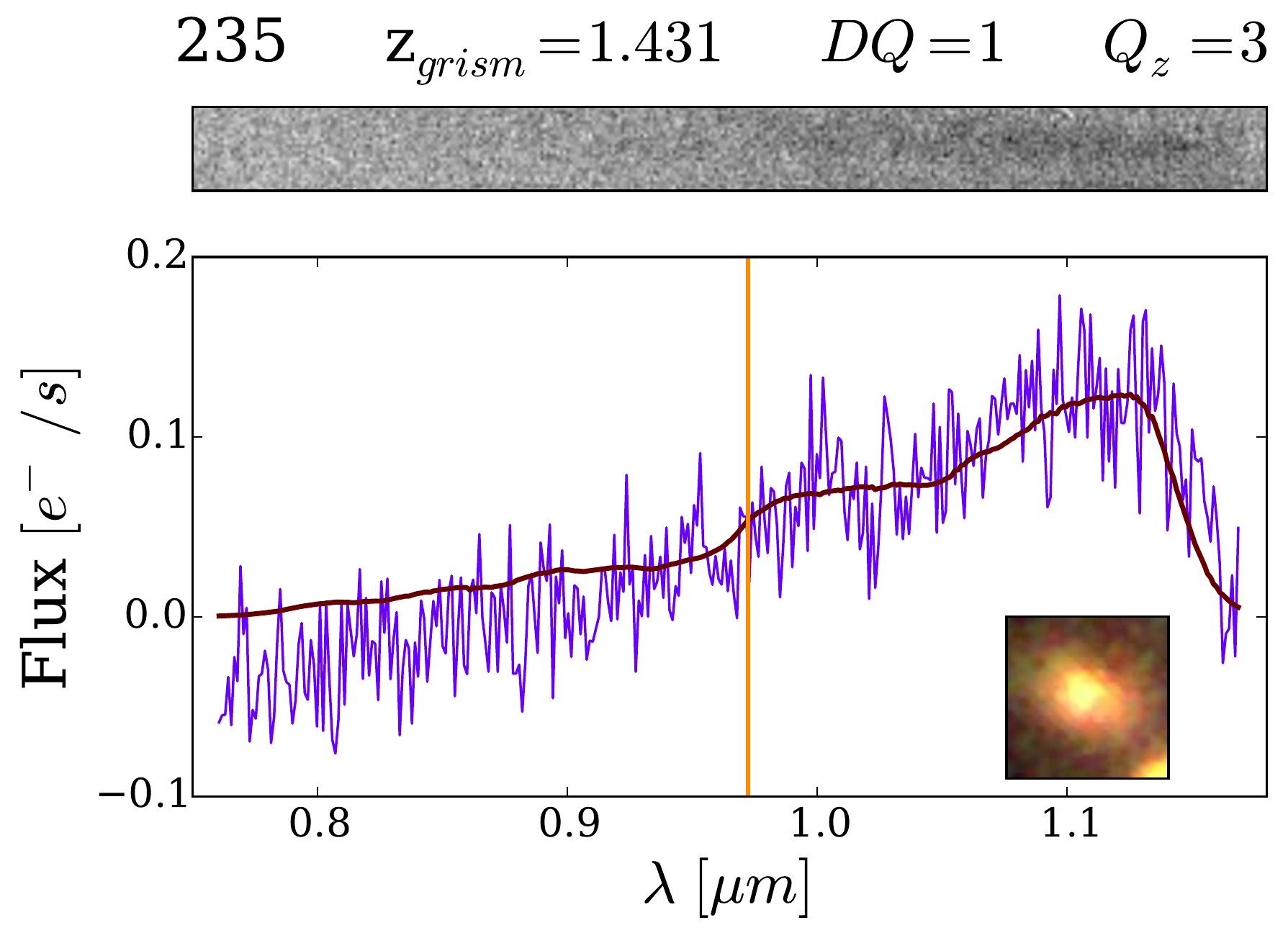} & 
\includegraphics[width = 0.33 \textwidth]{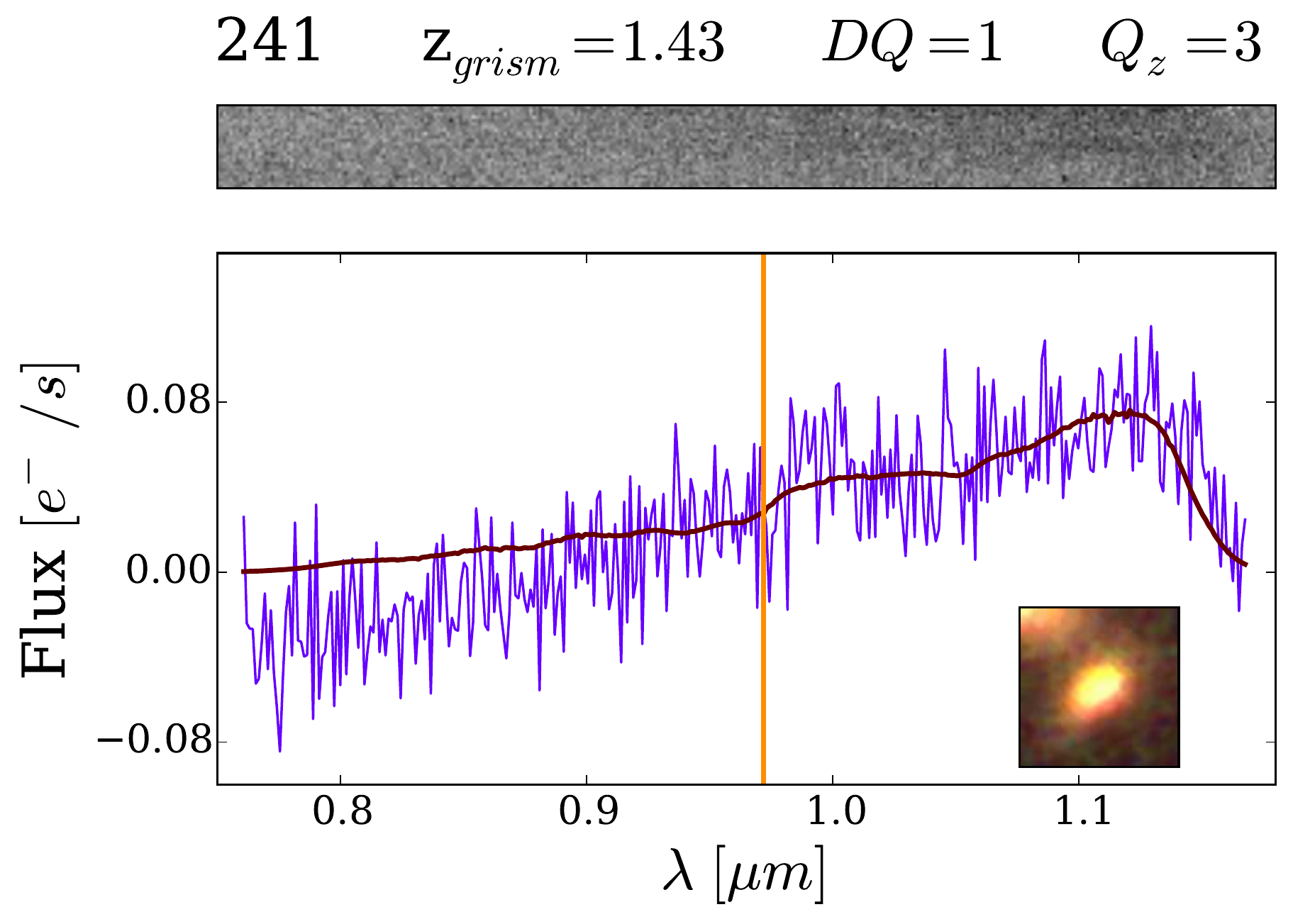} & 
\includegraphics[width = 0.33 \textwidth]{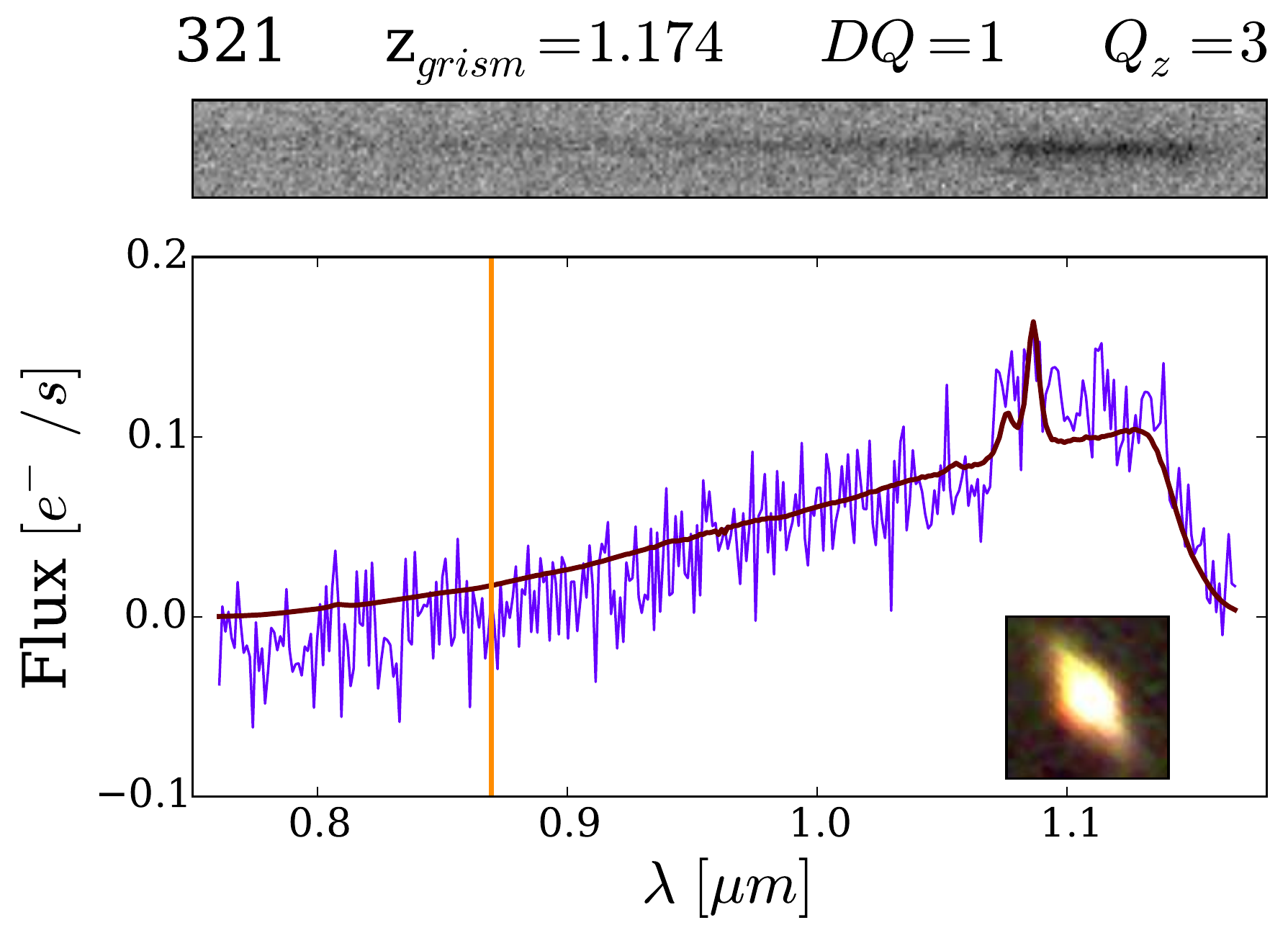} \\ 
\includegraphics[width = 0.33 \textwidth]{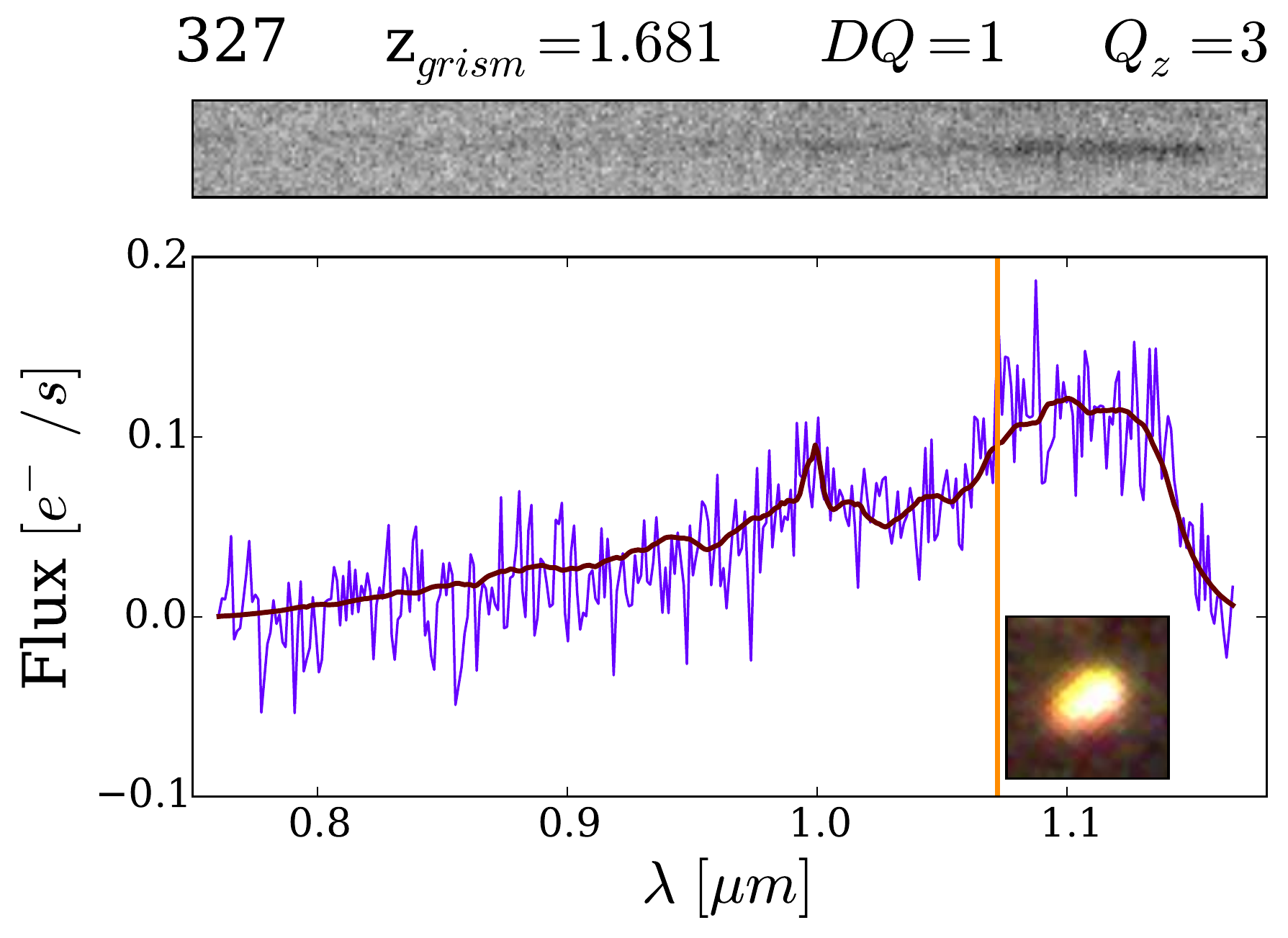} & 
\includegraphics[width = 0.33 \textwidth]{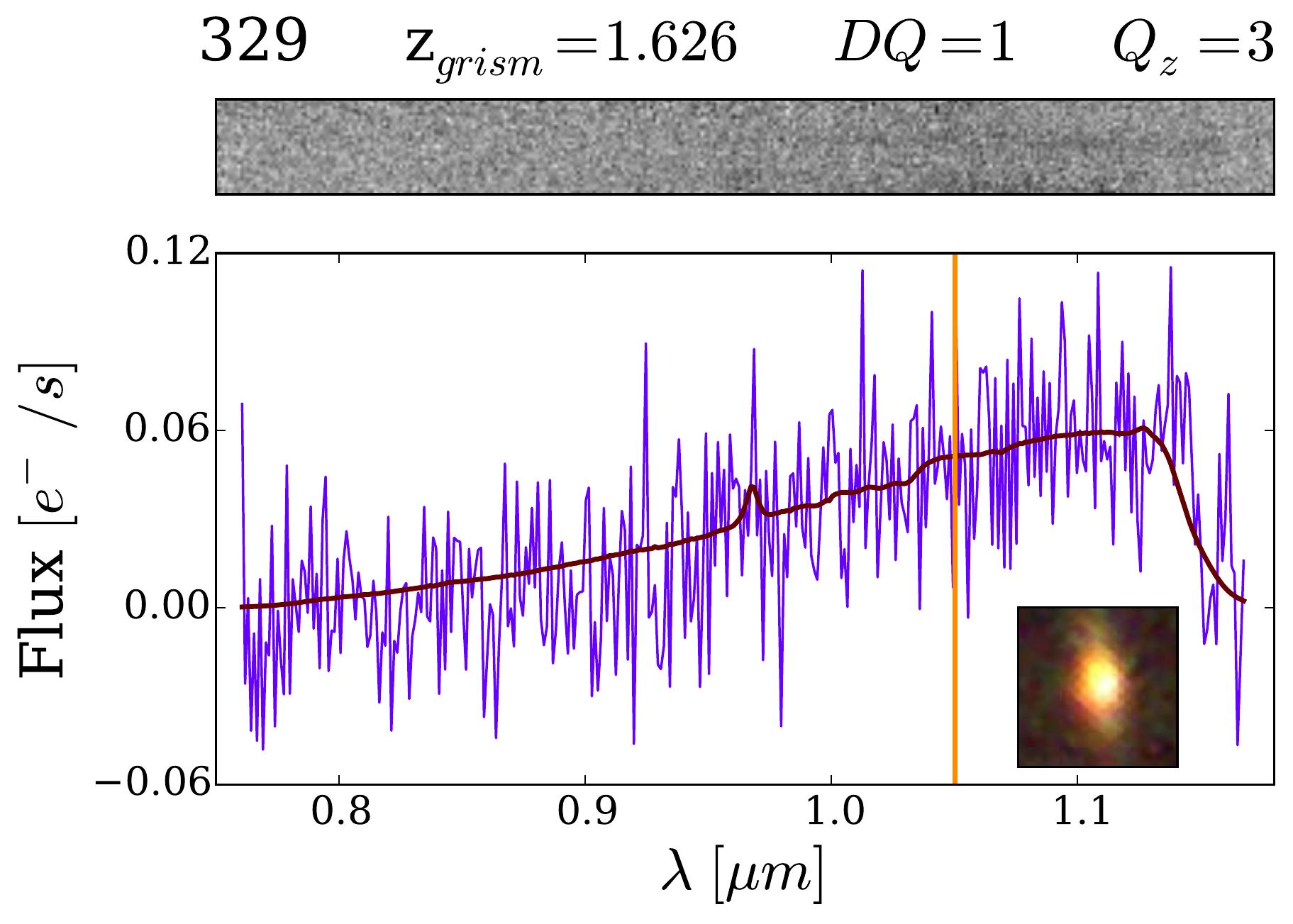} & 
\includegraphics[width = 0.33 \textwidth]{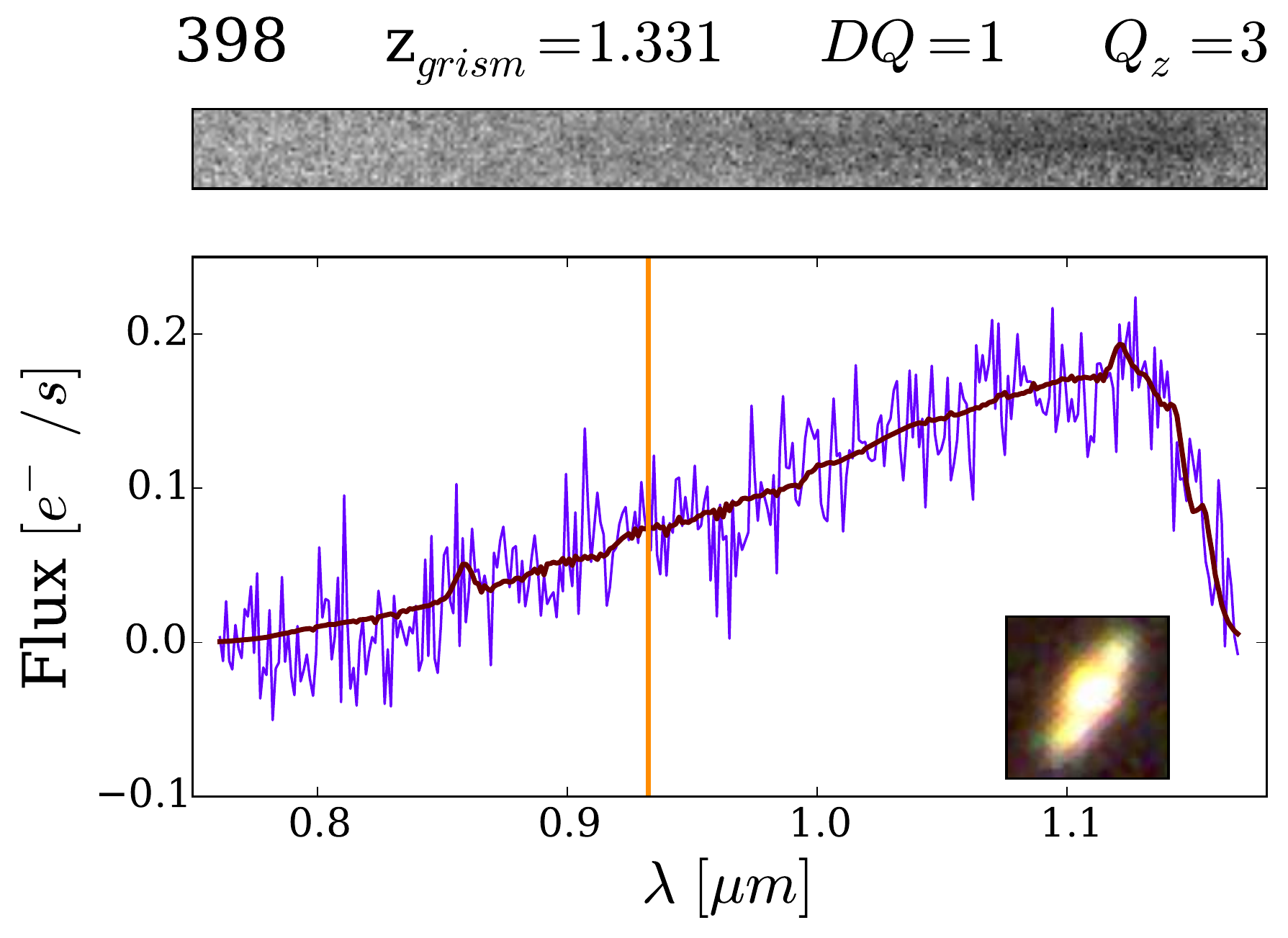} \\ 
\includegraphics[width = 0.33 \textwidth]{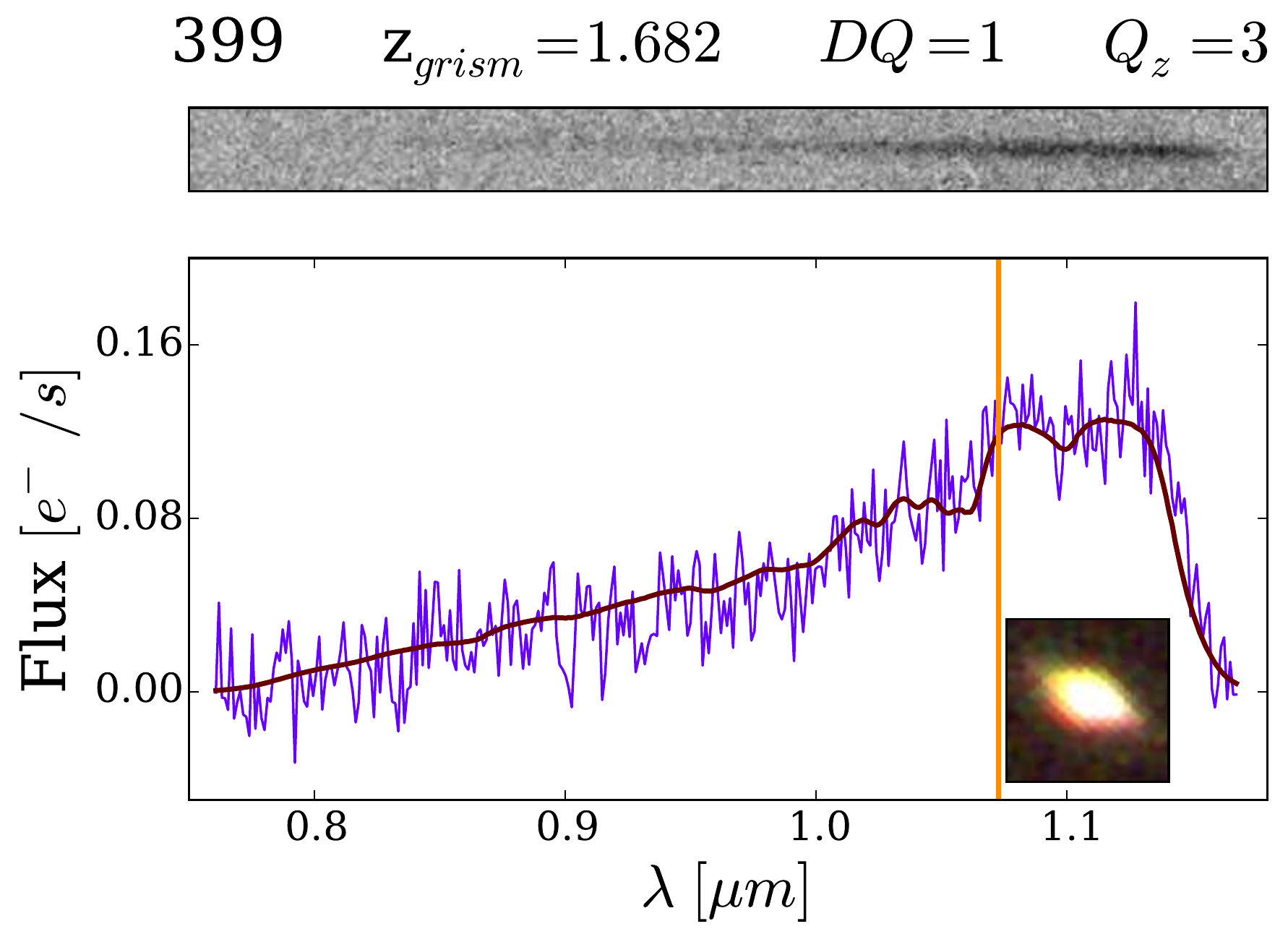} & 
\includegraphics[width = 0.33 \textwidth]{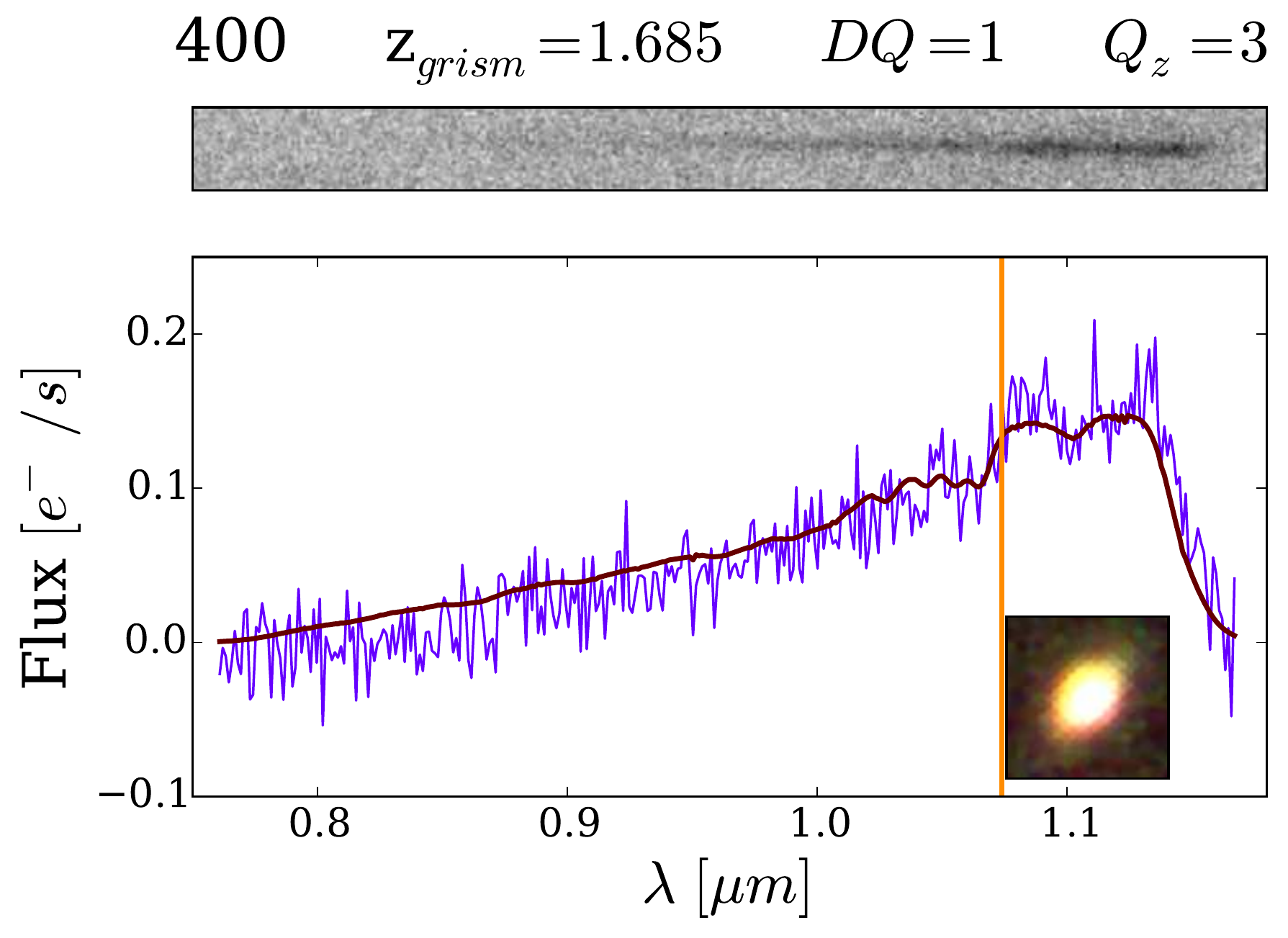} & 
\includegraphics[width = 0.33 \textwidth]{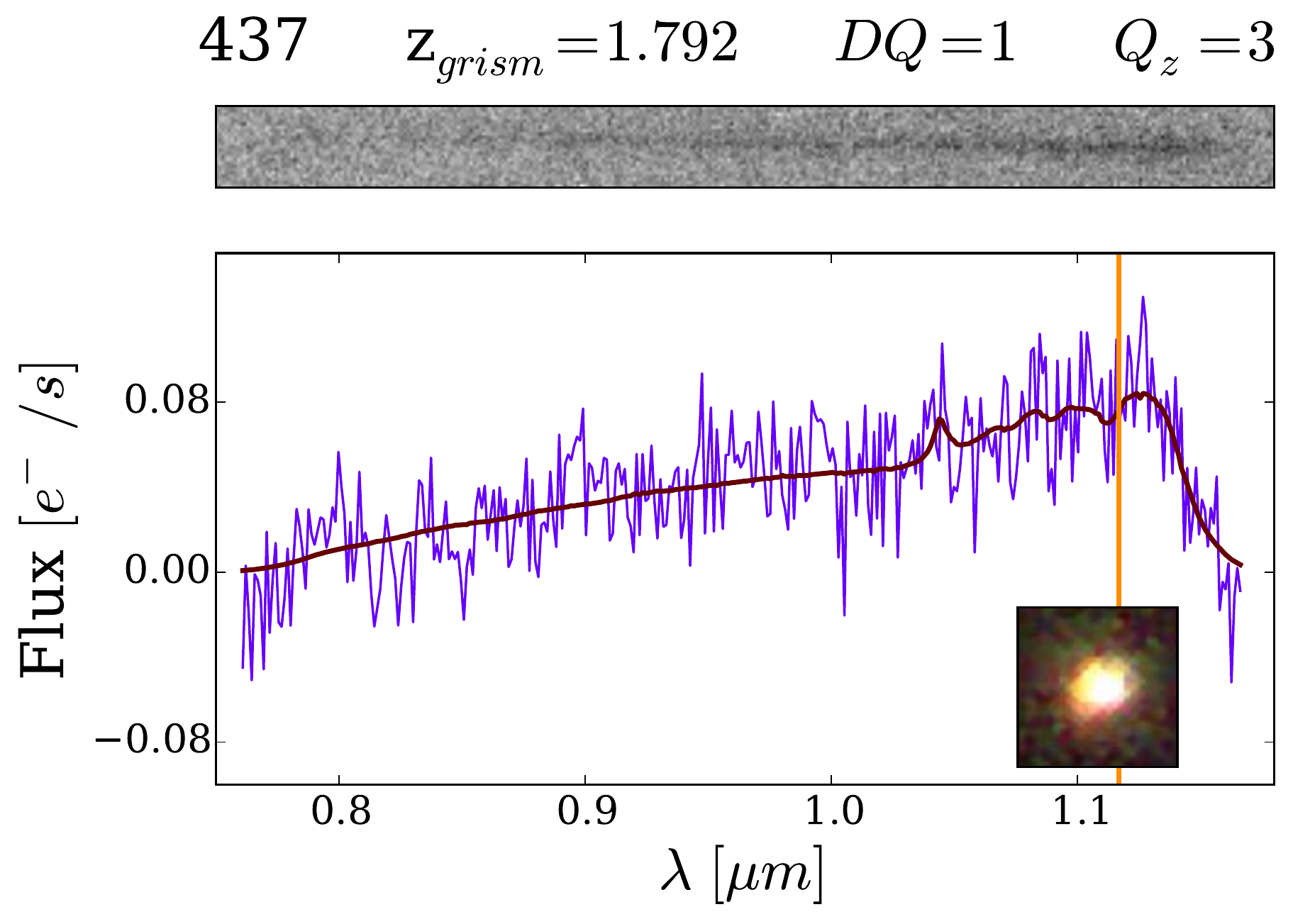} \\ 
\includegraphics[width = 0.33 \textwidth]{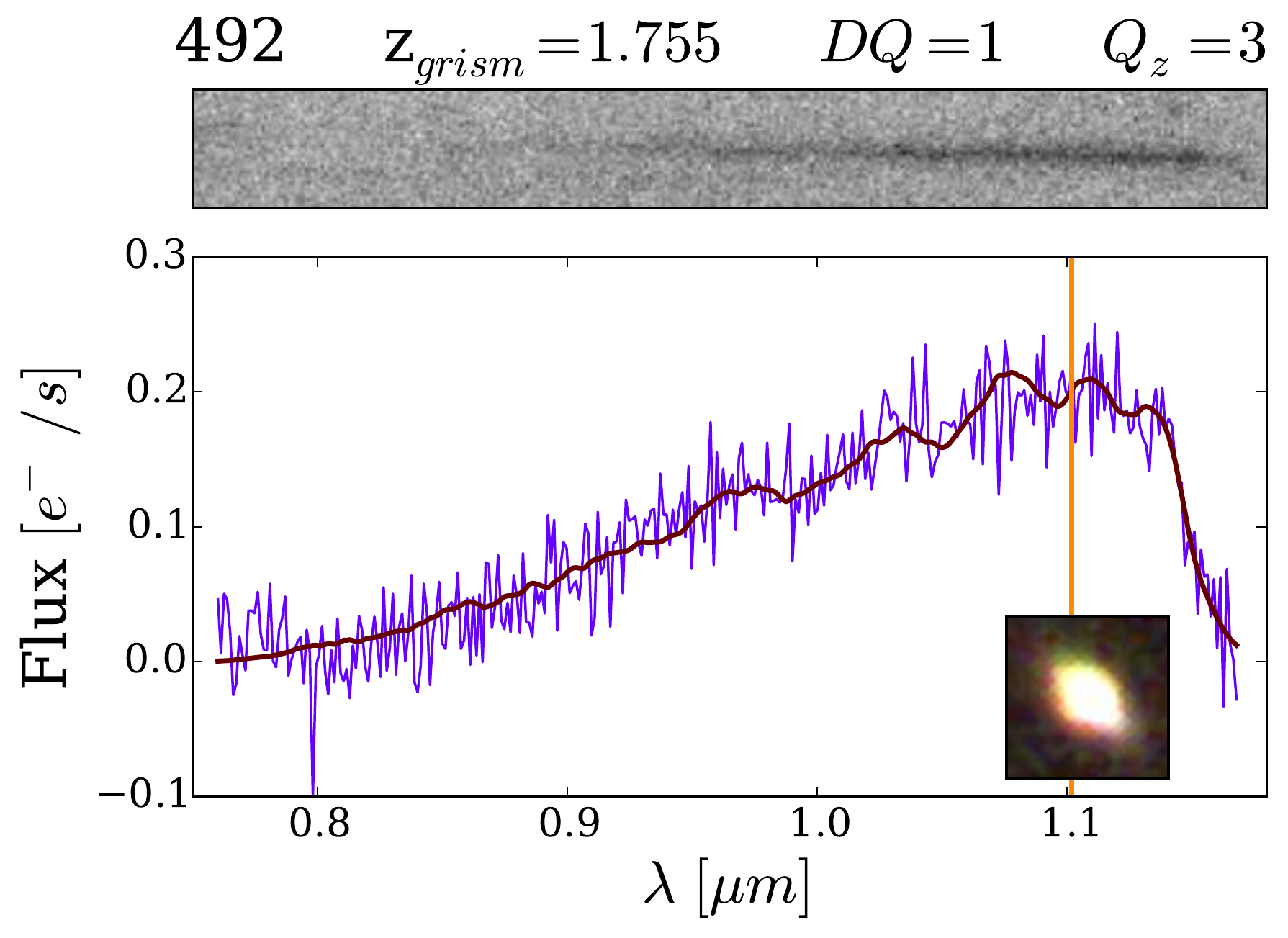} & 
\includegraphics[width = 0.33 \textwidth]{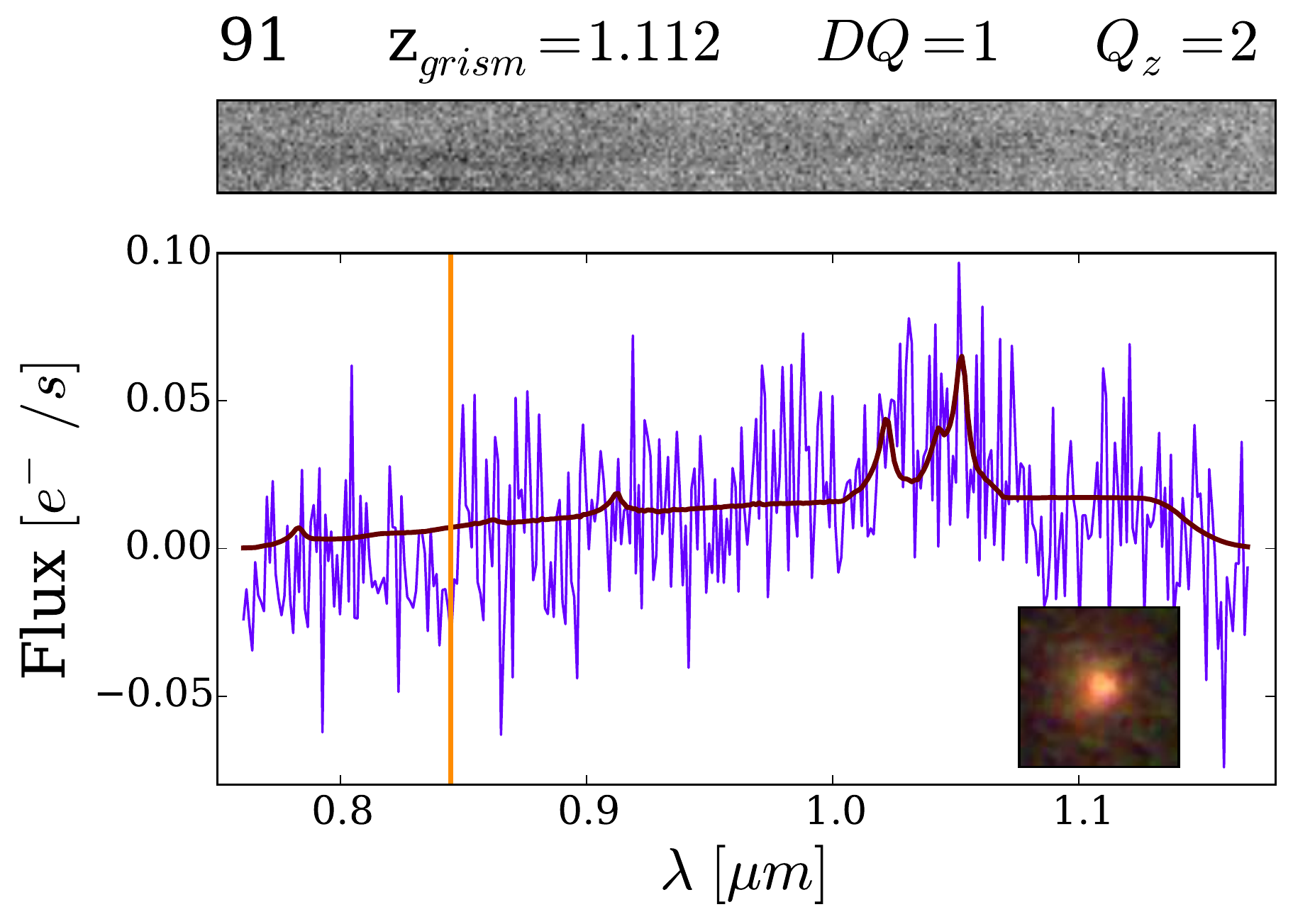} & 
\includegraphics[width = 0.33 \textwidth]{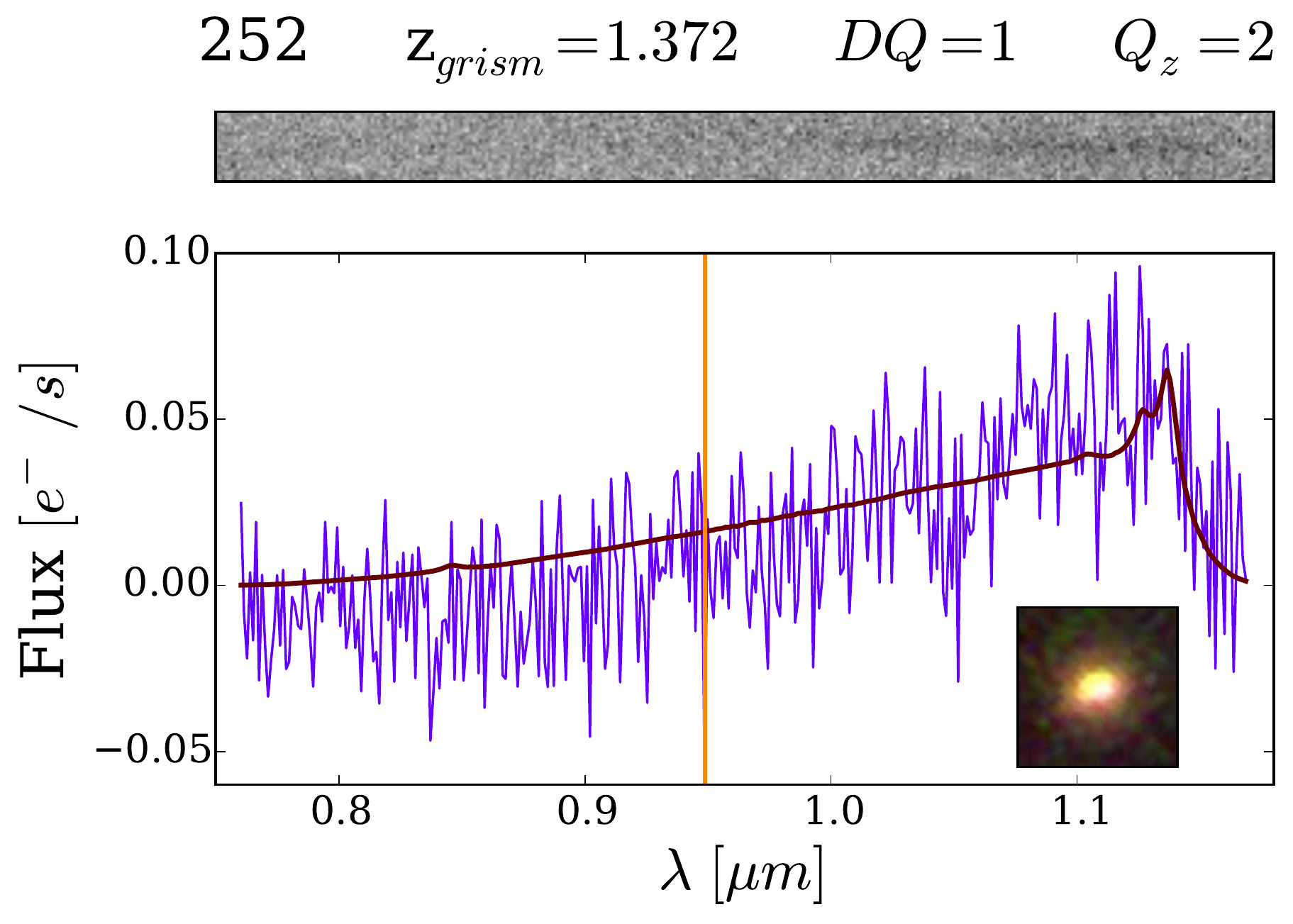} \\
\end{array}$
\caption{Continued on next page.}
\end{center}
\label{gris_pan_A1}
\end{figure}

\begin{figure} 
\figurenum{B1}
\begin{center}$
\begin{array}{ccc}
\includegraphics[width = 0.33 \textwidth]{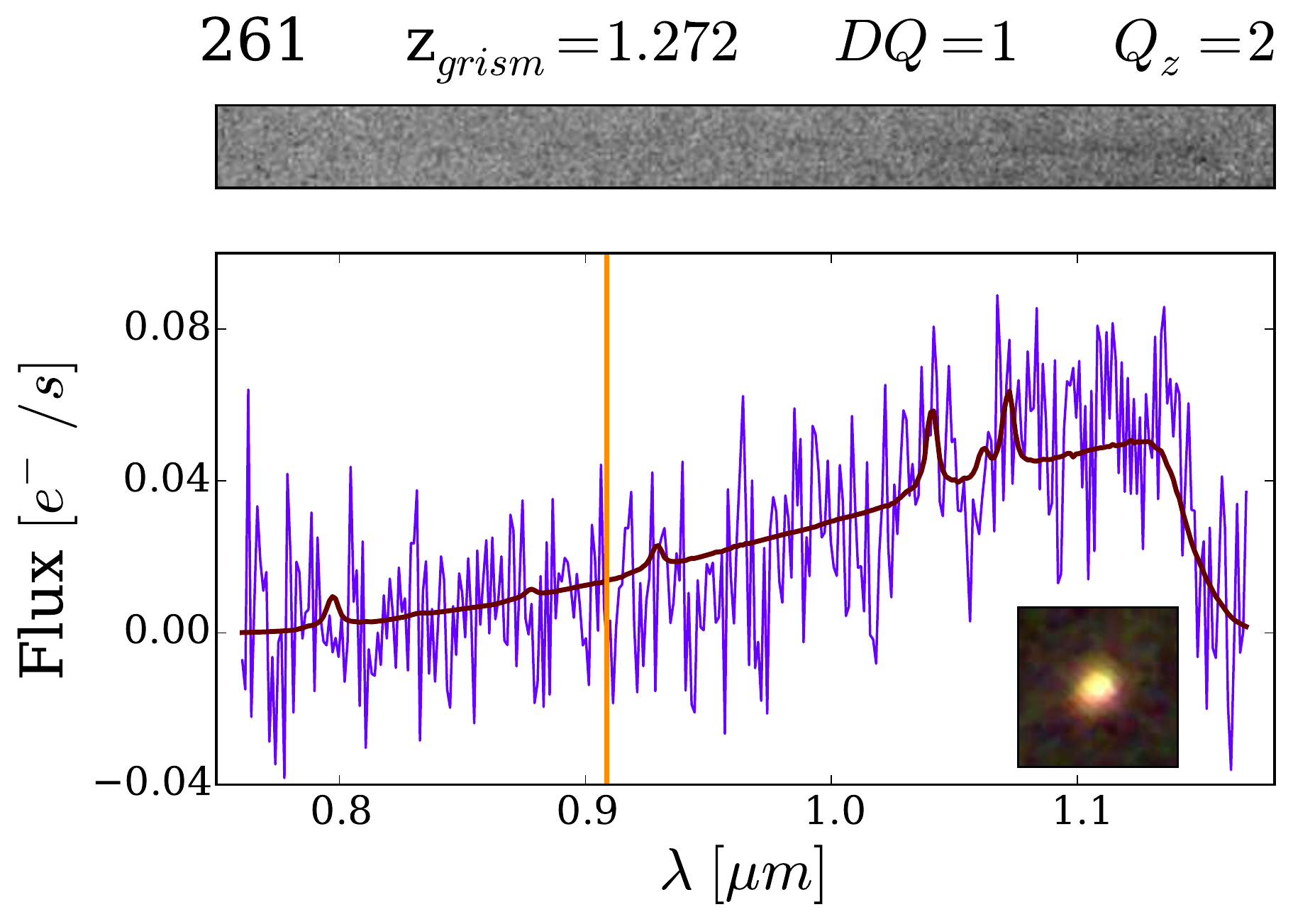} & 
\includegraphics[width = 0.33 \textwidth]{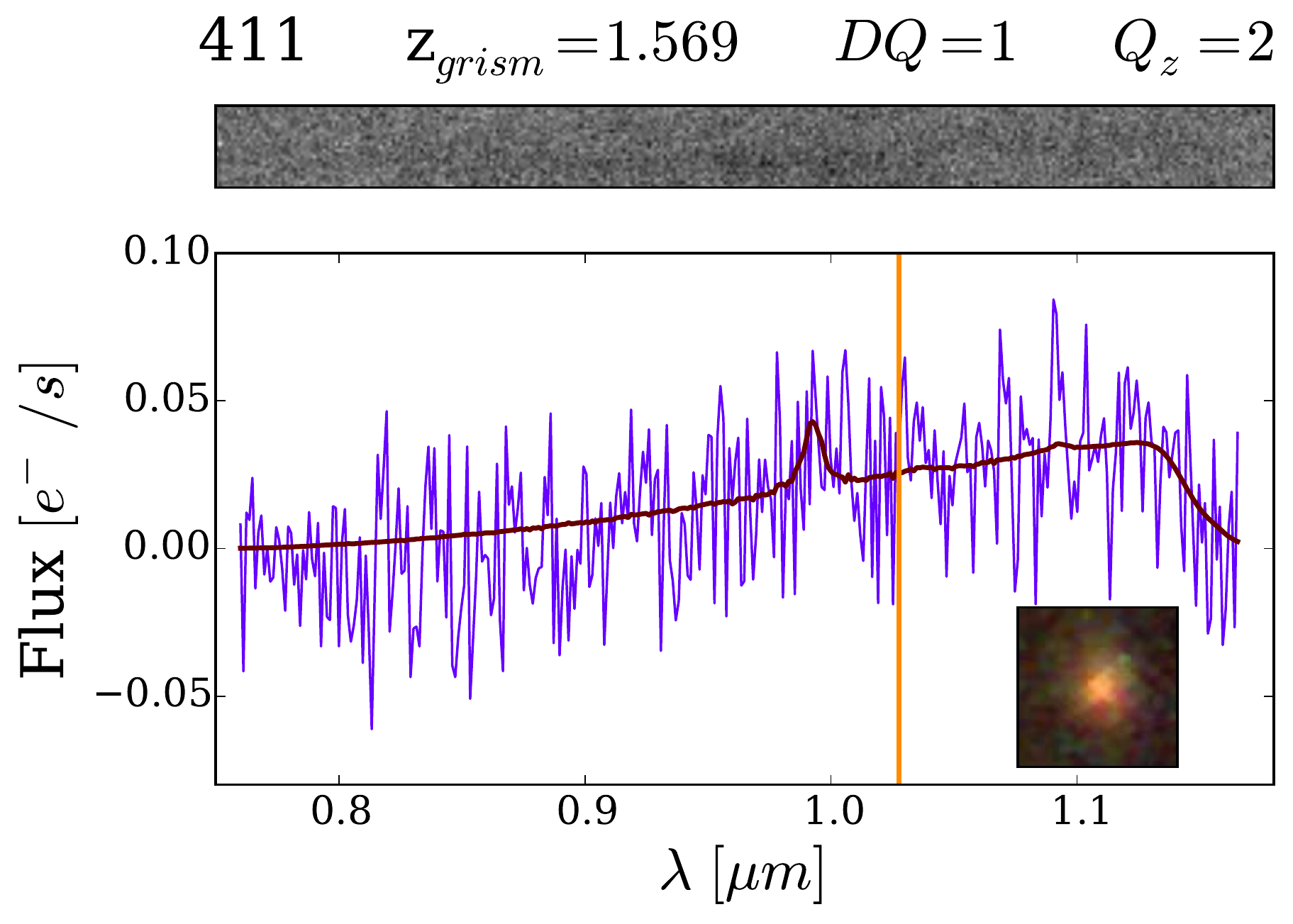} & 
\includegraphics[width = 0.33 \textwidth]{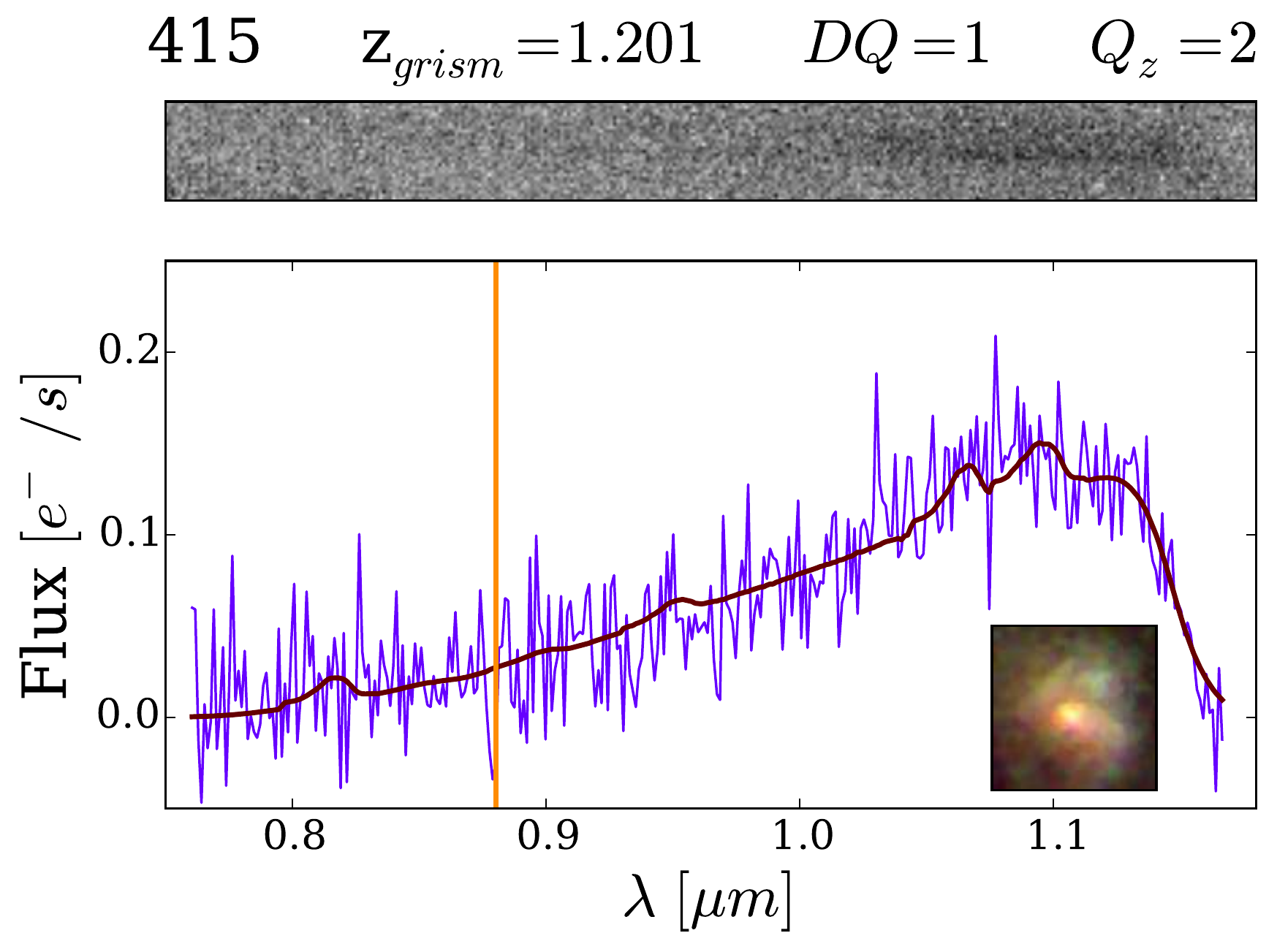} \\ 
\includegraphics[width = 0.33 \textwidth]{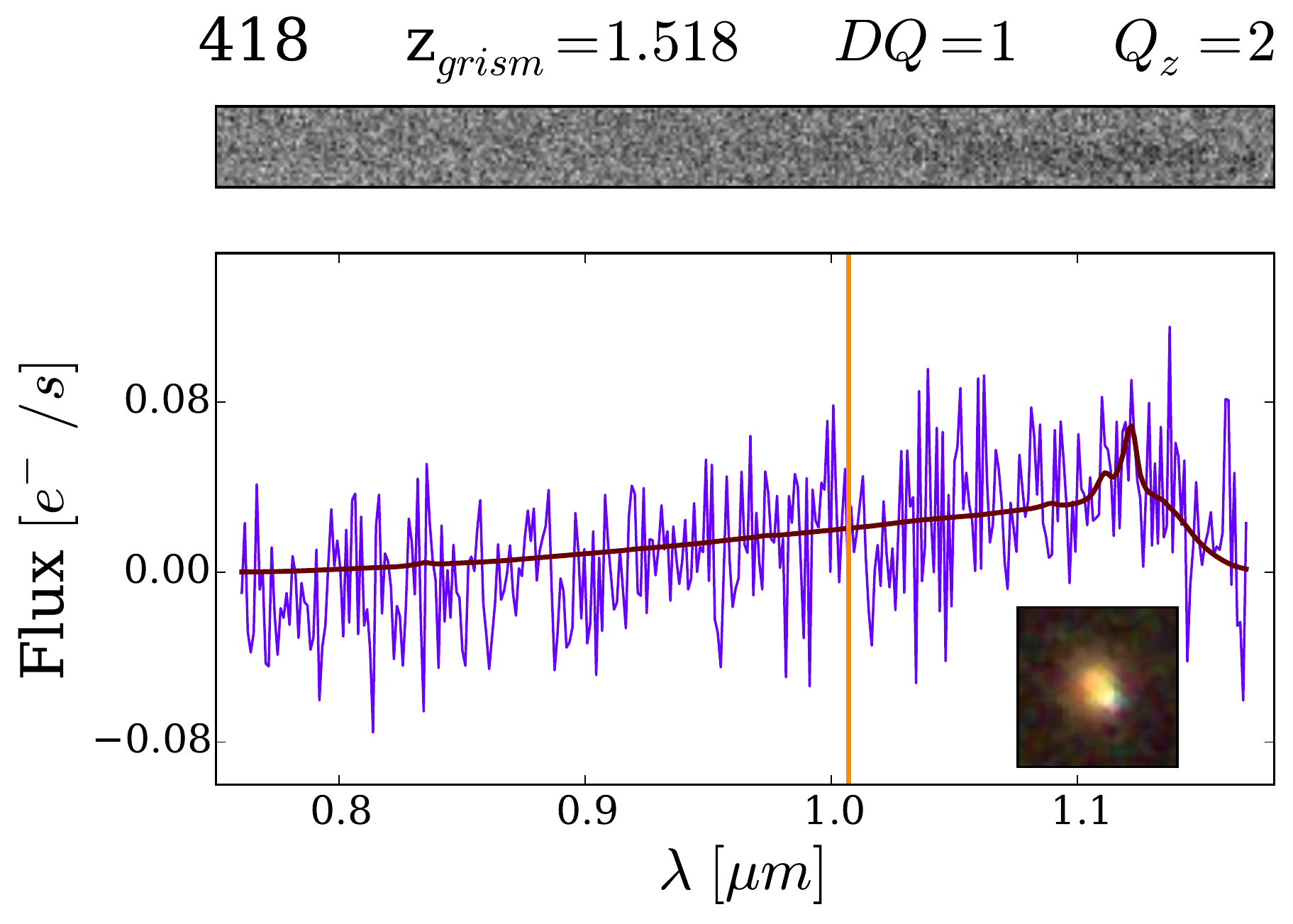} & 
\includegraphics[width = 0.33 \textwidth]{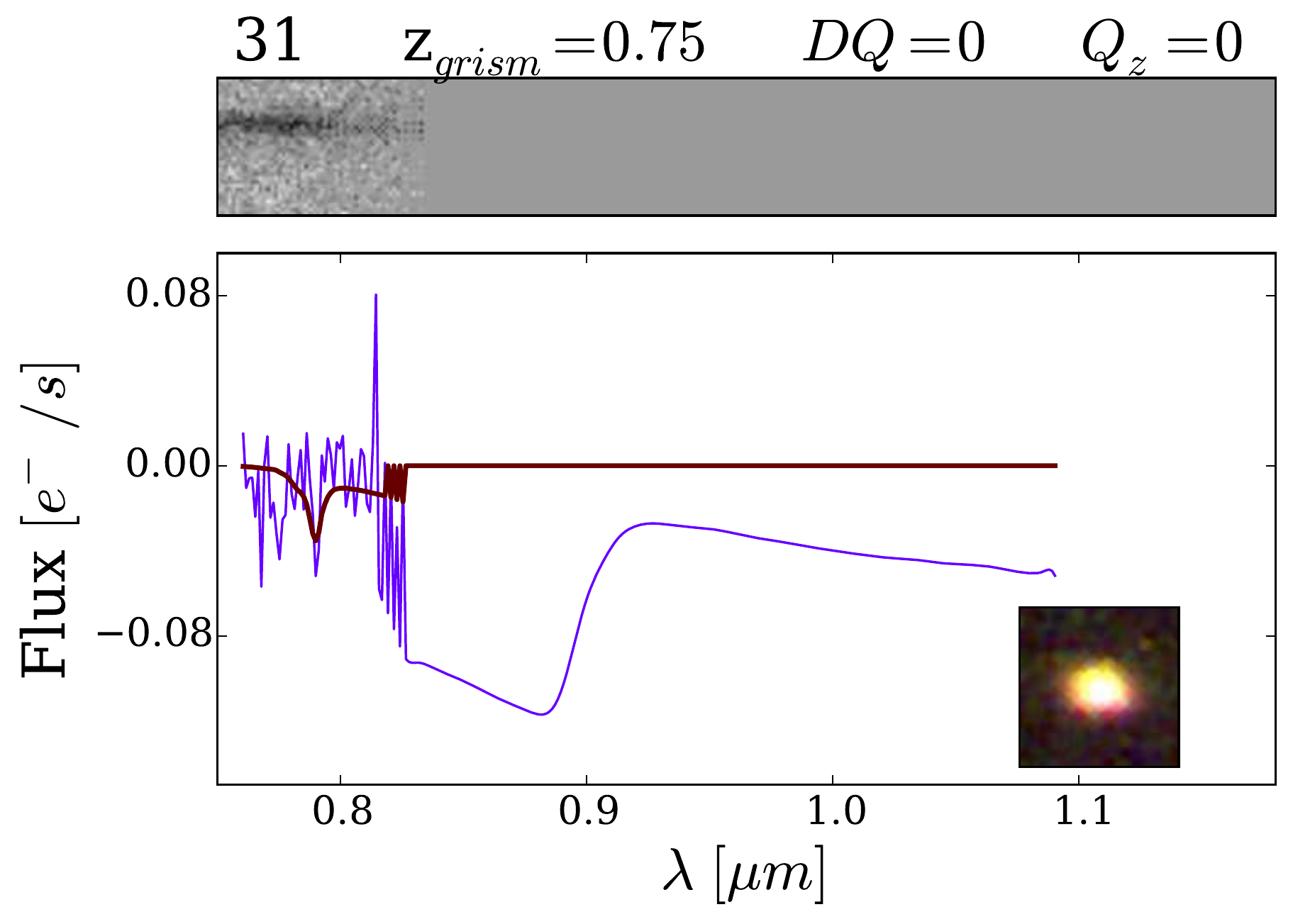} & 
\includegraphics[width = 0.33 \textwidth]{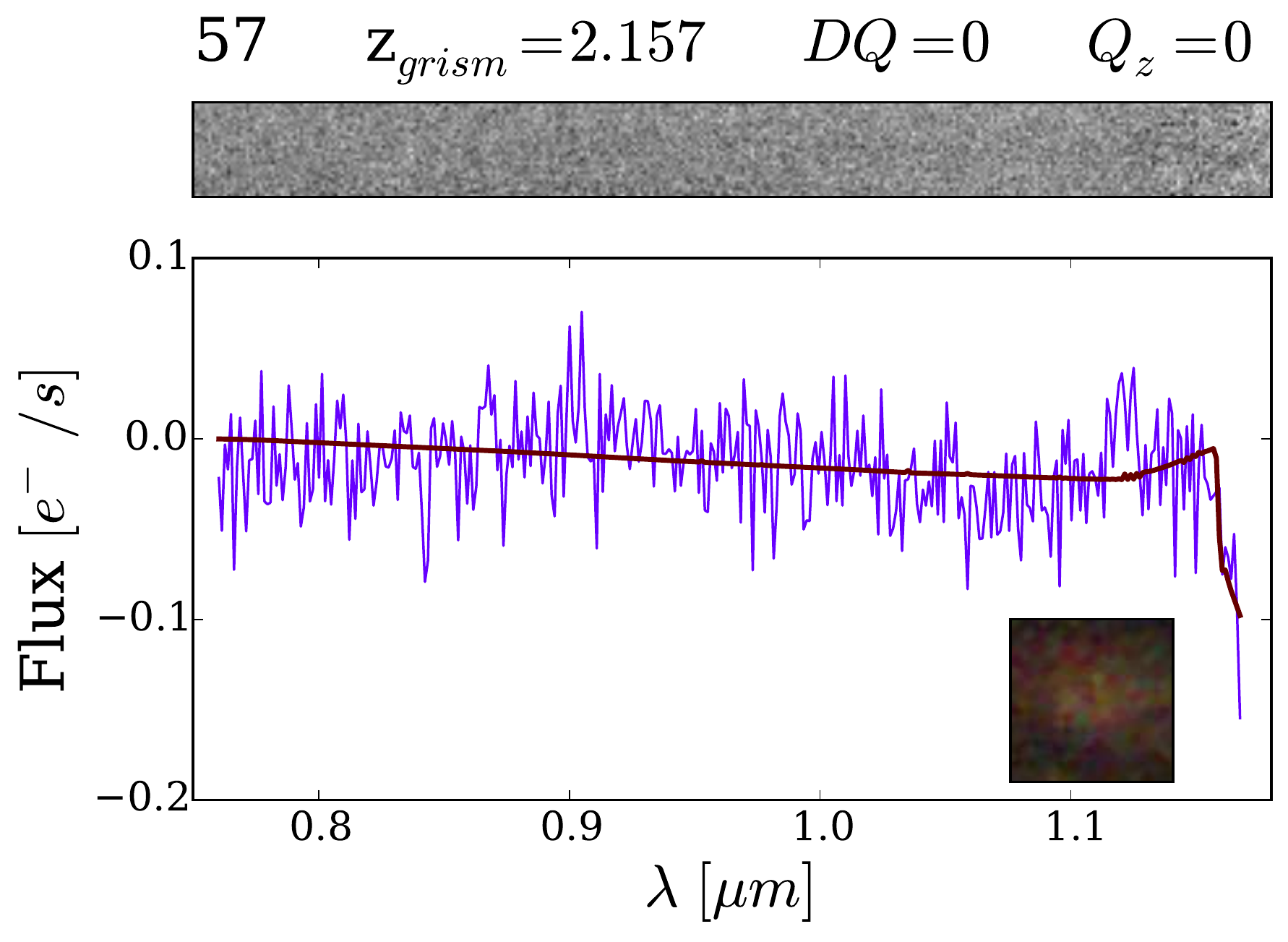} \\
\includegraphics[width = 0.33 \textwidth]{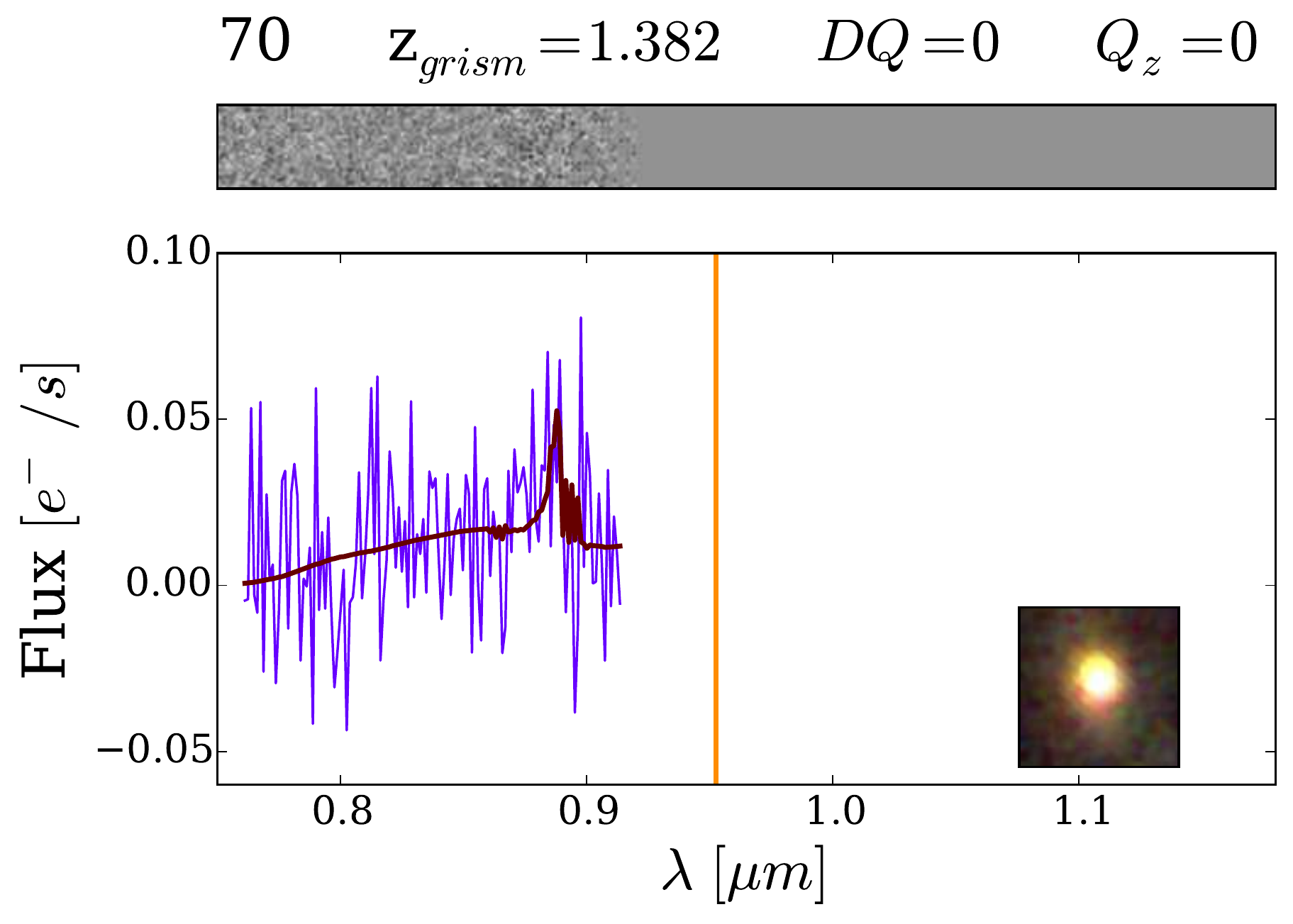} & 
\includegraphics[width = 0.33 \textwidth]{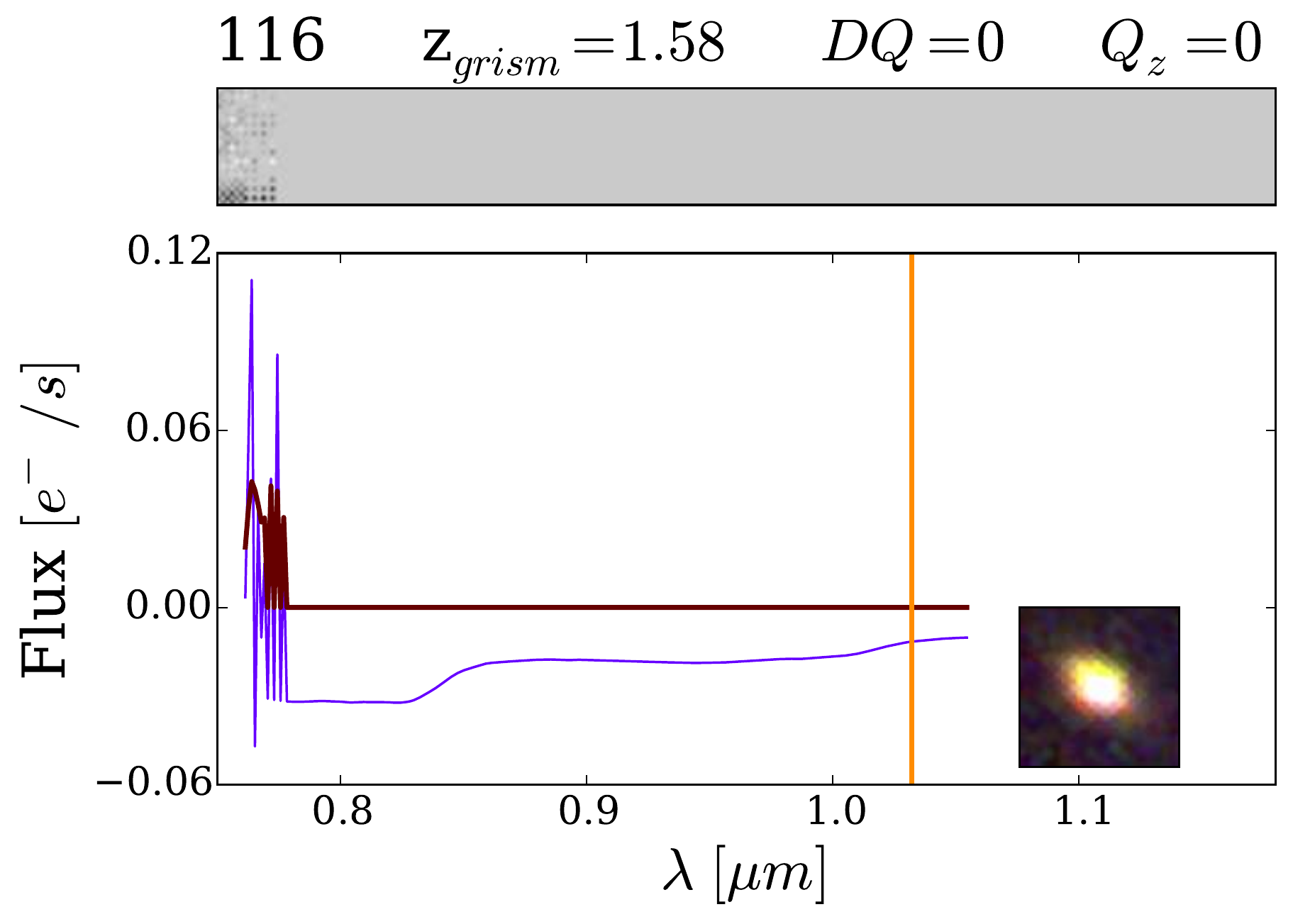} & 
\includegraphics[width = 0.33 \textwidth]{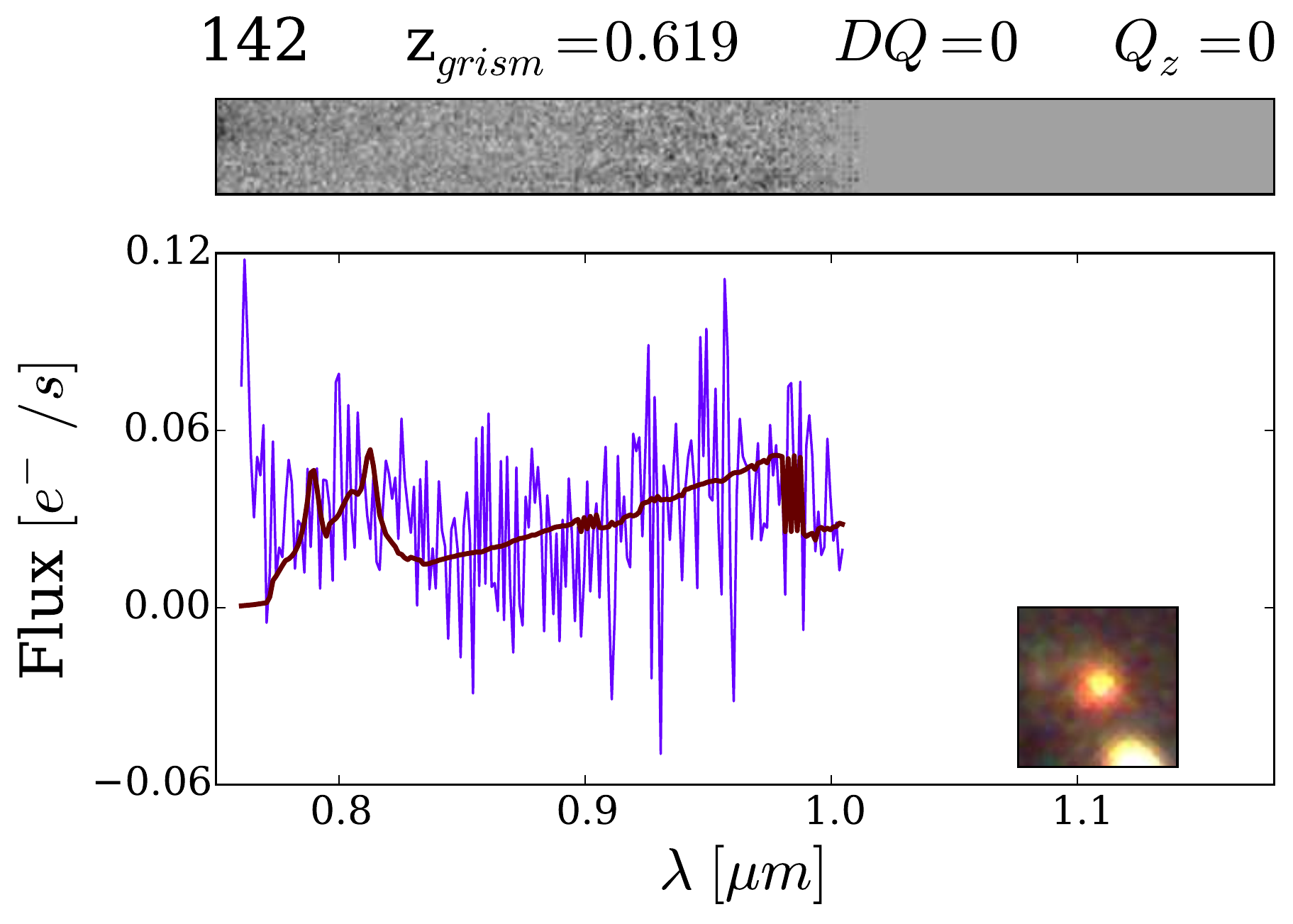} \\ 
\includegraphics[width = 0.33 \textwidth]{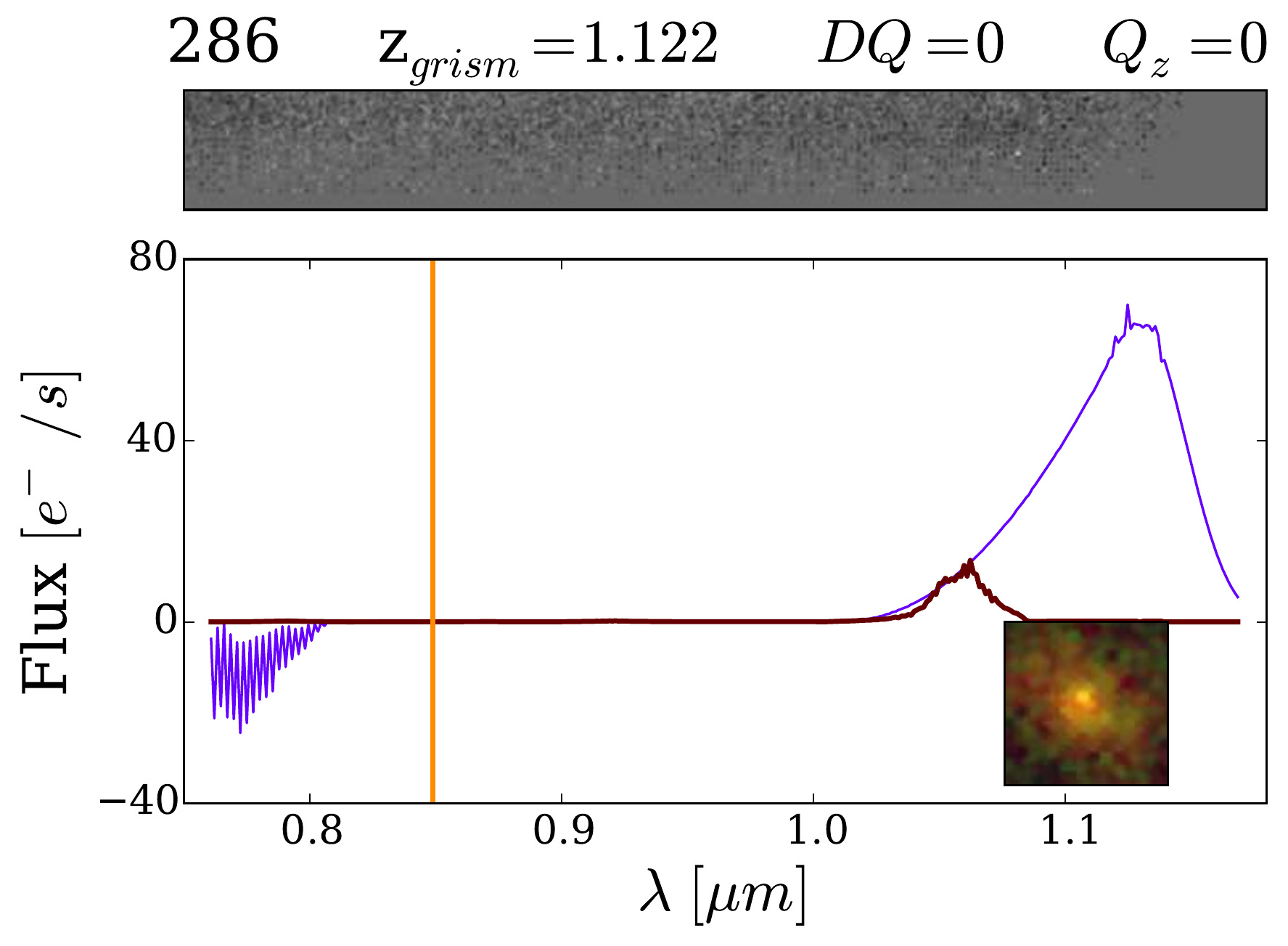} & 
\includegraphics[width = 0.33 \textwidth]{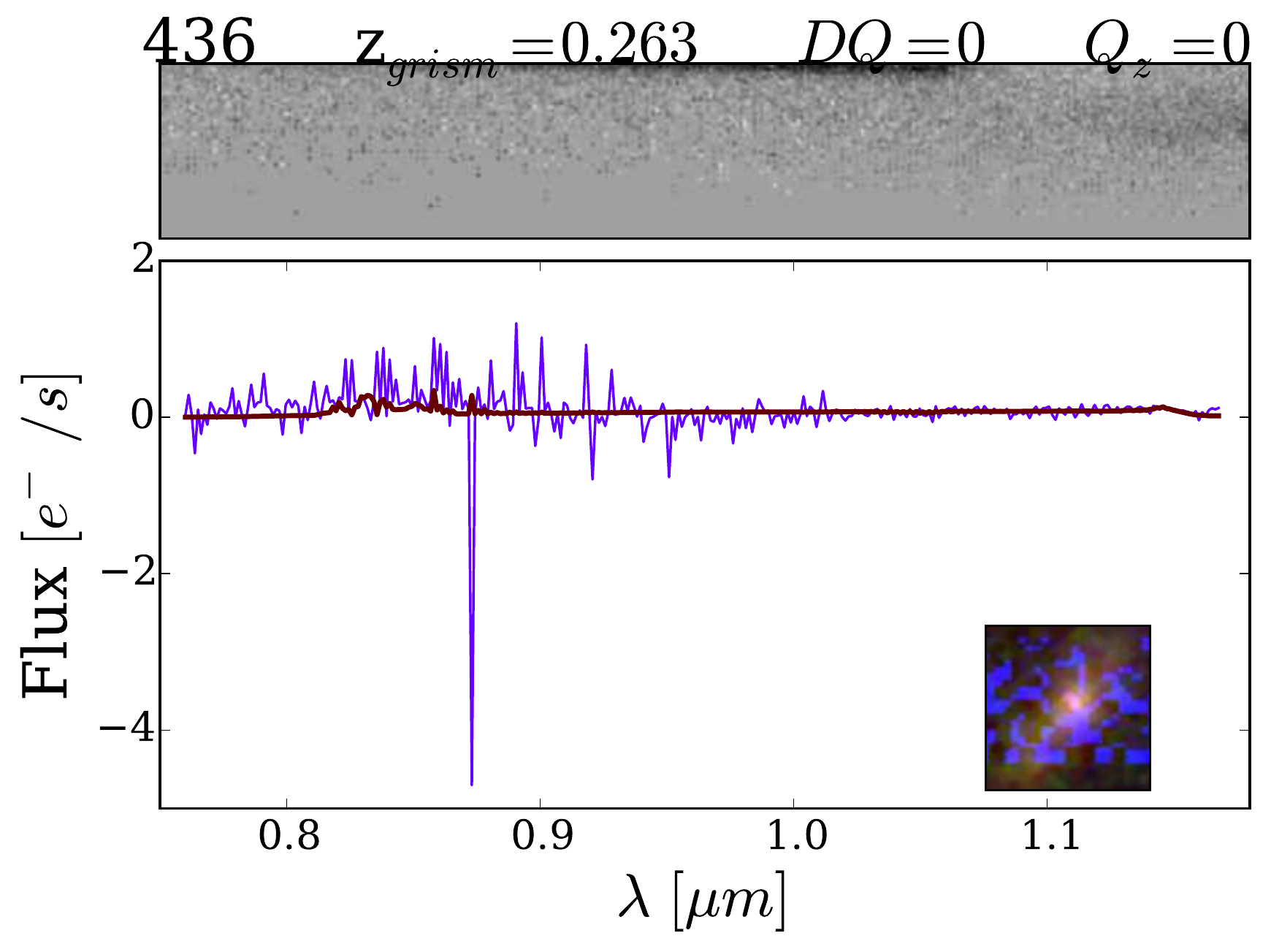} & 
\includegraphics[width = 0.33 \textwidth]{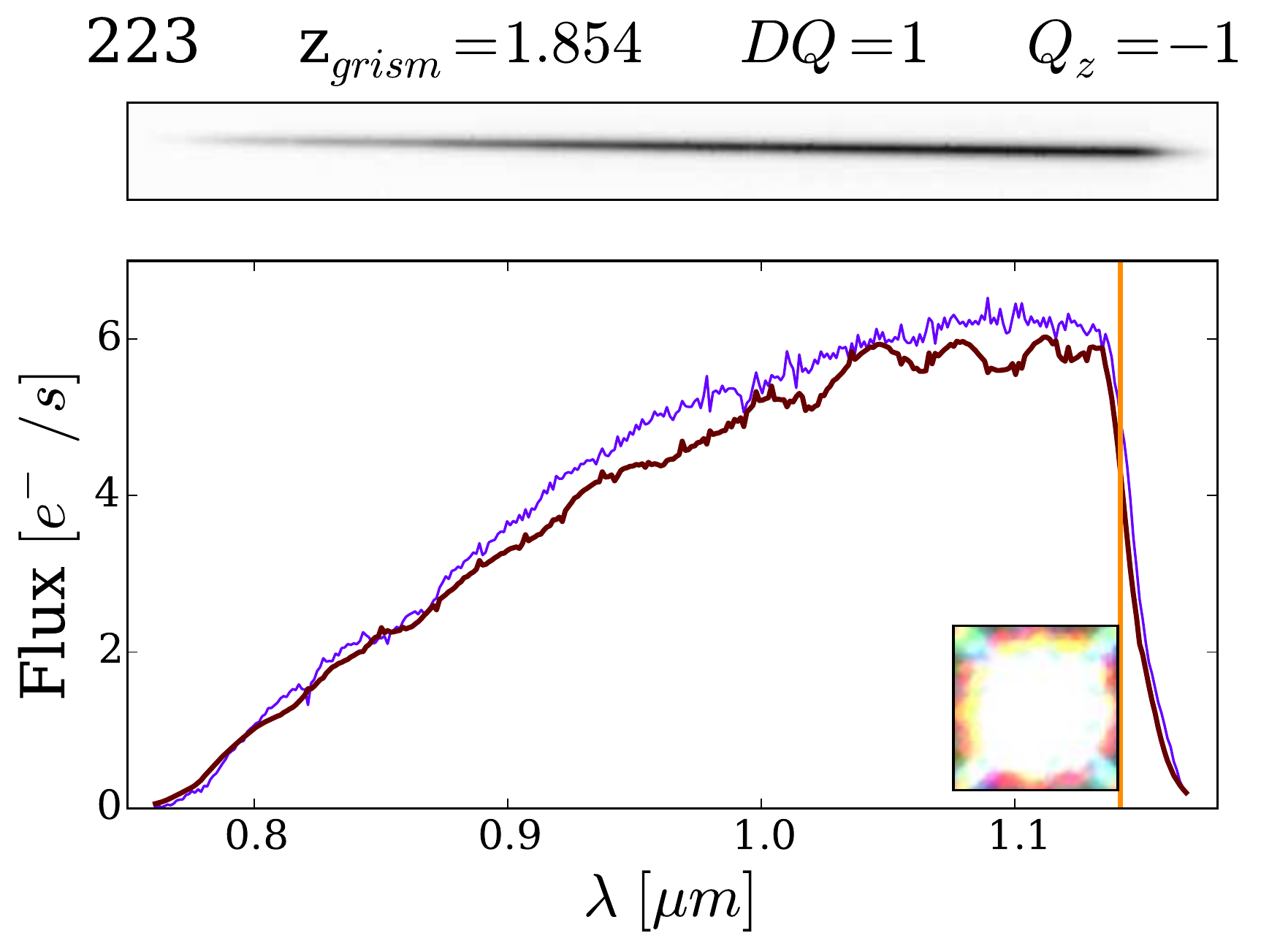} \\
\includegraphics[width = 0.33 \textwidth]{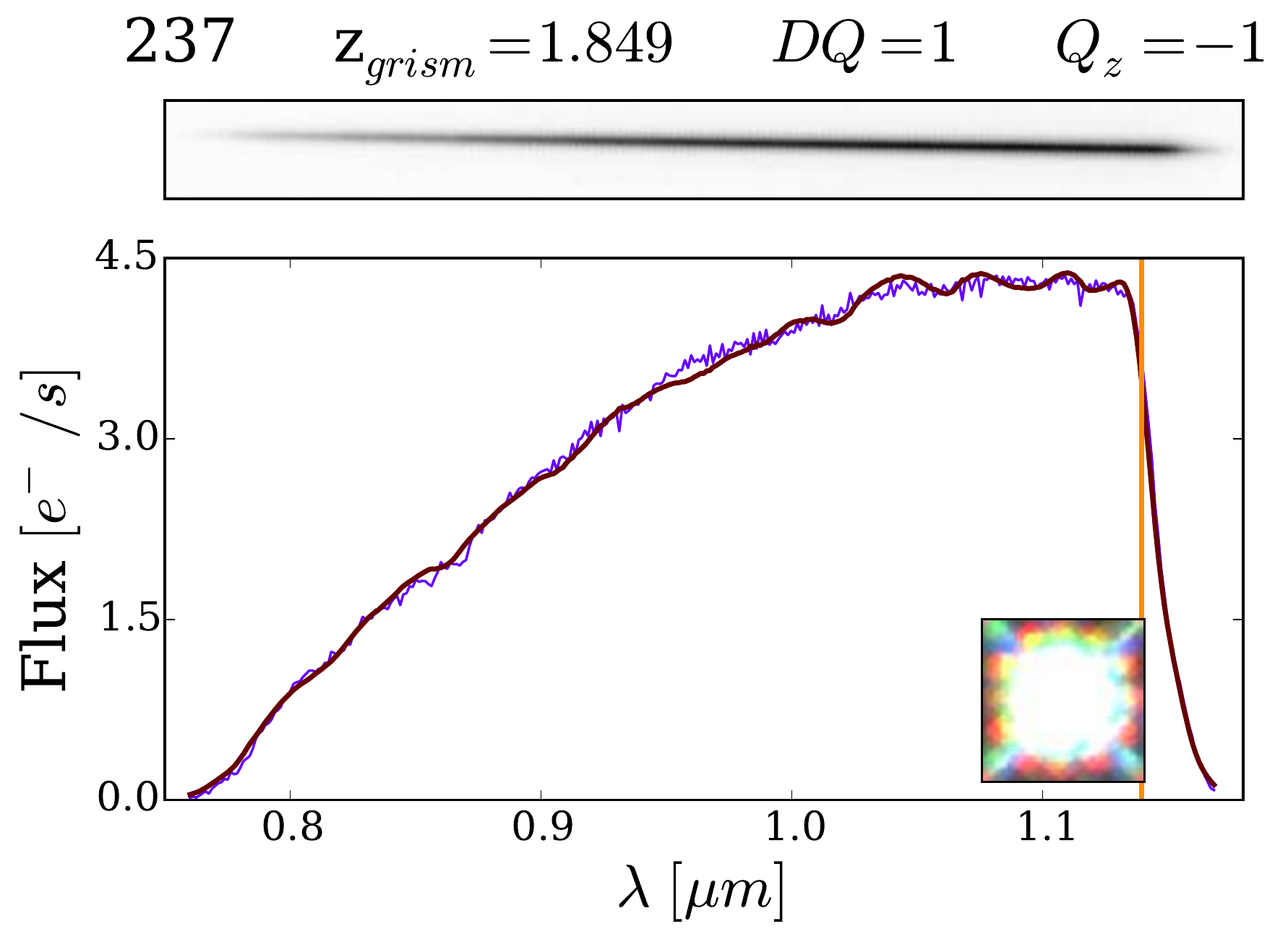} & 
\end{array}$
\caption{2D and 1D grism spectra for IRC-A candidate cluster members selected using method III (see \S \ref{sec:III}). The red curve represents the best-fit of the 1D grism spectra. The orange line denotes the redshifted 4000\AA\ line. Galaxies are ordered according to the value of the redshift quality flag (\qz) value, with \qz = 3 indicating the most robust grism measurement and \qz = 1 or 0 being the least reliable. Note that some objects shown here may not have been used in the calculation of the pair fraction, but are included for completeness (see \S \ref{sec:III}).}
\end{center}
\label{gris_pan_A2}
\end{figure}

\begin{figure} 
\figurenum{B2}
\begin{center}$
\begin{array}{ccc}
\includegraphics[width = 0.33 \textwidth]{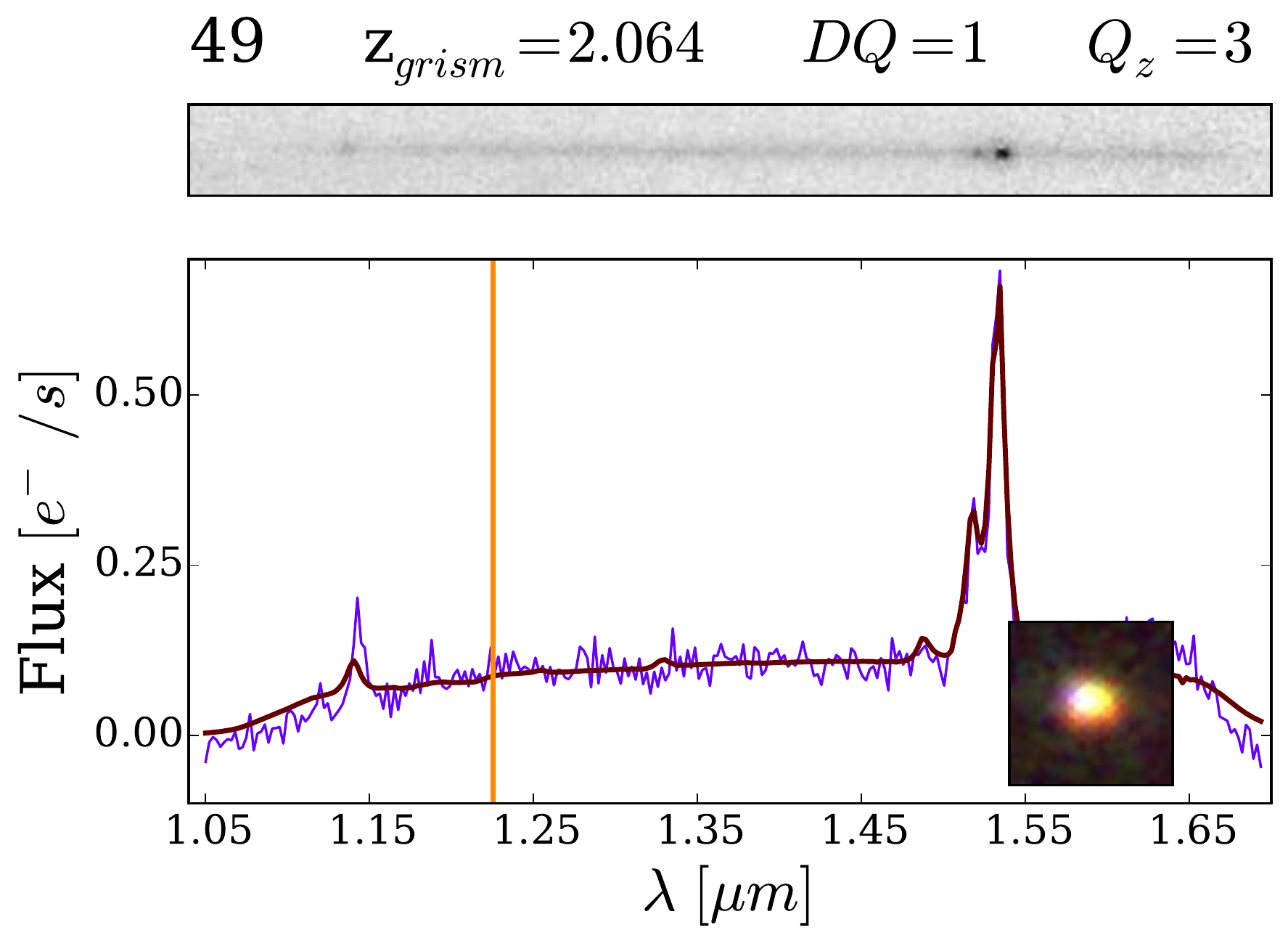} & 
\includegraphics[width = 0.33 \textwidth]{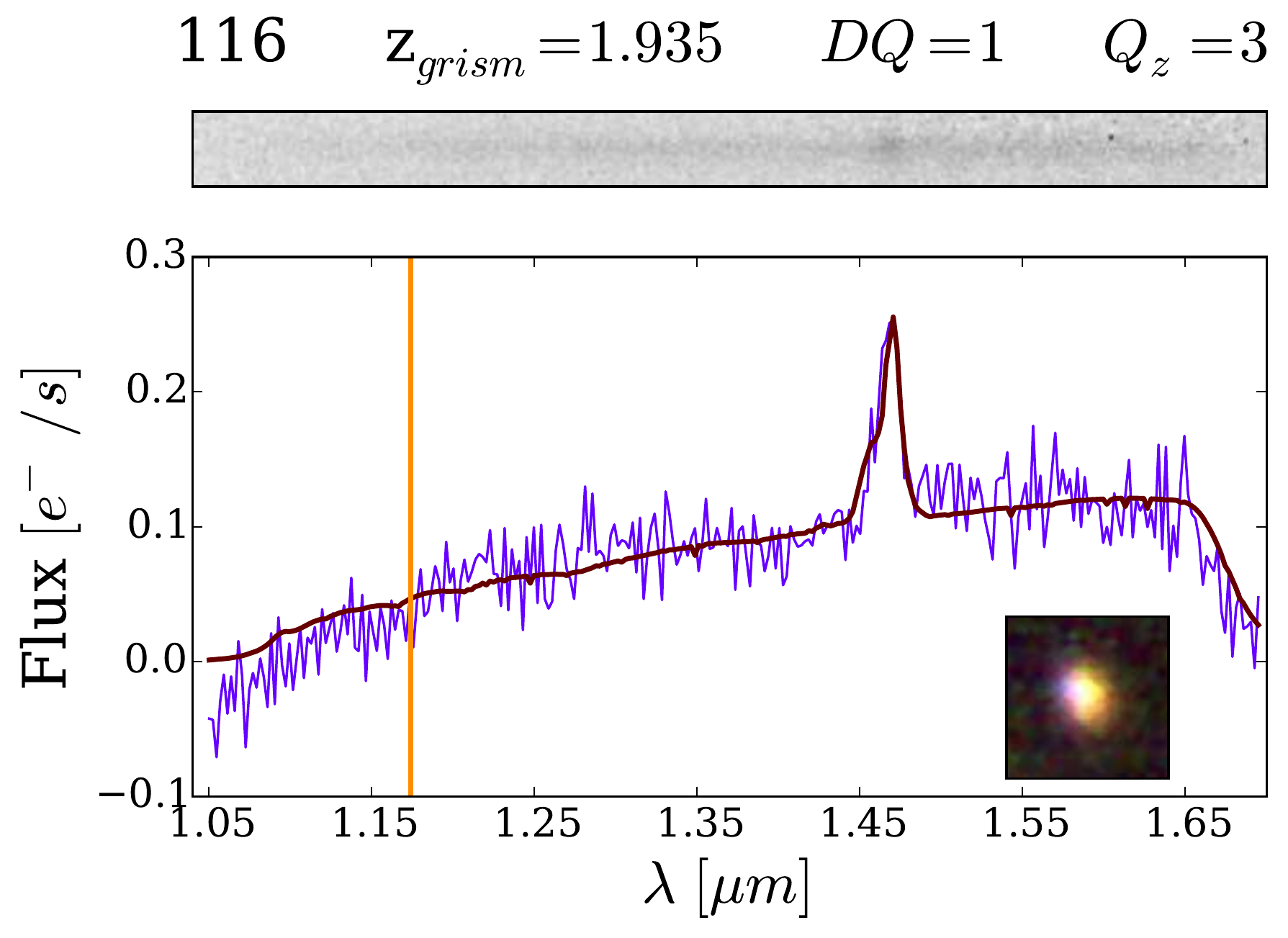} & 
\includegraphics[width = 0.33 \textwidth]{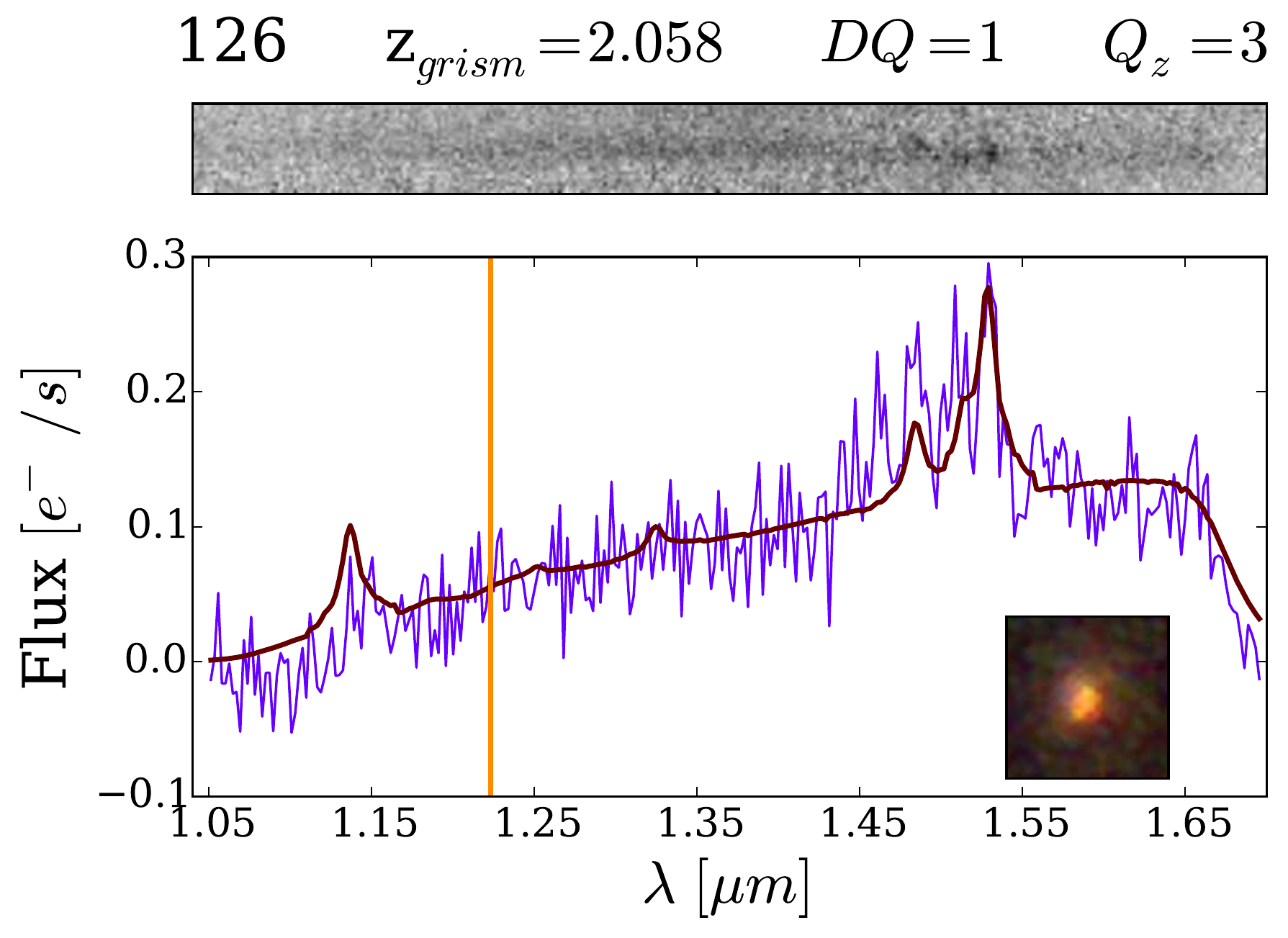} \\
\includegraphics[width = 0.33 \textwidth]{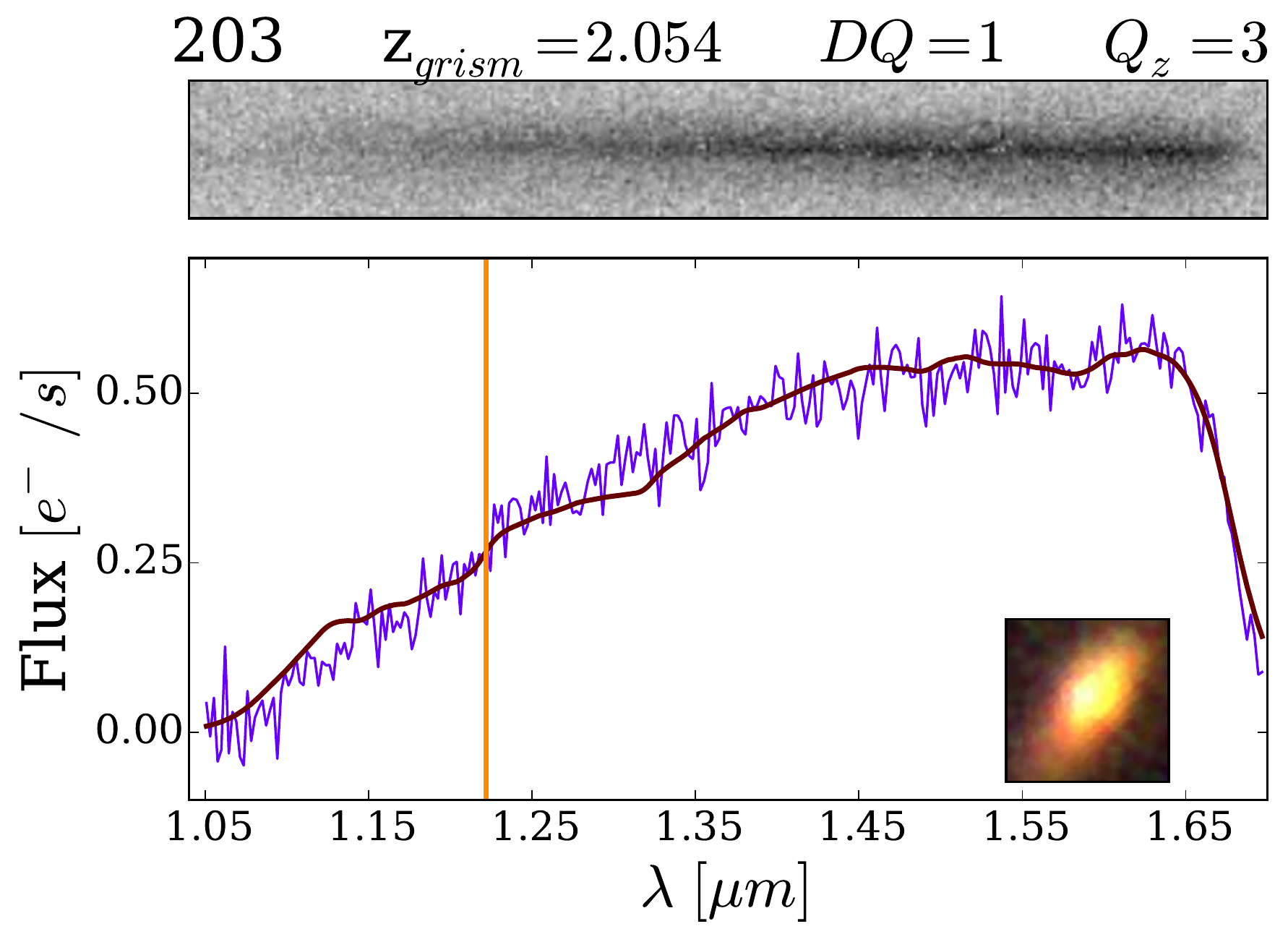} & 
\includegraphics[width = 0.33 \textwidth]{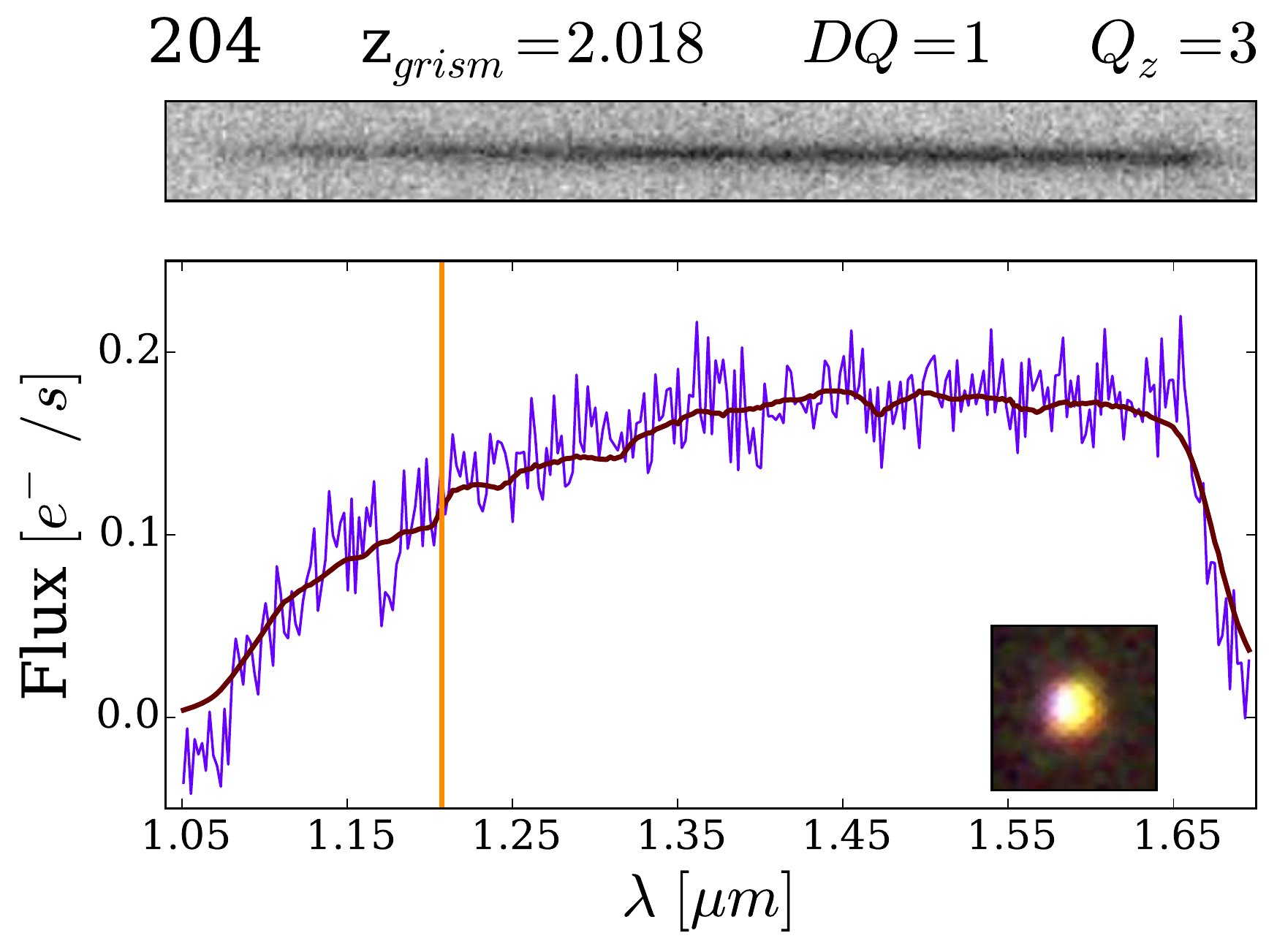} & 
\includegraphics[width = 0.33 \textwidth]{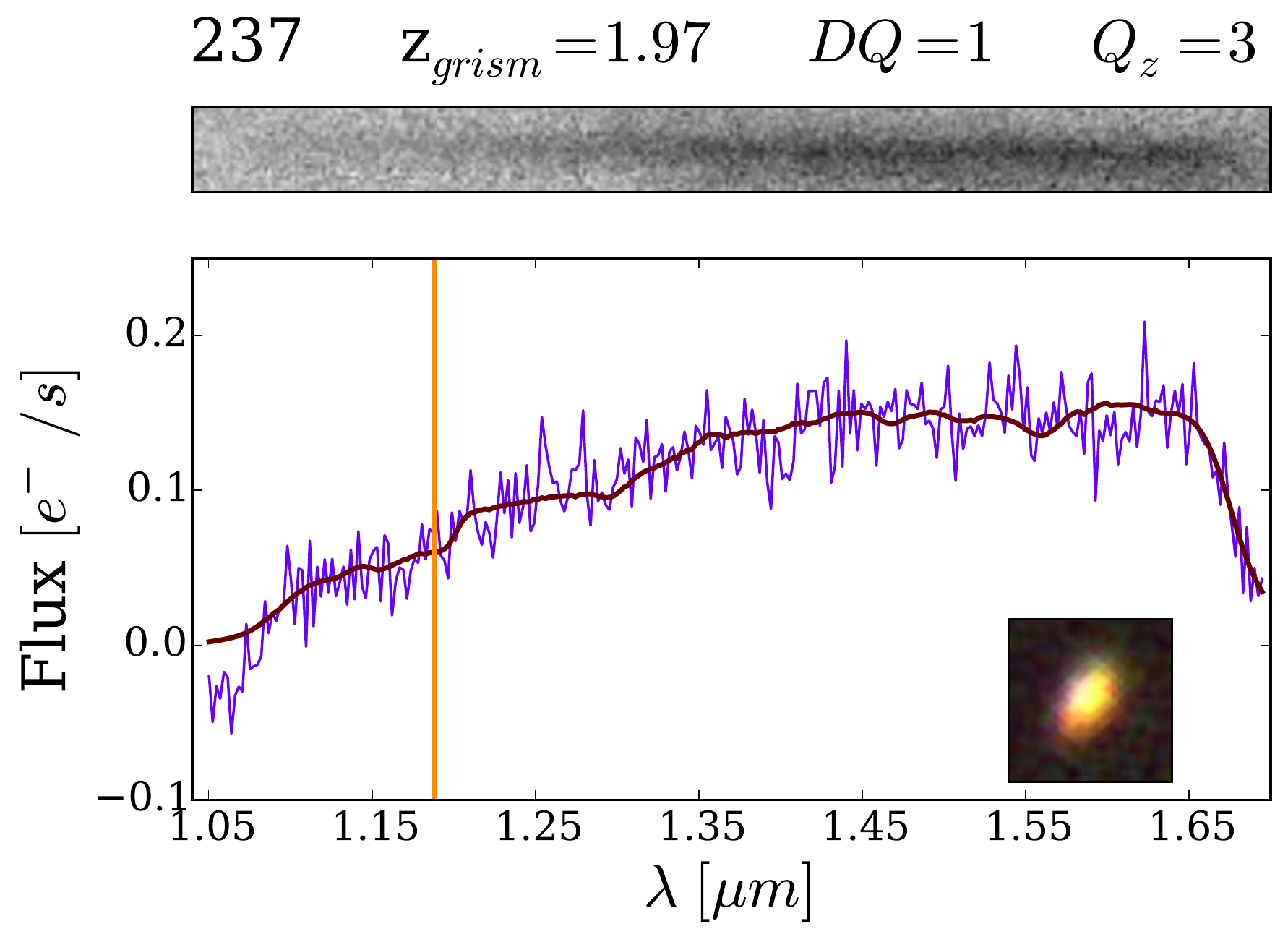} \\ 
\includegraphics[width = 0.33 \textwidth]{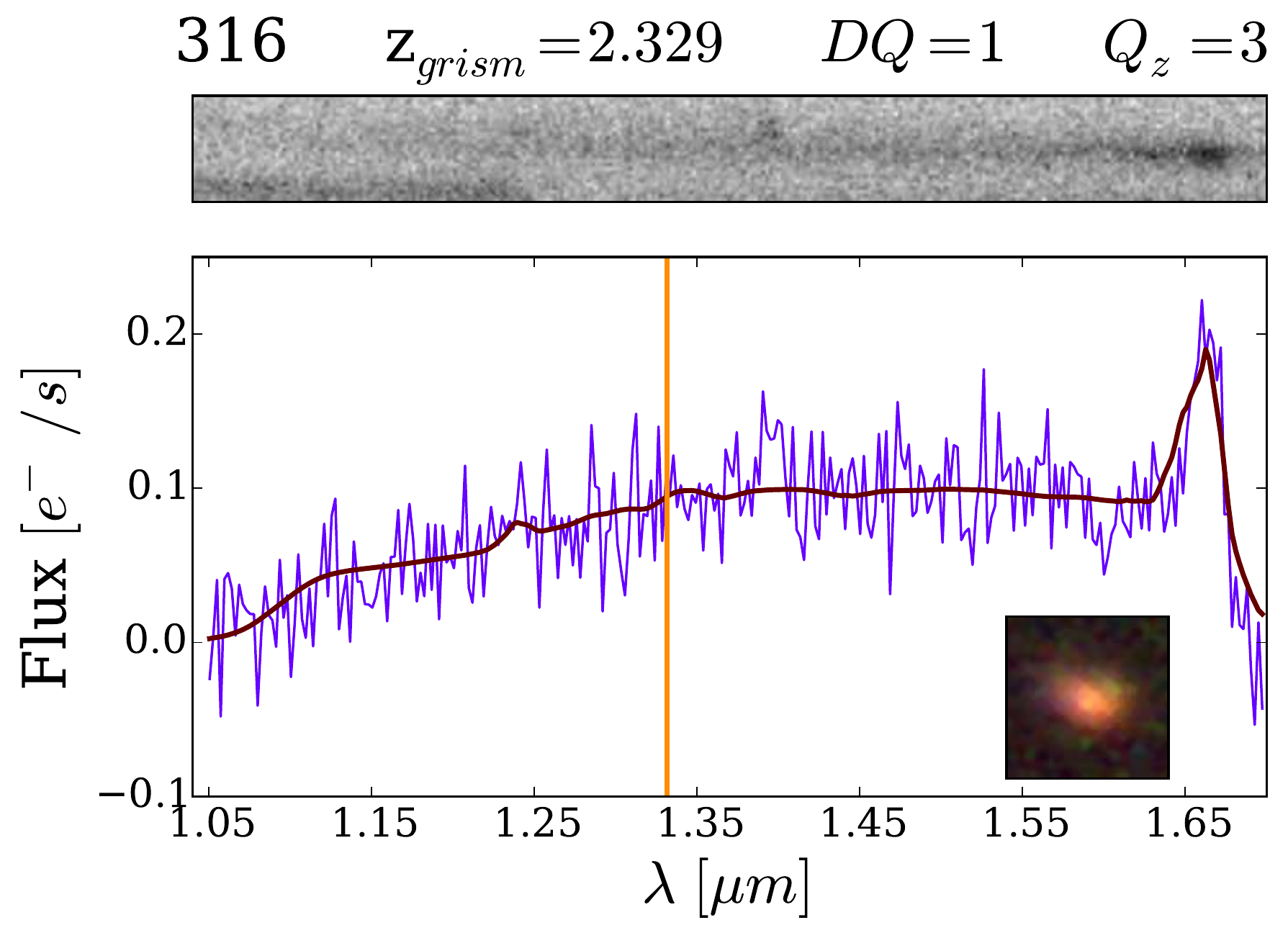} & 
\includegraphics[width = 0.33 \textwidth]{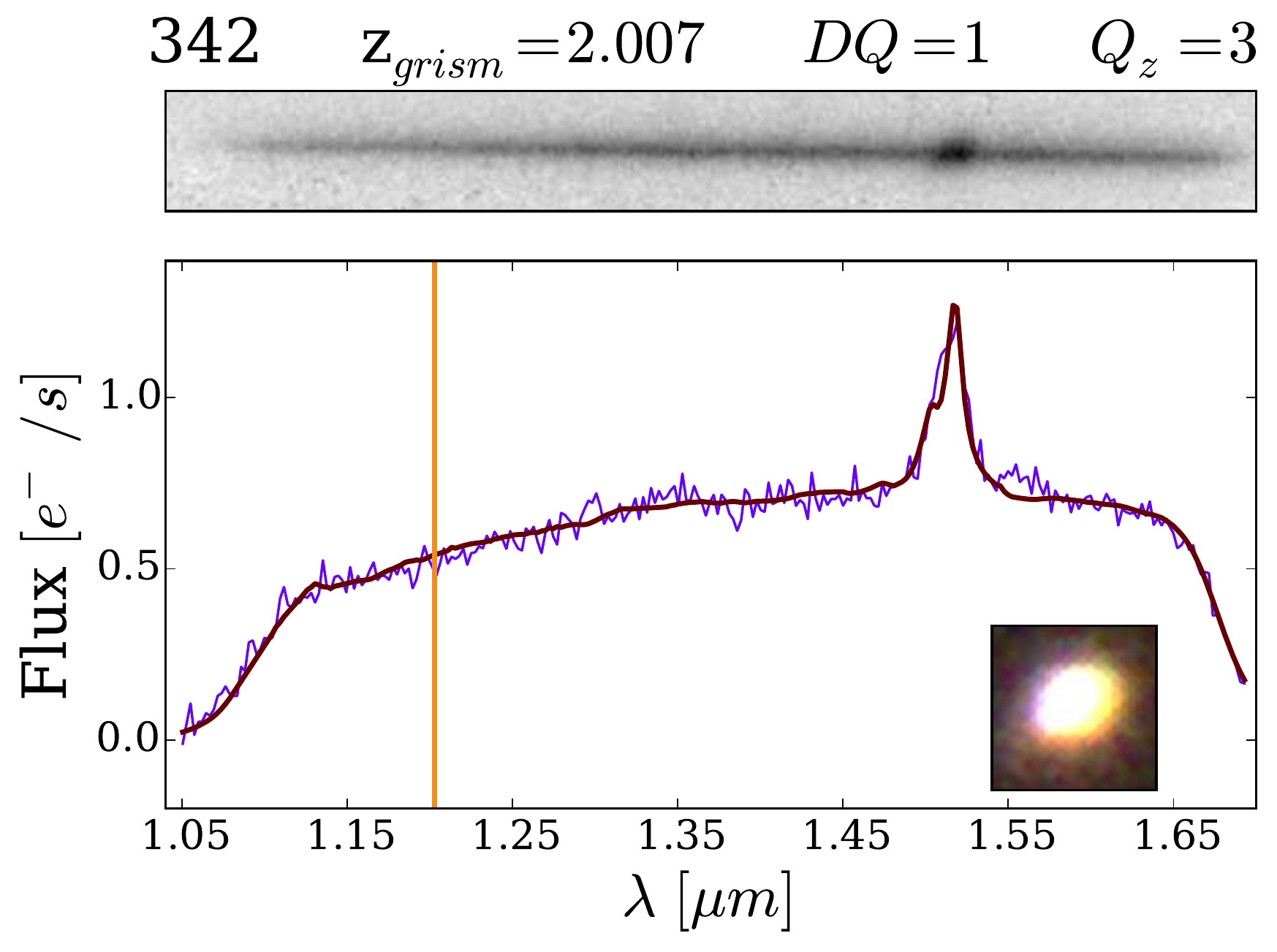} & 
\includegraphics[width = 0.33 \textwidth]{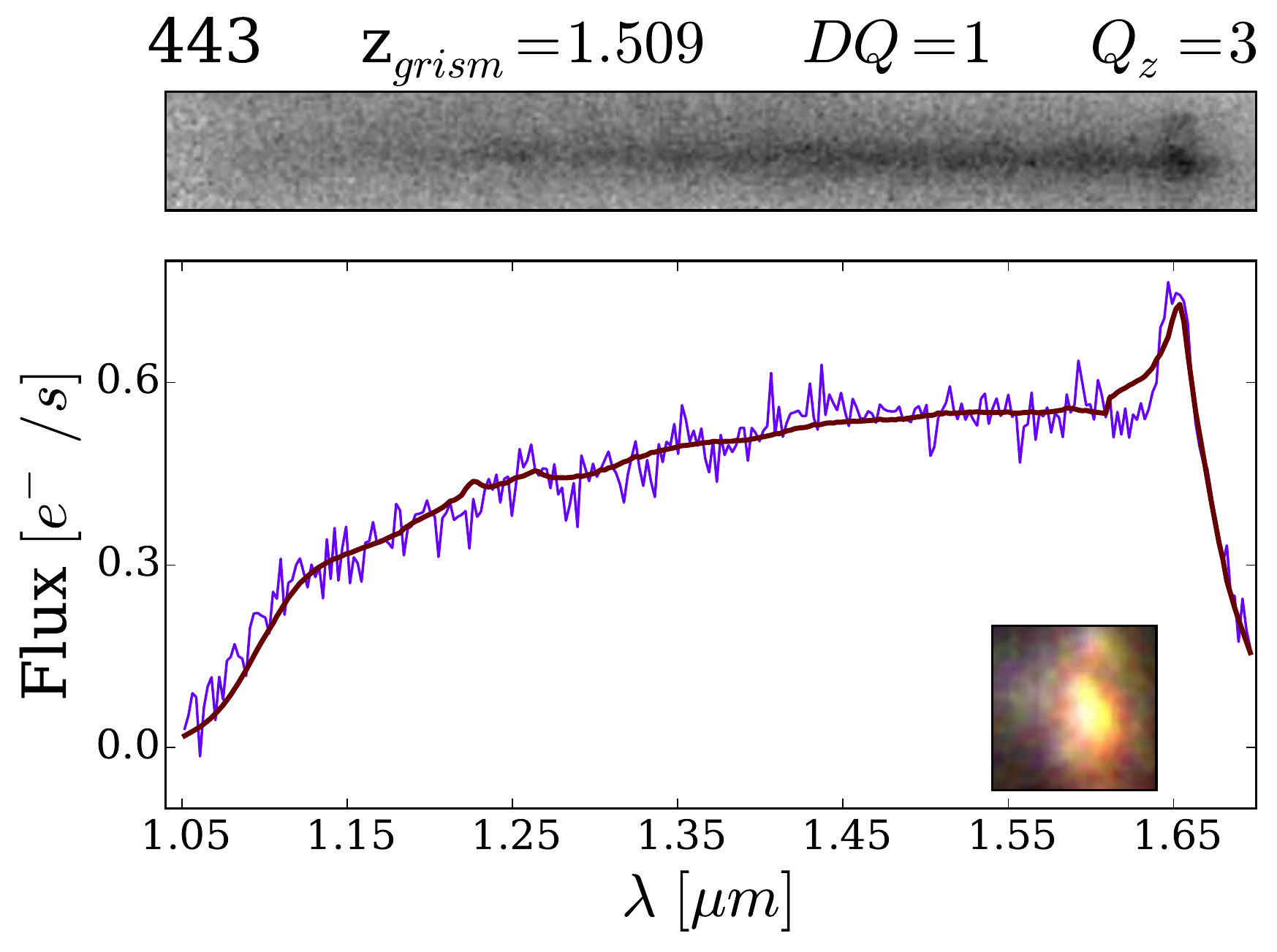} \\ 
\includegraphics[width = 0.33 \textwidth]{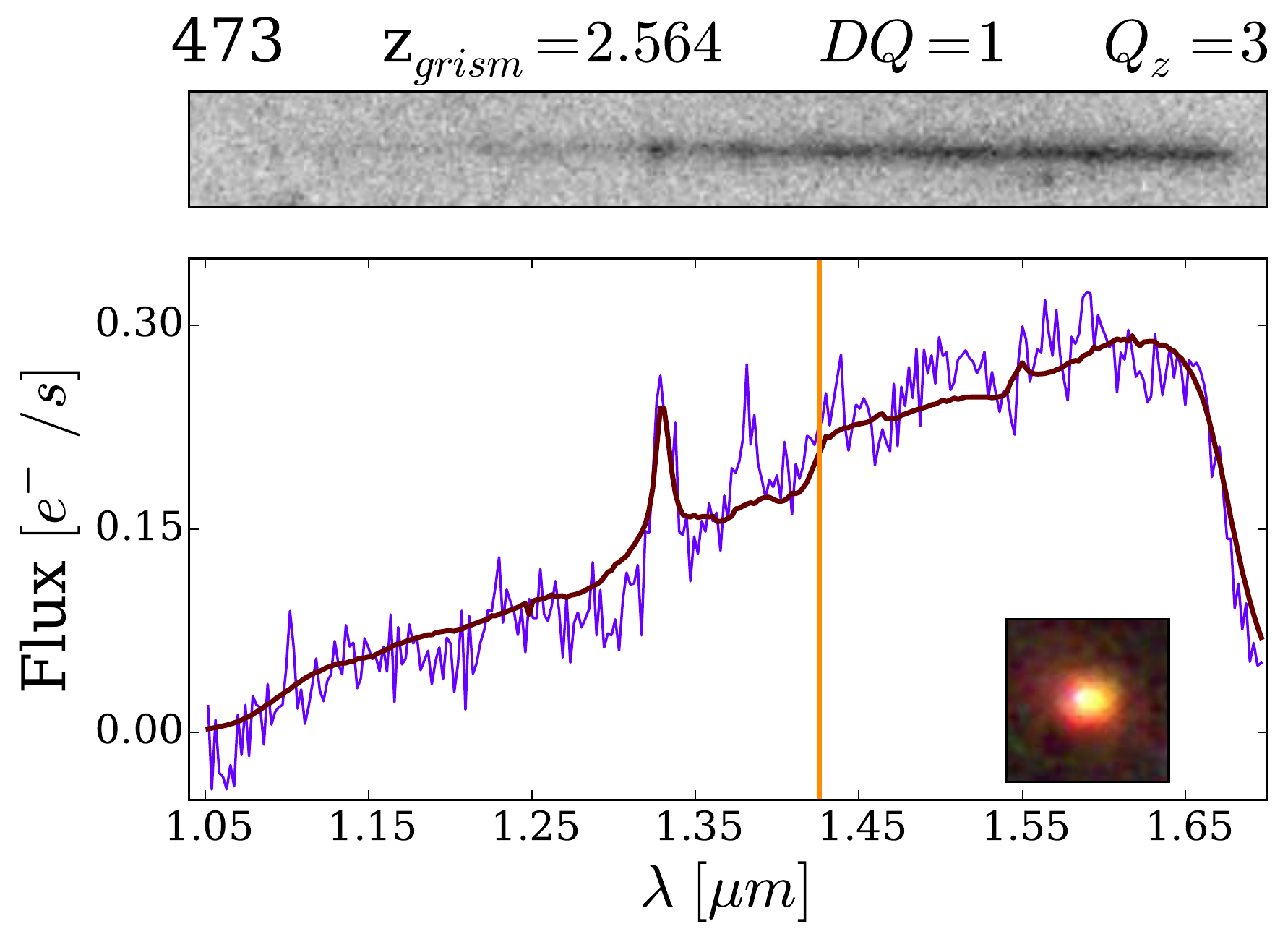} & 
\includegraphics[width = 0.33 \textwidth]{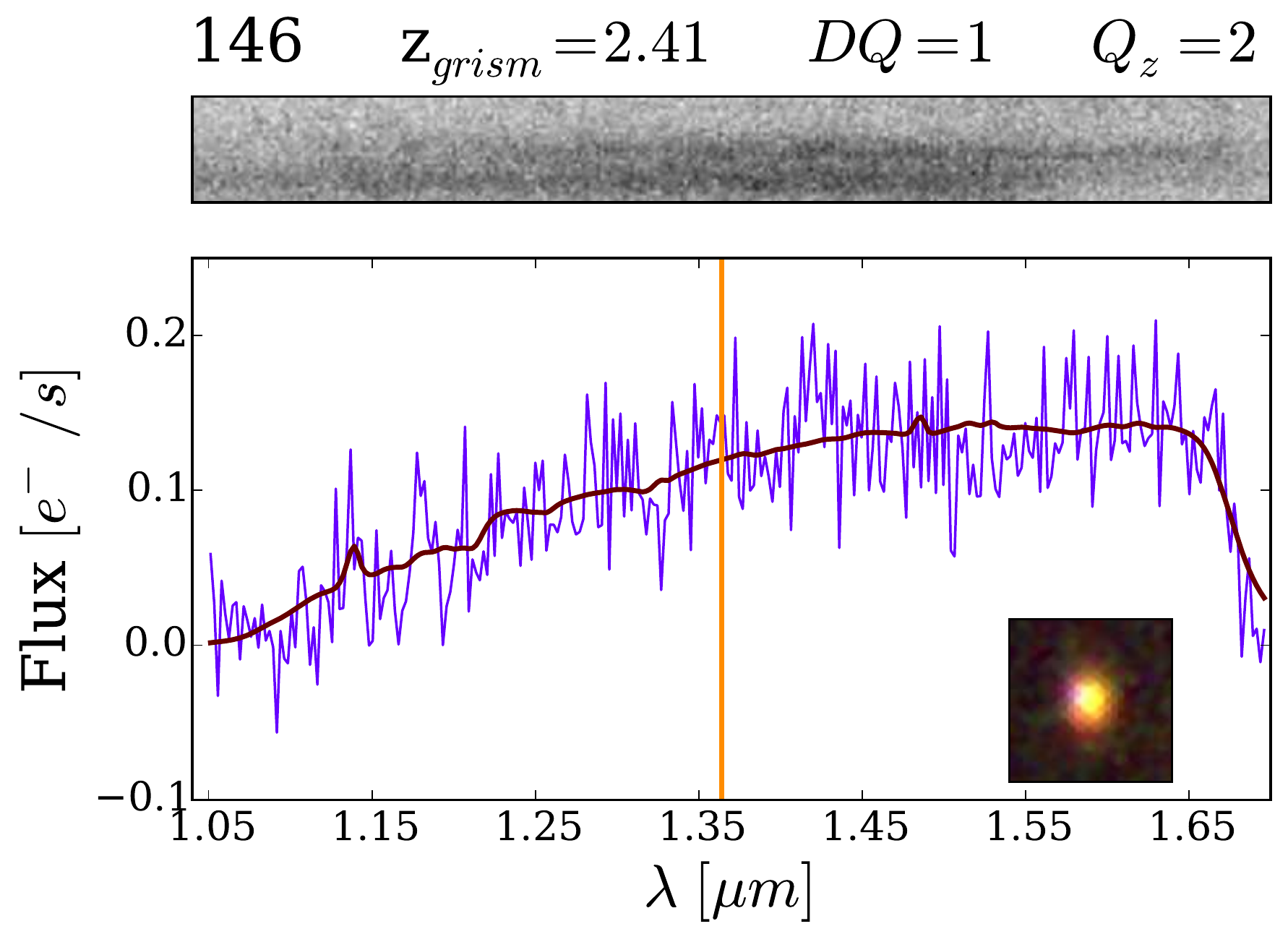} & 
\includegraphics[width = 0.33 \textwidth]{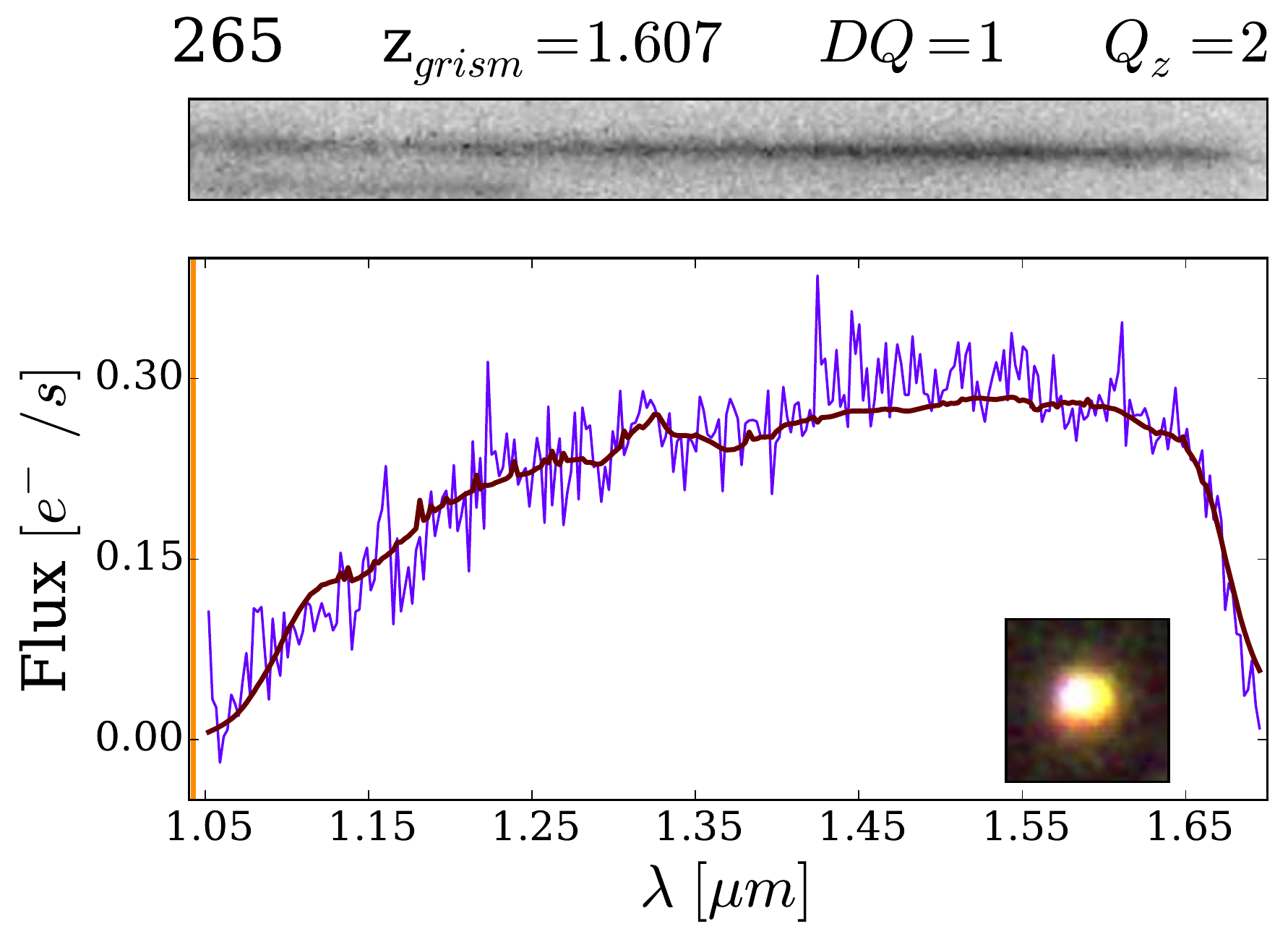} \\ 
\includegraphics[width = 0.33 \textwidth]{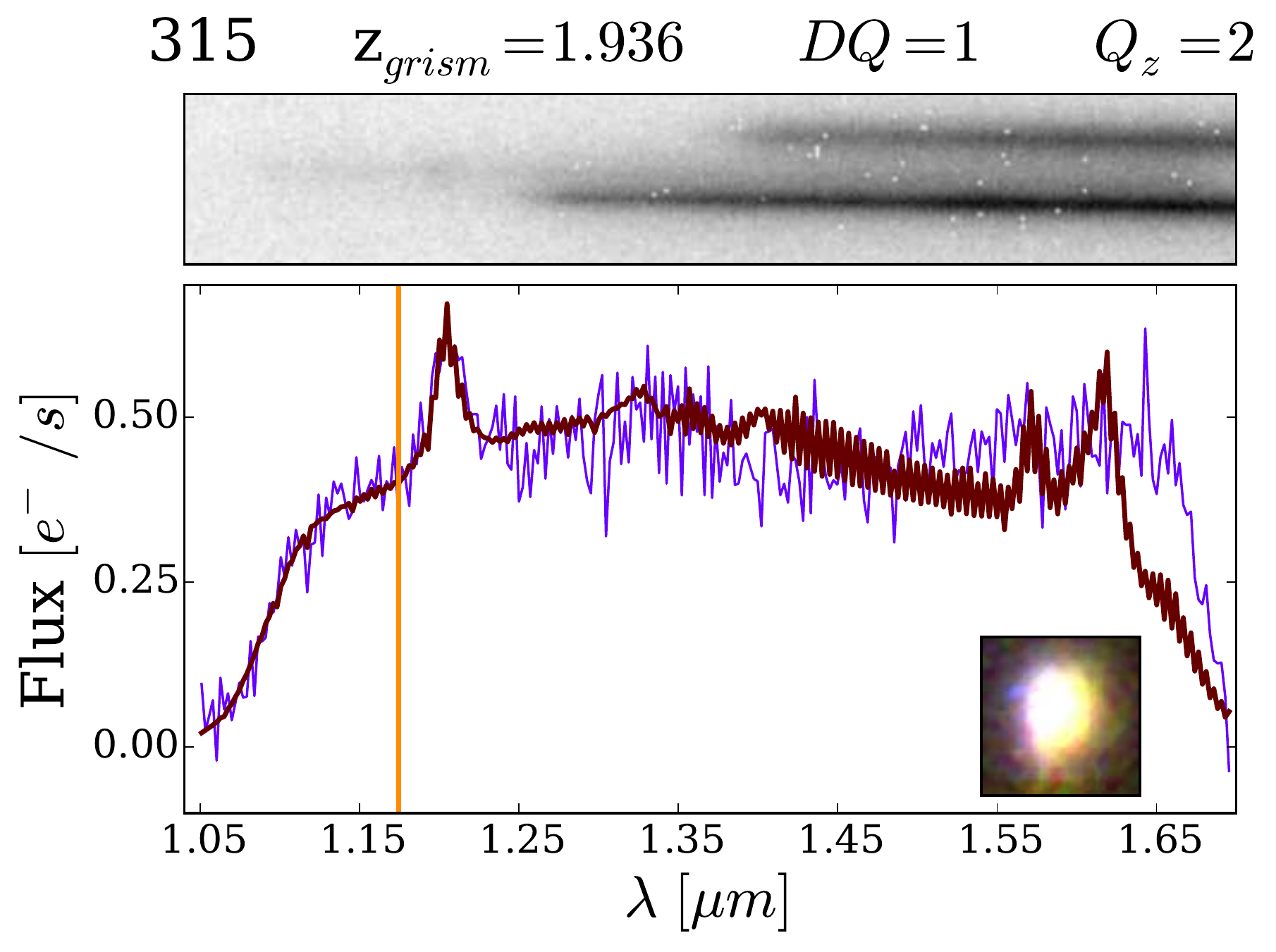} & 
\includegraphics[width = 0.33 \textwidth]{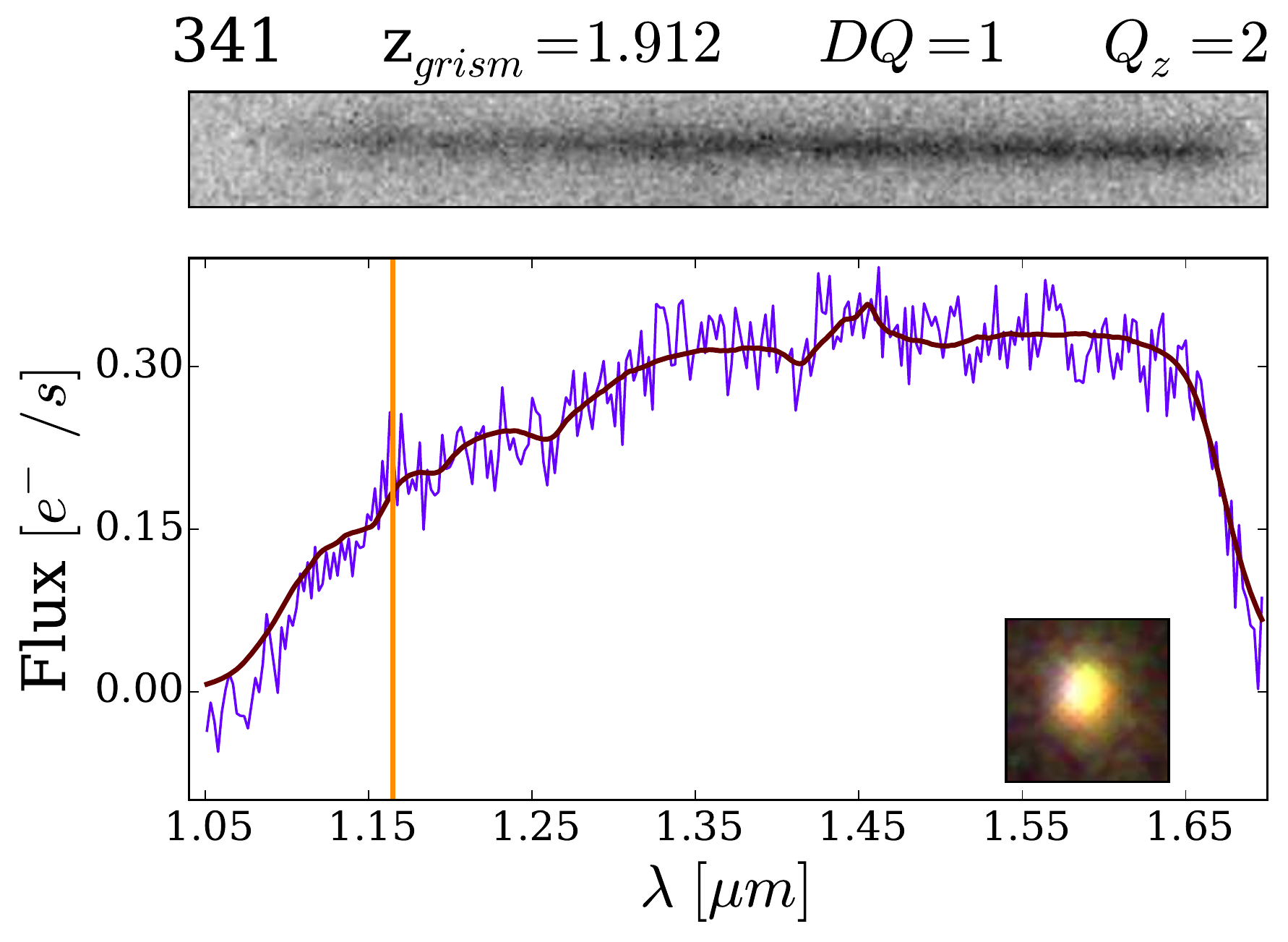} & 
\includegraphics[width = 0.33 \textwidth]{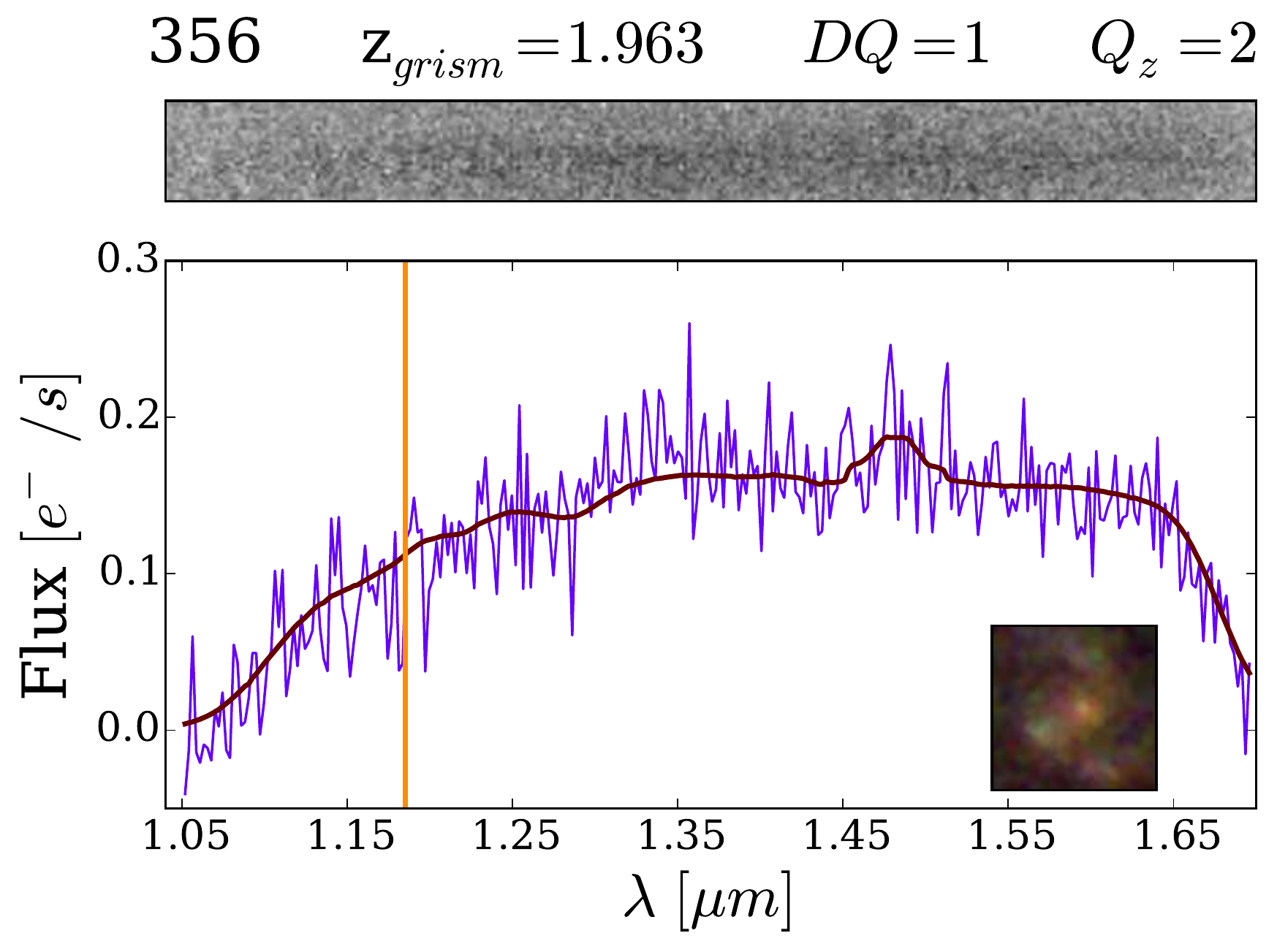} \\
\end{array}$
\caption{Continued on next page.}
\end{center}
\label{gris_pan_B1}
\end{figure}

\begin{figure} 
\figurenum{B2}
\begin{center}$
\begin{array}{ccc}
\includegraphics[width = 0.33 \textwidth]{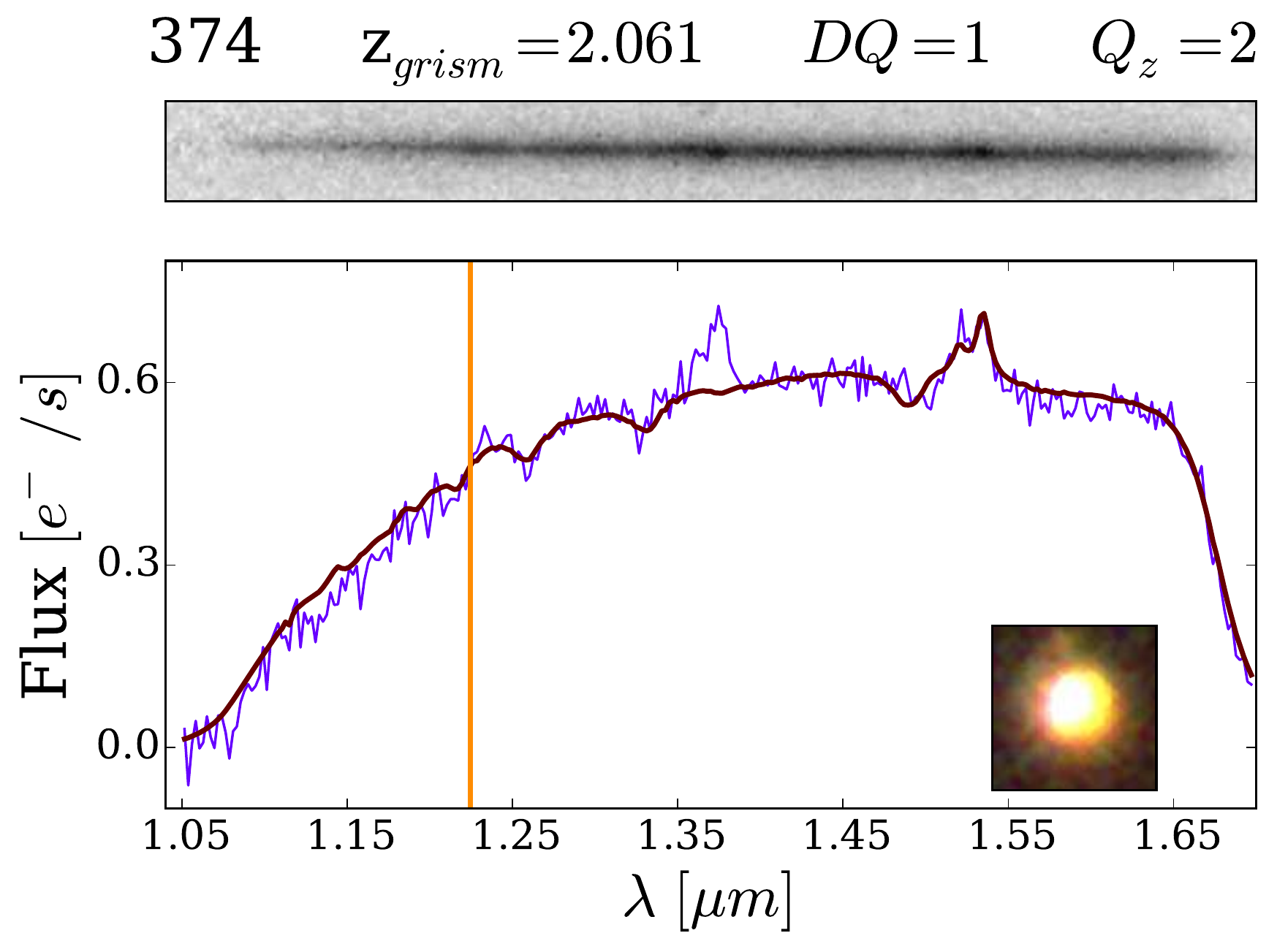} & 
\includegraphics[width = 0.33 \textwidth]{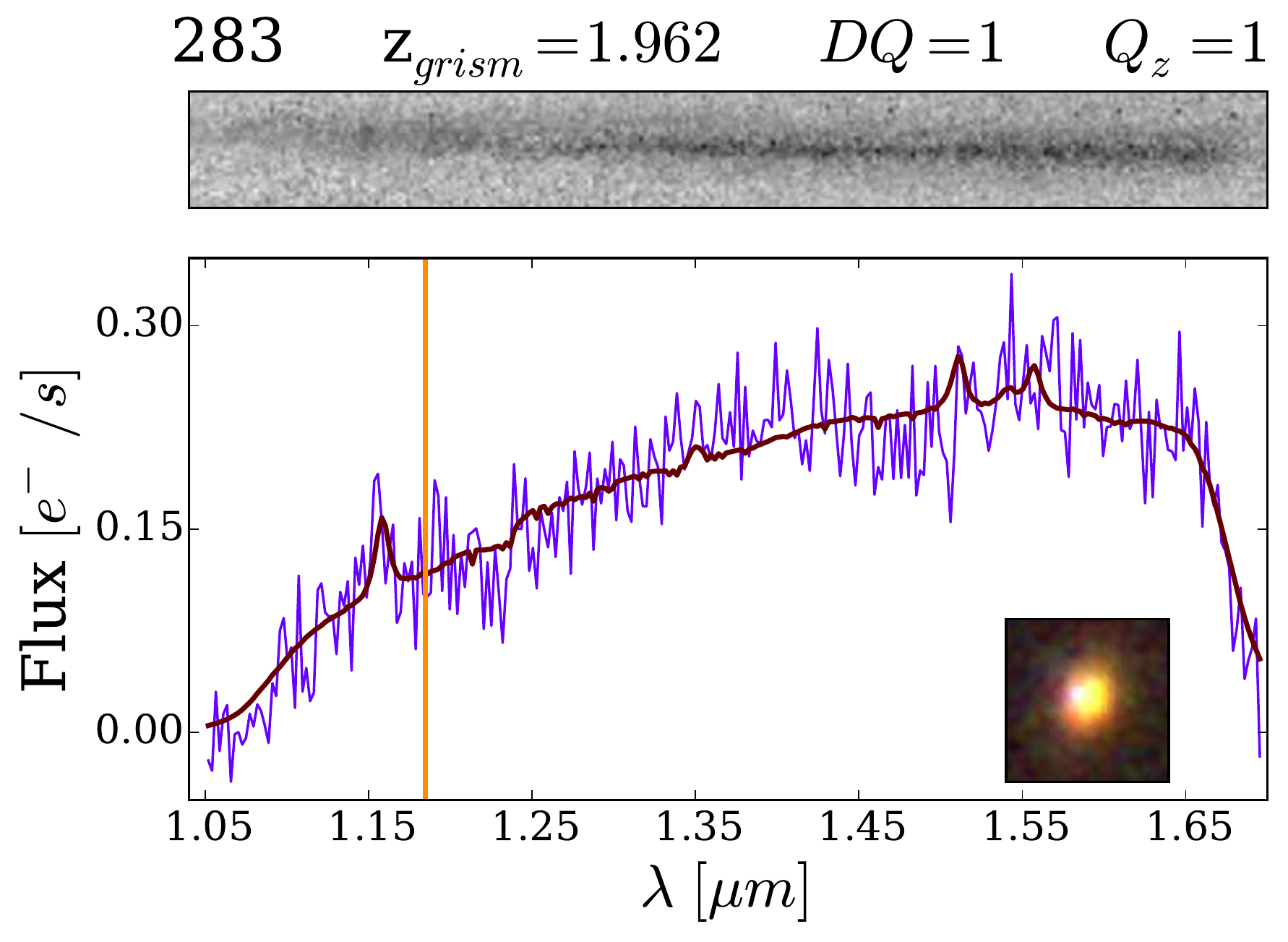} & 
\includegraphics[width = 0.33 \textwidth]{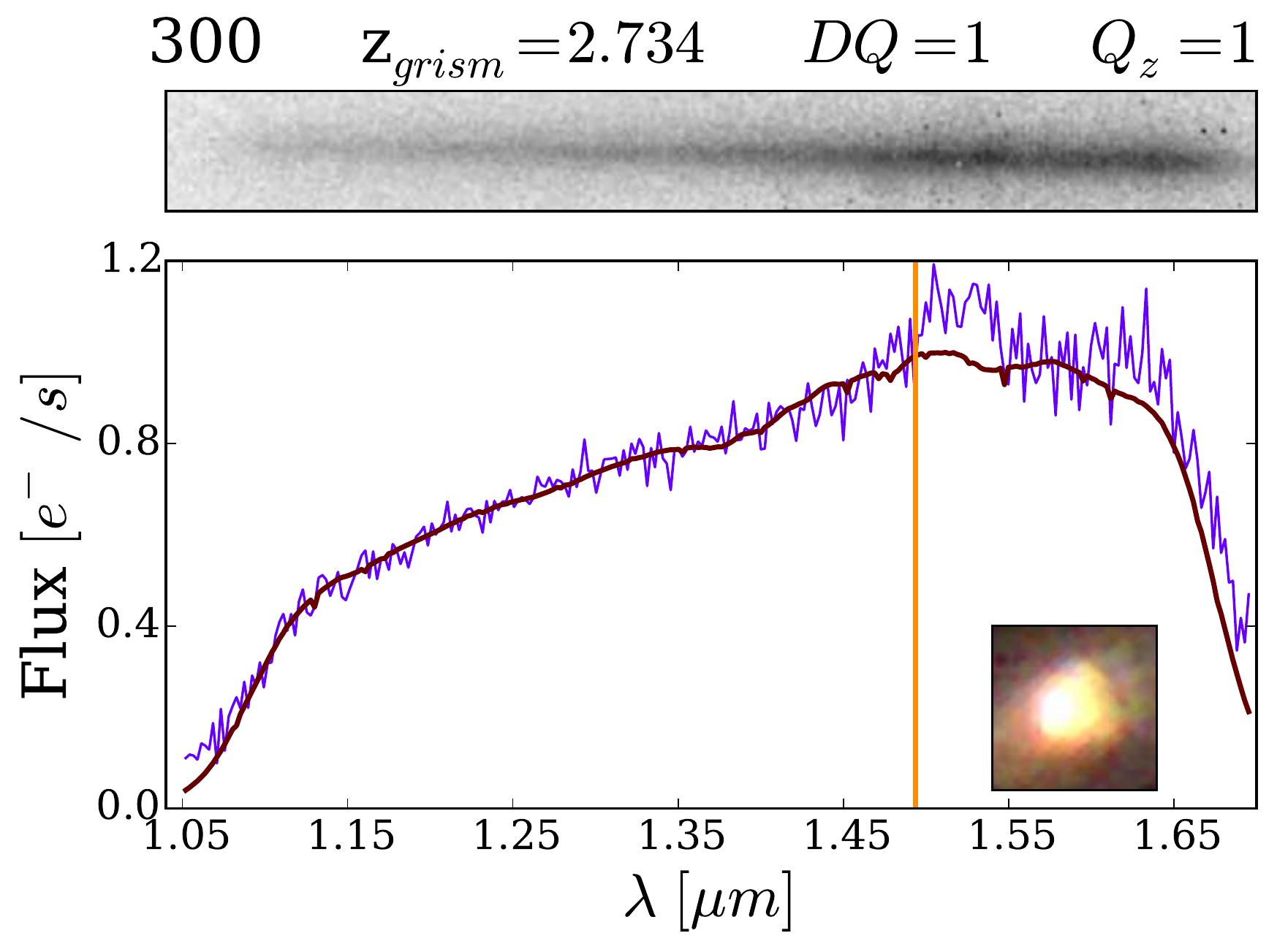} \\
\includegraphics[width = 0.33 \textwidth]{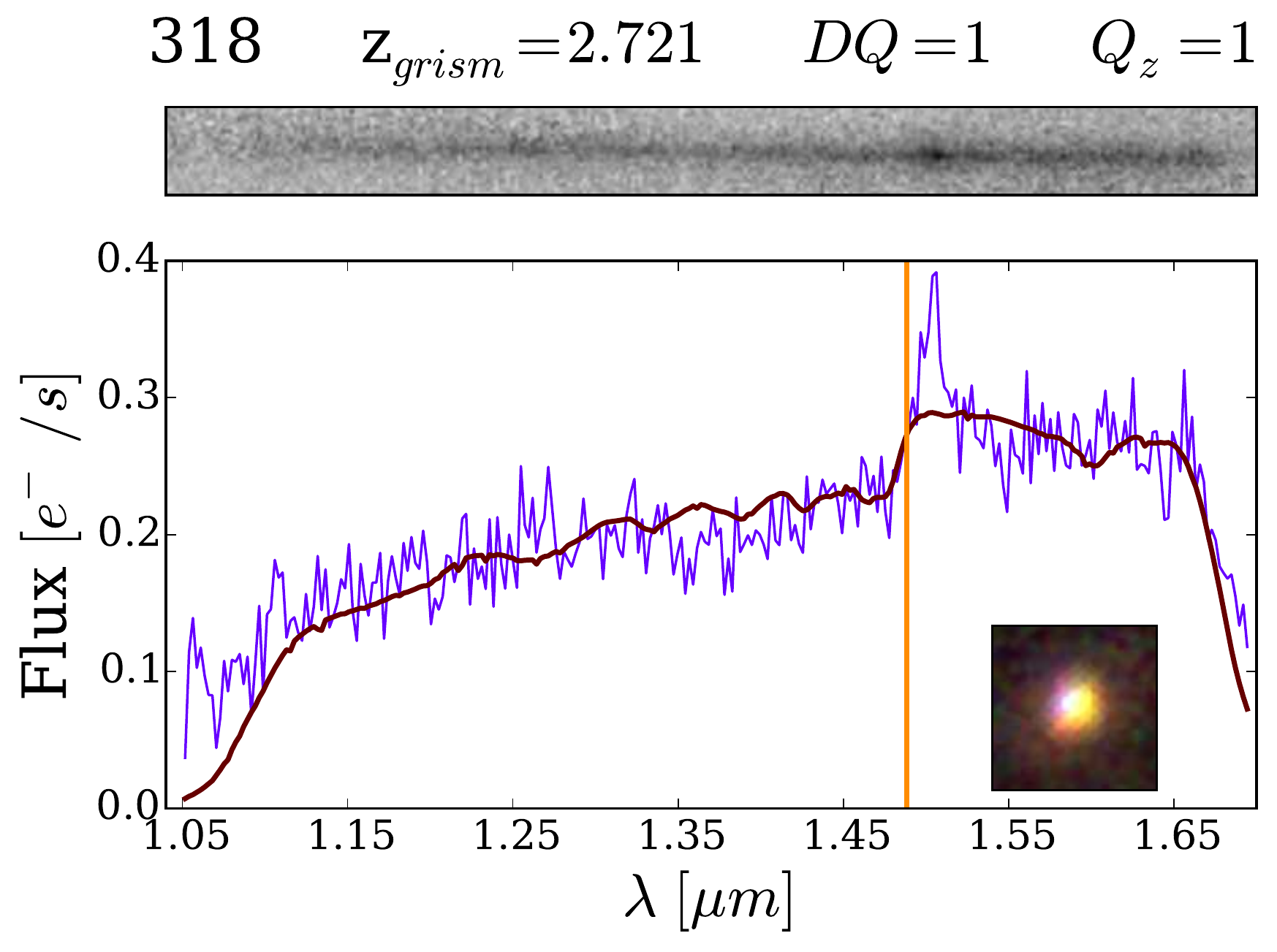} & 
\includegraphics[width = 0.33 \textwidth]{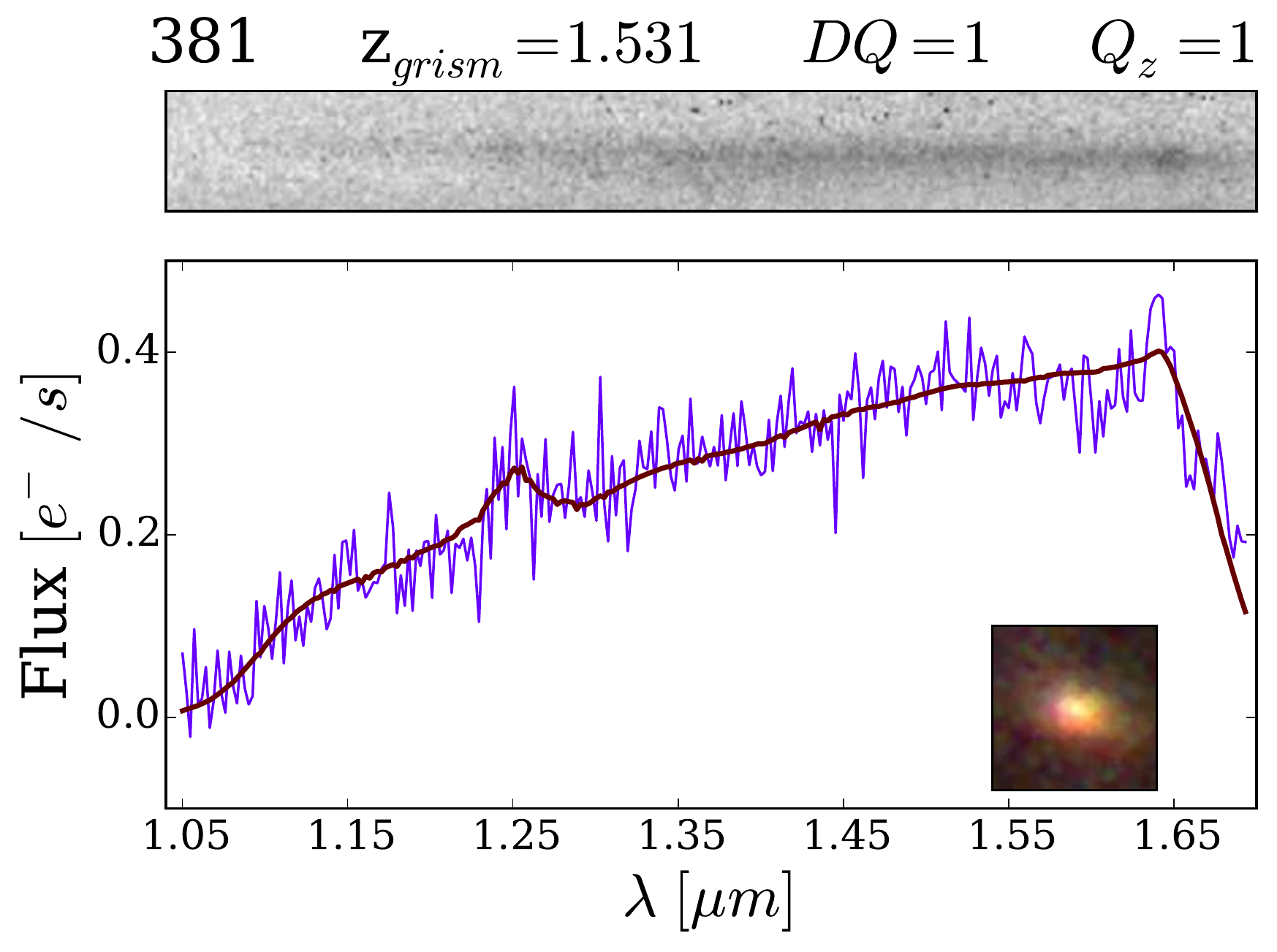} & 
\includegraphics[width = 0.33 \textwidth]{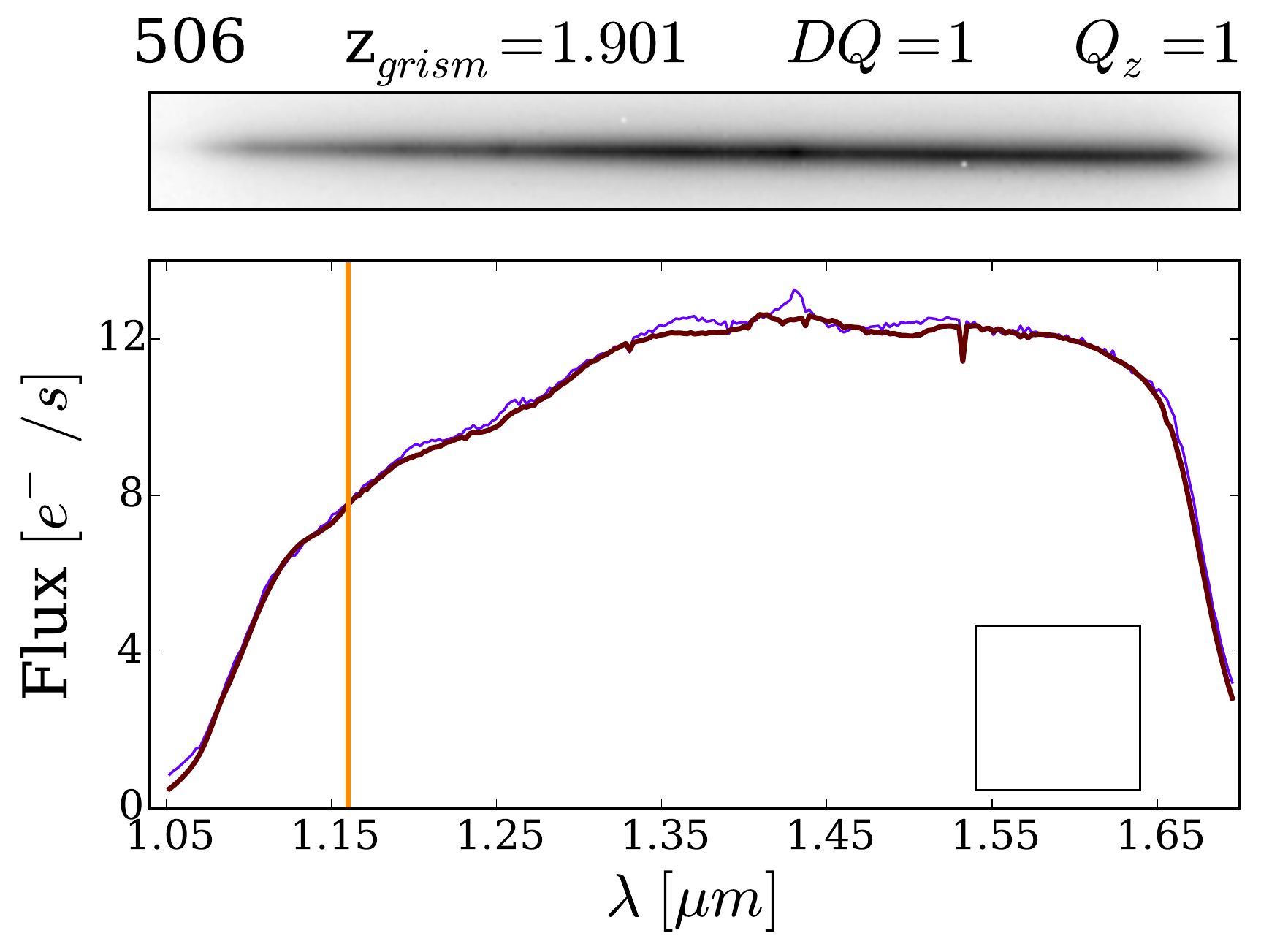} \\ 
\includegraphics[width = 0.33 \textwidth]{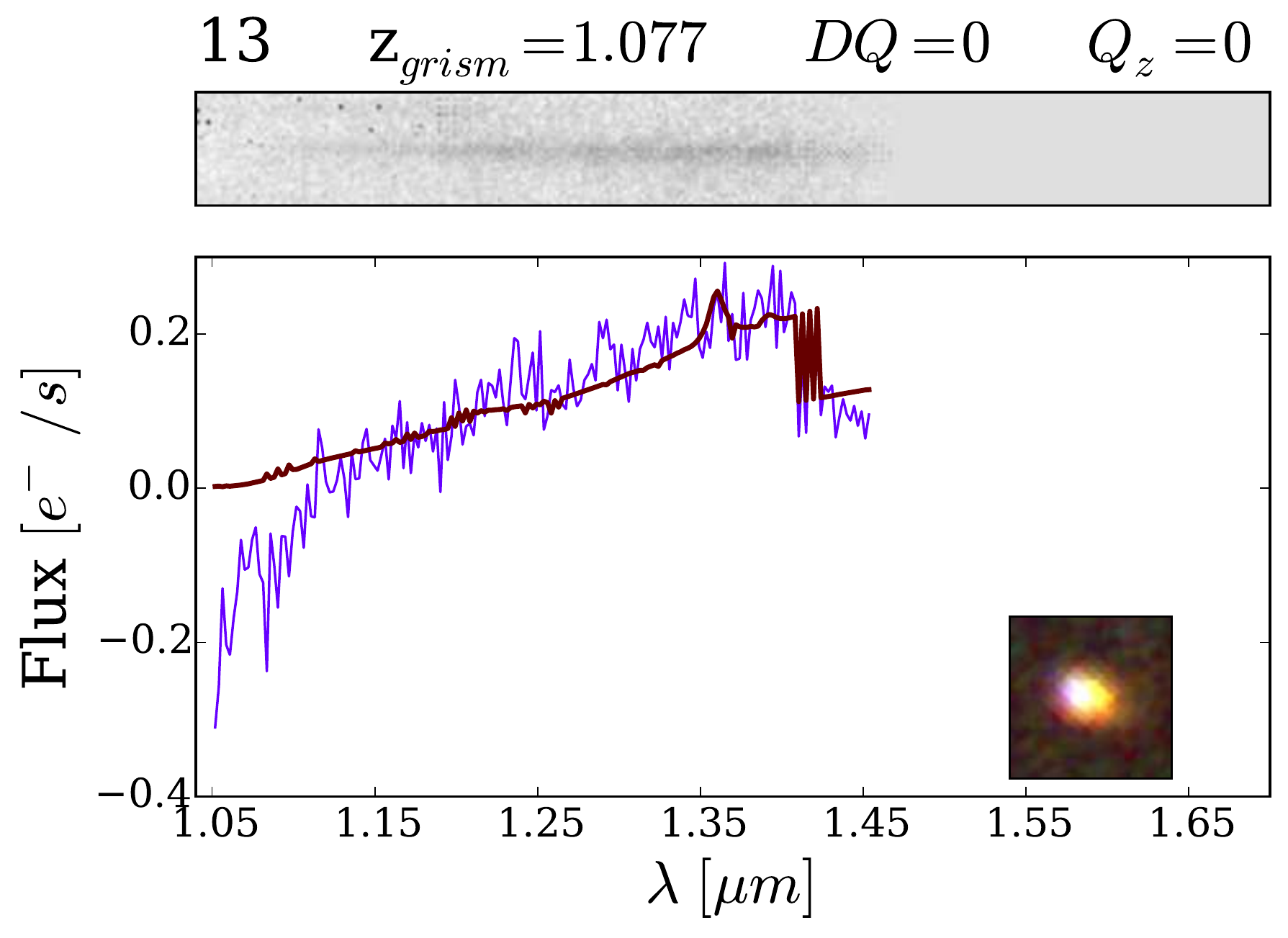} & 
\includegraphics[width = 0.33 \textwidth]{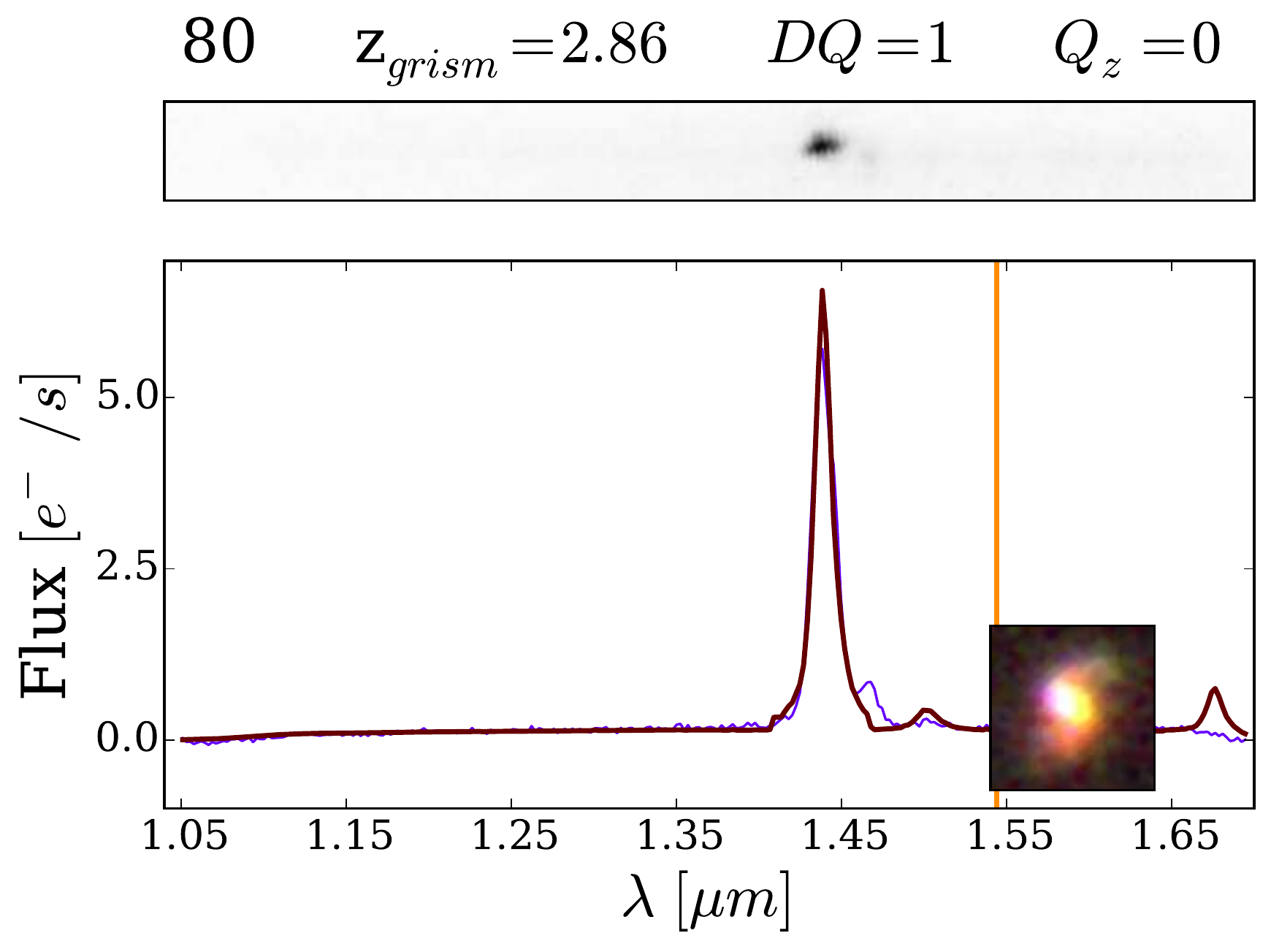} & 
\includegraphics[width = 0.33 \textwidth]{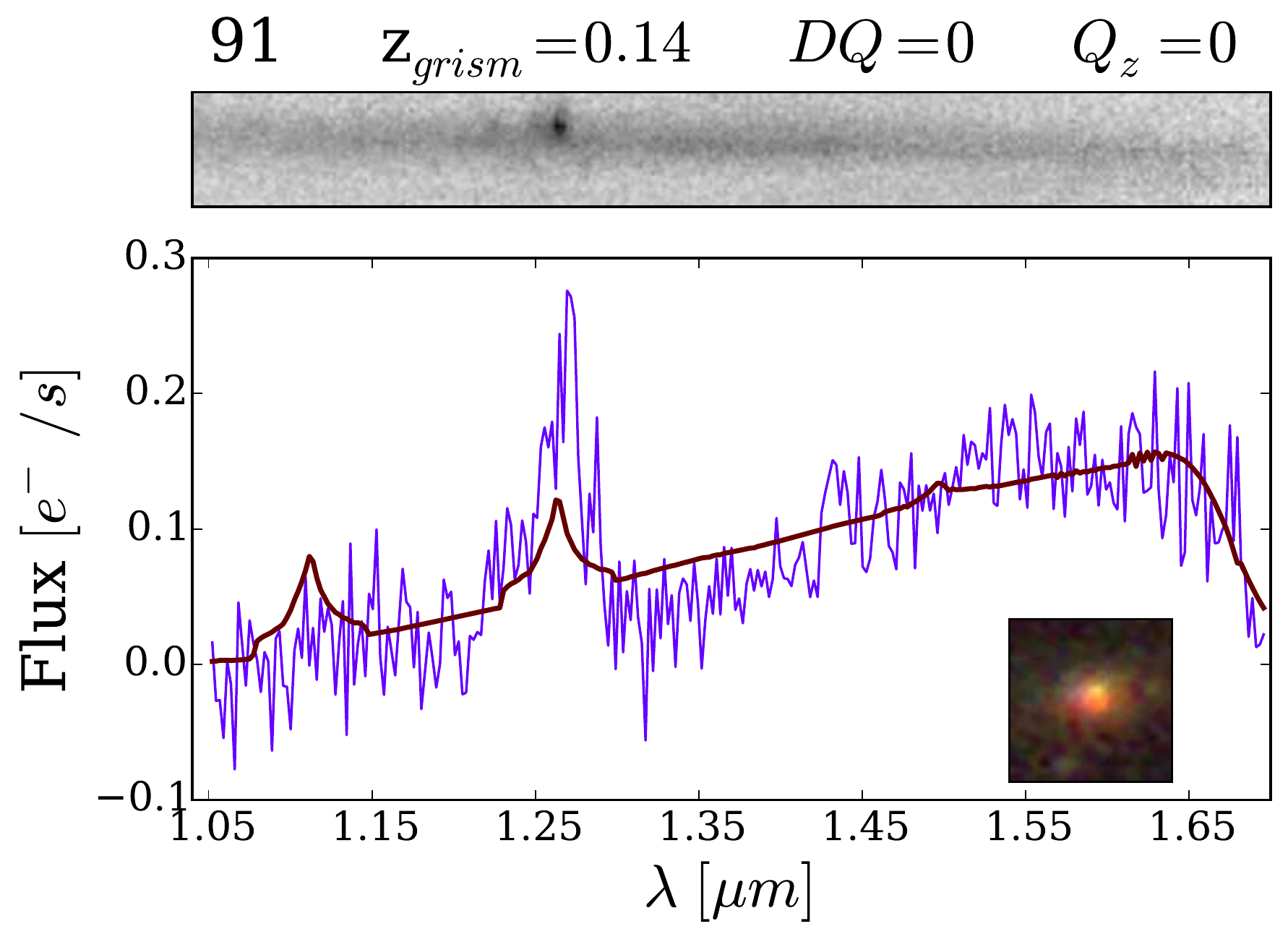} \\
\includegraphics[width = 0.33 \textwidth]{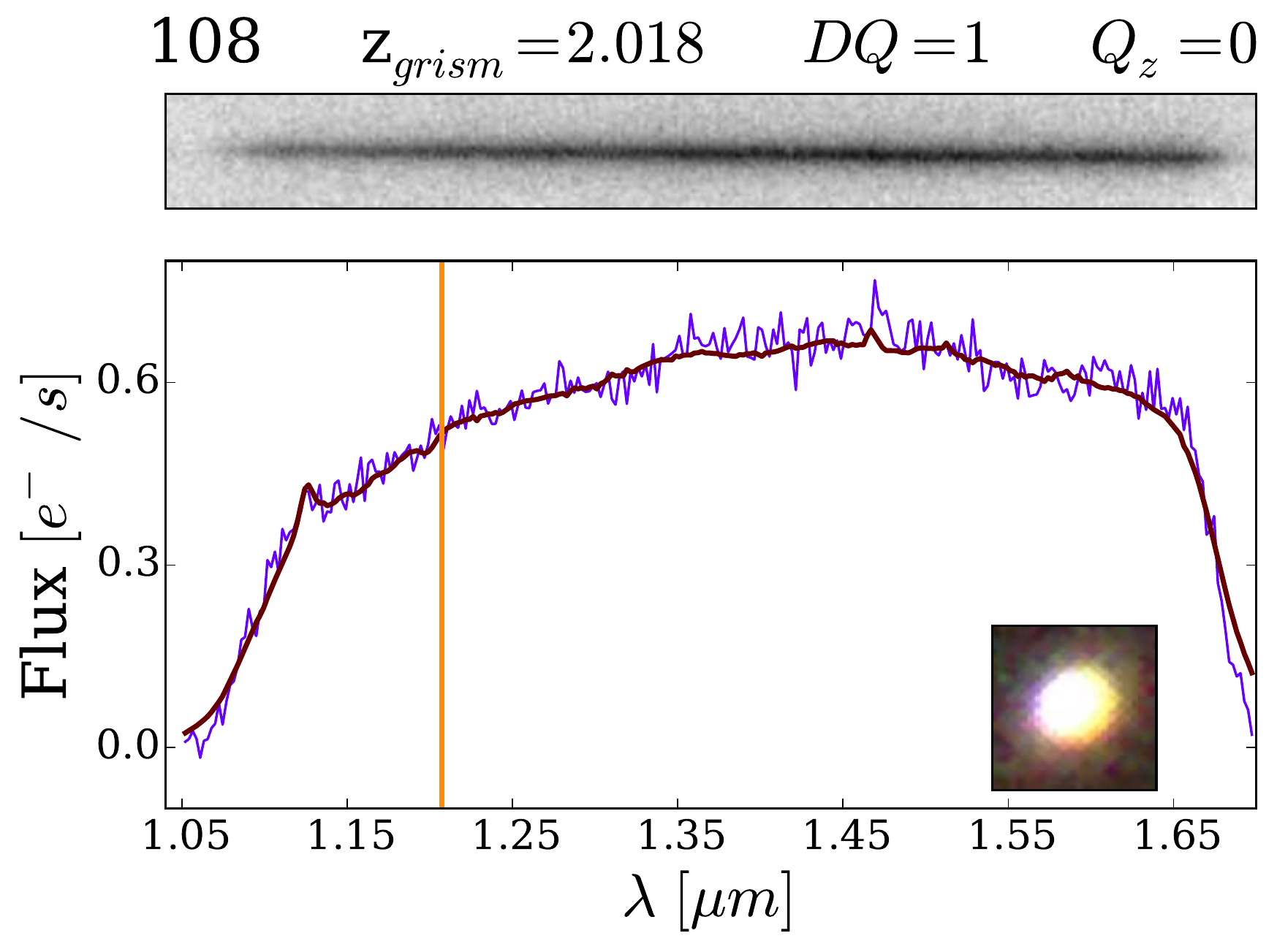} & 
\includegraphics[width = 0.33 \textwidth]{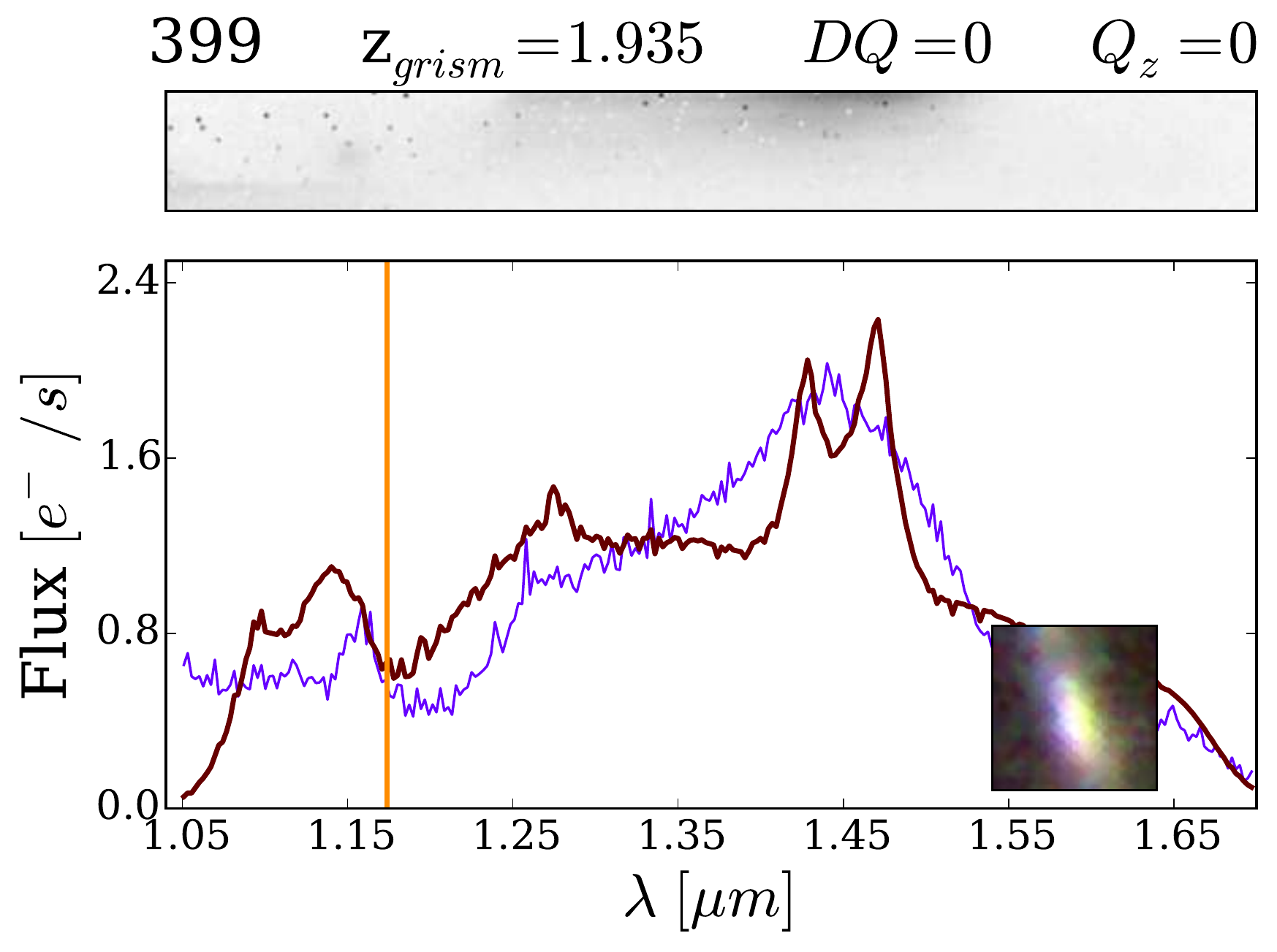} & 
\includegraphics[width = 0.33 \textwidth]{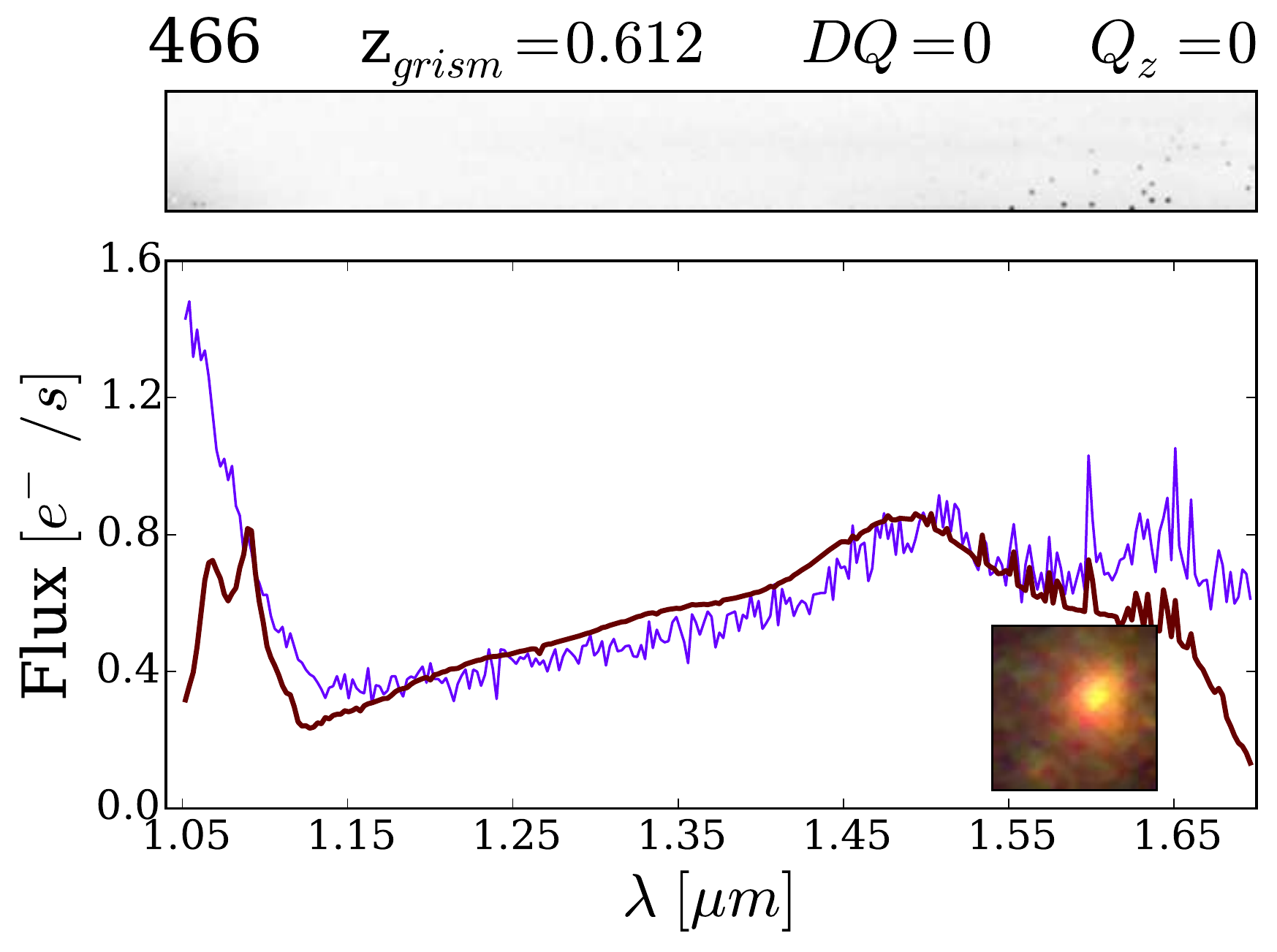} \\
\includegraphics[width = 0.33 \textwidth]{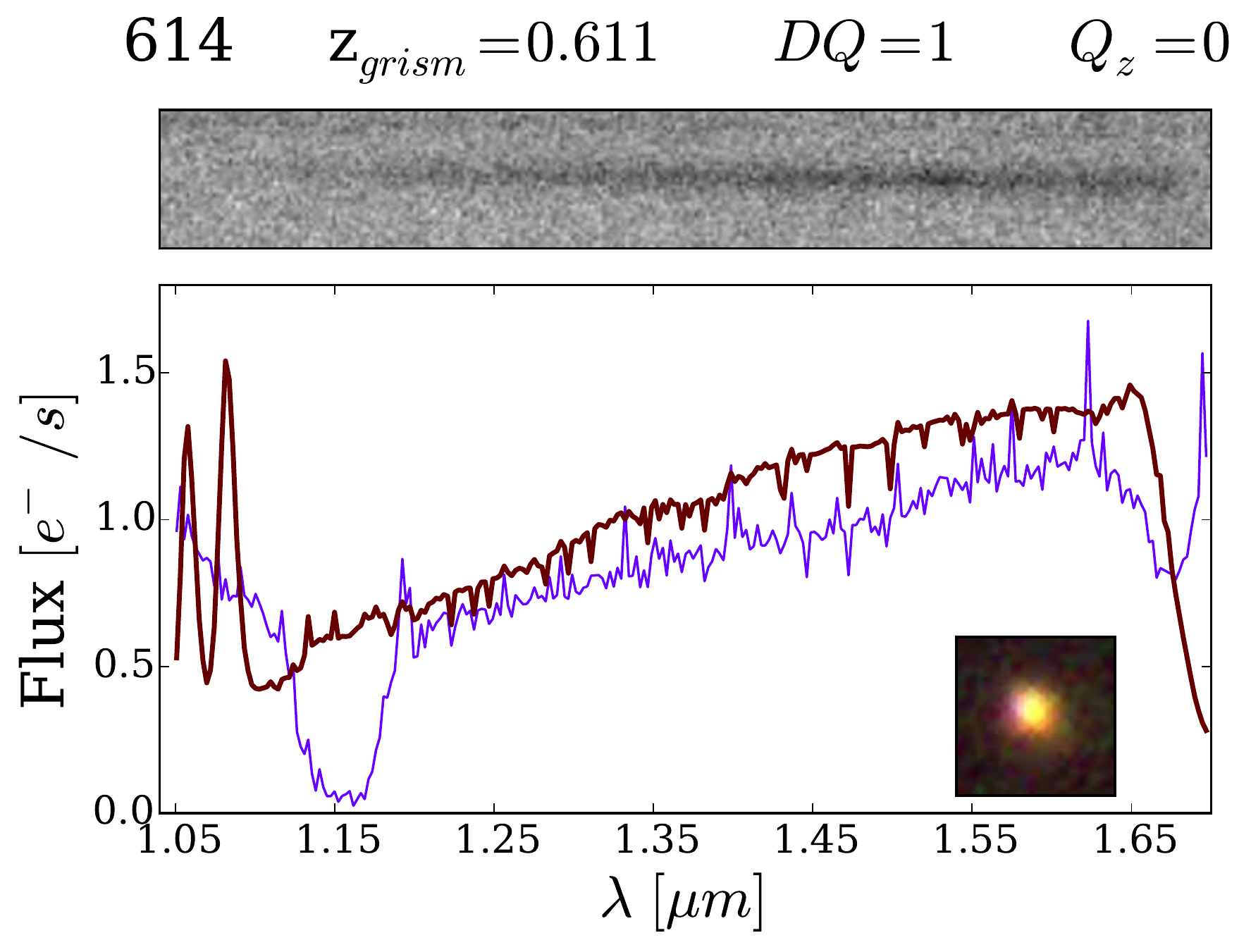} & 
\includegraphics[width = 0.33 \textwidth]{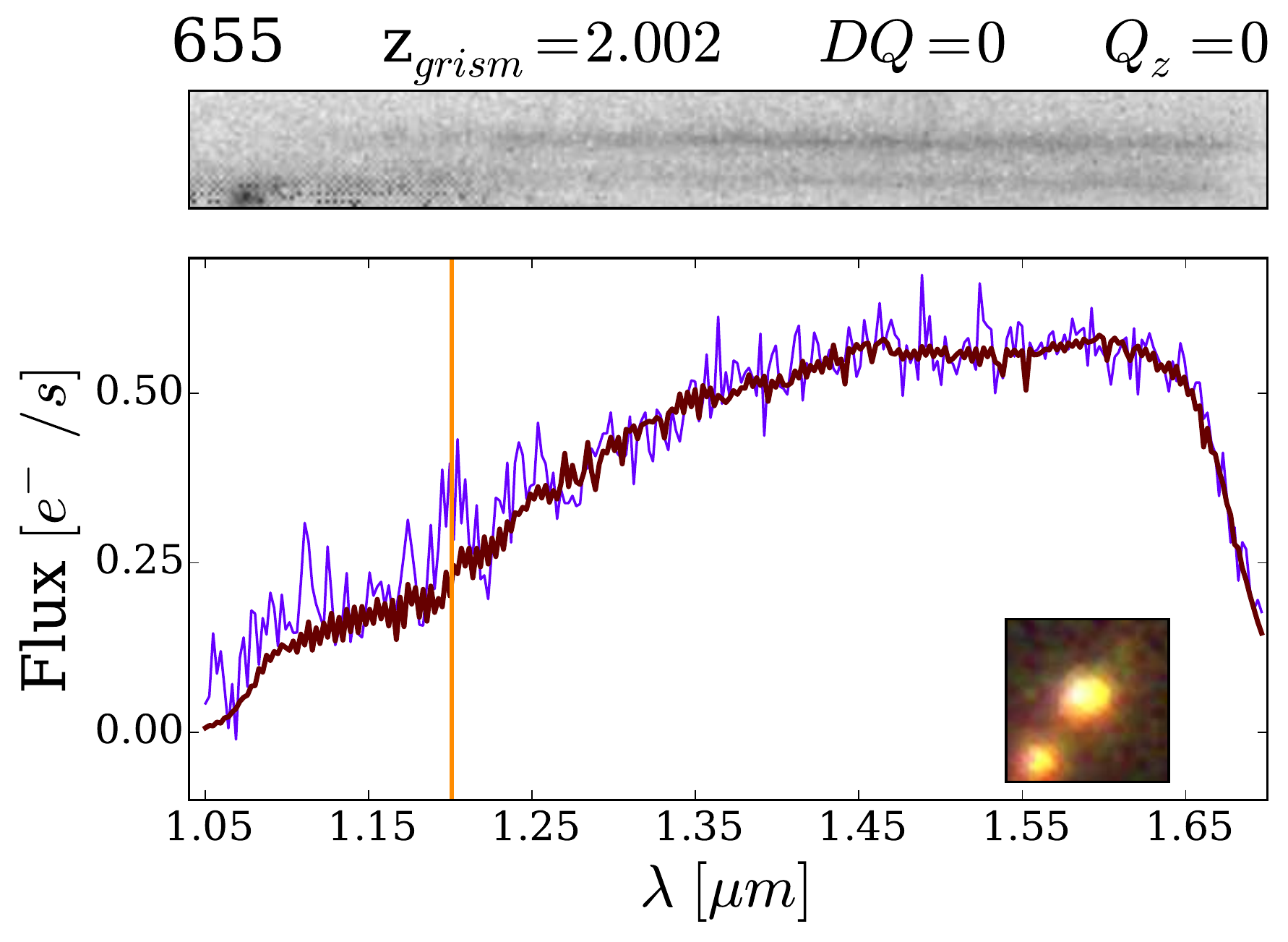} & 
\end{array}$
\caption{2D and 1D grism spectra for IRC-B candidate cluster members selected using method III (see \S \ref{sec:III}). The red curve represents the best-fit of the 1D grism spectra. The orange line denotes the redshifted 4000\AA\ line. Galaxies are ordered according to the value of the redshift quality flag (\qz) value, with \qz = 3 indicating the most robust grism measurement and \qz = 1 or 0 being the least reliable. Note that some objects shown here may not have been used in the calculation of the pair fraction, but are included for completeness (see \S \ref{sec:III}).}
\end{center}
\label{gris_pan_B3}
\end{figure}

\end{document}